\documentclass[12pt]{article}
\usepackage{epsfig}
\usepackage{graphicx}
\usepackage{latexsym}
\usepackage{ifpdf}
\ifpdf
\usepackage{textcomp}
\usepackage{amssymb}

\usepackage{amsfonts,amsthm,amstext,amscd}
\usepackage{slashed}
\usepackage{amsfonts,amssymb,amsthm,amstext,amscd}

\usepackage{amsmath}
\usepackage{ifthen}
\usepackage[]{inputenc}
\usepackage{subfigure}
\usepackage{fancyhdr}
\usepackage{enumerate}
\usepackage{hyperref}
\usepackage{mathrsfs}
\usepackage{float}
\usepackage{simplewick}
\usepackage{cancel}
\usepackage{pdfpages}
\usepackage{color}
\usepackage{feynmp}

\usepackage{dsfont}

\definecolor{grayt}{rgb}{.95,.95,.95}

\newcommand{\beqa}{\begin{eqnarray}}
\newcommand{\eeqa}{\end{eqnarray}}

\def\tr{\,{\rm Tr}}

\newcommand{\be}{\begin{equation}}
\newcommand{\ee}{\end{equation}}
\newcommand{\beq}{\begin{equation}}
\newcommand{\eeq}{\end{equation}}
\newcommand{\bea}{\begin{eqnarray}}
\newcommand{\eea}{\end{eqnarray}}

\newcommand{\bear}{\begin{eqnarray}}
\newcommand{\eear}{\end{eqnarray}}

\newcommand{\unit}{\mathds{1}}
\newcommand{\Tr}{\mathrm{Tr}\,}
\newcommand{\Str}{\mathrm{STr}\,}

\newcommand{\bvec}[1]{{\boldsymbol{#1}}}
\newcommand{\mvec}[1]{\vec{#1}}

\newcommand{\hodge}{\!\ ^\star}

\theoremstyle{plain}

\theoremstyle{plain}

{\mdfsetup{linecolor=black!30,linewidth=1.3pt,%
backgroundcolor=white,rightline=false,topline=false,%
bottomline=true,usetwoside=false,leftmargin=.3cm,%
skipbelow=.5cm,roundcorner=10pt}%
\begin{mdframed}\begin{footnotesize}}
{\end{footnotesize}\end{mdframed}\par }

\newcommand{\figurenamev}{Figure\;}  

\newcommand{\tablenamev}{Table\;}

\newcommand{\U}{\mathcal{U}}

\begin{document}
\baselineskip=15.5pt
\pagestyle{plain}
\setcounter{page}{1}

\def\r{\rho}
\def\CC{{\mathchoice
{\rm C\mkern-8mu\vrule height1.45ex depth-.05ex
width.05em\mkern9mu\kern-.05em}
{\rm C\mkern-8mu\vrule height1.45ex depth-.05ex
width.05em\mkern9mu\kern-.05em}
{\rm C\mkern-8mu\vrule height1ex depth-.07ex
width.035em\mkern9mu\kern-.035em}
{\rm C\mkern-8mu\vrule height.65ex depth-.1ex
width.025em\mkern8mu\kern-.025em}}}

\newfont{\namefont}{cmr10}
\newfont{\addfont}{cmti7 scaled 1440}
\newfont{\boldmathfont}{cmbx10}
\newfont{\headfontb}{cmbx10 scaled 1728}
\renewcommand{\theequation}{{\rm\thesection.\arabic{equation}}}

\vspace{1cm}

\begin{center}
{\huge{\bf Theta dependence in\\ Holographic QCD}}
\end{center}

\vskip 10pt

\begin{center}
{\large Lorenzo Bartolini$^{a}$, Francesco Bigazzi$^{a,b}$, Stefano Bolognesi$^{b}$, \\ Aldo L. Cotrone$^{a}$ and Andrea Manenti$^{b,c}$}
\end{center}

\vskip 10pt
\begin{center}
\textit{$^a$ Dipartimento di Fisica e Astronomia, Universit\`a di
Firenze and INFN, Sezione di Firenze; Via G. Sansone 1, I-50019 Sesto Fiorentino (Firenze), Italy.}\\
\textit{$^b$ Dipartimento di Fisica ``E. Fermi", Universit\'a di Pisa and INFN, Sezione di Pisa; Largo B. Pontecorvo 3, I-56127 Pisa, Italy.}\\
\textit{$^c$ Institute of Physics, EPFL; Rte de la Sorge, BSP 728, CH-1015 Lausanne, Switzerland.}\\

\vskip 10pt
{\small lorenzobartolini89@gmail.com, bigazzi@fi.infn.it, stefanobolo@gmail.com, cotrone@fi.infn.it, andrea.manenti@epfl.ch}
\end{center}

\vspace{25pt}

\begin{center}
 \textbf{Abstract}
\end{center}
  
\noindent
We study the effects of the CP-breaking topological $\theta$-term in the large $N_c$ QCD model by Witten, Sakai and Sugimoto with $N_f$ degenerate light flavors. We first compute the ground state energy density, the topological susceptibility and the masses of the lowest lying mesons, finding agreement with expectations from the QCD chiral effective action. Then, focusing on the $N_f=2$ case, we consider the baryonic sector and determine, to leading order in the small $\theta$ regime, the related holographic instantonic soliton solutions. We find that while the baryon spectrum does not receive ${\cal O}(\theta)$ corrections, this is not the case for observables like the electromagnetic form factor of the nucleons. In particular, it exhibits a dipole term, which turns out to be vector-meson dominated. The resulting neutron electric dipole moment, which is exactly the opposite as that of the proton, is of the same order of magnitude of previous estimates in the literature. Finally, we compute the CP-violating pion-nucleon coupling constant ${\bar g}_{\pi N N}$, finding that it is zero to leading order in the large $N_c$ limit.
\newpage

\tableofcontents

\section{Introduction}
In the electroweak sector of the Standard Model, parity (P), time reversal (T) and charge conjugation (C) can be separately broken, while their combination (CPT) is preserved. Whether some of these discrete symmetries are separately broken also in QCD remains to be experimentally verified. Instantons in the model naturally induce a P- and T-violating topological term proportional to $\theta\tr F\wedge F$, where $F$ is the $SU(3)$ field strength and $\theta$ is a parameter. In principle, nothing forbids $\theta$ from taking a generic value. However, experiments tell us that it should be extremely small. The strongest bound on its value comes from measurements of the neutron electric dipole moment (NEDM) $d_n$. Recent experiments \cite{pendlebury,NEDMexp} give $|d_n|\le 2.9\times10^{-26} e\cdot \mathrm{cm}\, (90\%\, \rm{CL})$. The topological $\theta$ angle in QCD could provide the main contribution to the NEDM, since CP-violating effects from the electroweak sector give rise to a dipole moment which is orders 
of magnitude 
smaller than the above mentioned experimental bound. A tentative order-of-magnitude theoretical estimate \cite{baluni,weinberg} gives $|d_n|\approx |\theta| e\, m_{\pi}^2 M_N^{-3}\approx 10^{-16} \, |\theta| e\cdot \mathrm{cm}$ where $m_{\pi}$ (resp. $M_N$) is the pion (resp. nucleon) mass. Put together with the above mentioned experimental bound, this gives an unnaturally small value $|\theta|\le 10^{-10}$ for the topological parameter. This is the so called strong CP problem, a possible theoretical resolution of which (a $\theta$ angle relaxing to zero dynamically) is provided by the Peccei-Quinn mechanism \cite{PecceiQuinn} which would imply the existence of axions \cite{axions}.

From a theoretical perspective, studying how the $\theta$ parameter affects the physics of QCD requires going beyond perturbation theory. 
Lattice techniques find some limitations in this case, since the topological term is imaginary in the Euclidean Lagrangian and a sign problem arises. 
While relevant results have been obtained expanding, up to few terms, around $\theta=0$ in the pure Yang-Mills case (see e.g. \cite{VicariQCD} for a detailed review on the subject), lattice estimates of CP-breaking observables in full QCD, notably estimates of the NEDM (see e.g. \cite{NEDMLattice}), are still plagued by quite large systematic and statistical errors.

In this perspective it is important to compare lattice results with model calculations. 
Famous results arise within chiral perturbation theory, where both the $\theta$-dependent ground state energy density \cite{WittenLargeNChi} and the NEDM - which turns out to be proportional to the non-derivative CP-violating pion-nucleon coupling ${\bar g}_{\pi\,N\,N}$ \cite{CdVVW-NEDM} - have been computed. 
Within this approach only the pion cloud contributes to the NEDM, since massive (axial) vector mesons have been integrated out. 

Another model approach, complementary to the one above, consists in taking 't Hooft's large $N_c$ limit where $N_c$ is the number of colors. 
This limit is known not to commute in general with the small quark mass one in which chiral perturbation theory is organized. 
In the unflavored Yang-Mills case, relevant features of the $\theta$-dependent ground state energy density have been first discussed in \cite{WittenLargeNChi} and then explicitly realized, to leading order in $\theta/N_c$, in a holographic Yang-Mills model in \cite{Wittentheta}.\footnote{An extension of these results to any order in $\theta/N_c$ and an analysis of the $\theta$-dependent behavior of various relevant Yang-Mills observables can be found in \cite{wymtheta}.}

When $N_c=\infty$ mesons (and glueballs) are non-interacting and stable. At large, finite $N_c$, meson-meson couplings are found to be of order $1/\sqrt{N_c}$, while baryon masses scale as $N_c$. 
This suggests that baryons can be seen as solitons in the effective large $N_c$ mesonic Lagrangian \cite{Witten1NBaryons}. 
This picture is actually realized within the chiral effective theory (the Skyrme model \cite{skyrme}), whose solitons are identified with the baryons.  
Static properties of nucleons with $N_f=2$ massless (resp. massive) flavors have been studied in the seminal paper \cite{ANWSkyrme} (resp. \cite{ANWMassSkyrme}). 
In this context the NEDM has been computed both with $N_f=2+1$ massive flavors \cite{Dixon} and in the $N_f=2$ mass degenerate case \cite{Salomonson}. 
Differently from the chiral Lagrangian approach, in the Skyrme model virtual pion contributions to the NEDM are subleading in $1/N_c$. This could be related to the large $N_c$ scaling of the CP-breaking pion-nucleon coupling. A first estimate 
gave ${\bar g}_{\pi N N}\sim N_c^{1/2}$ \cite{schnitzer}, but a more careful analysis \cite{gpiNNSkyrme} suggested a neatly different scaling, ${\bar g}_{\pi N N}\sim N_c^x$, with $x\le - 1/2$.  A complementary check of the latter suggestion is clearly an interesting issue. 

Both the chiral Lagrangian and the Skyrme approach miss the effects induced by the whole massive (axial) vector meson tower. 
To overcome this and other limitations of the effective approach, we consider the large $N_c$ QCD model by Witten, Sakai and Sugimoto (WSS) \cite{witten,SS1} where the $\theta$-dependence can be studied from first principle computations using the holographic correspondence. 
The WSS model is a non-supersymmetric $SU(N_c)$ gauge theory in $3+1$ dimensions, coupled to $N_f$ quarks and a tower of massive (Kaluza-Klein) matter fields transforming in the adjoint representation of $SU(N_c)$. 
In the regime where a classical dual gravity description is available, these massive fields cannot be decoupled and the UV behavior of the model neatly departs from that of real QCD. 
Despite this limitation, the WSS model exhibits, at low energy, all the crucial features, like confinement, chiral symmetry breaking and the formation of a mass gap, which appear in QCD. The WSS model provides analytic control on, 
as well as simple geometrical descriptions of, these highly non-trivial non perturbative effects. It remarkably contains, within a unique framework, different effective QCD models which have been built to describe specific sectors of the theory. This unifying perspective allows, at least qualitatively, to go far beyond the limits of the various effective descriptions. 

In the original version of the model the quarks are massless. 
In this case, as it is expected from field theory, any $\theta$-dependence is washed out by a chiral rotation of the quarks. 
We will discuss in the following how this is realized in the holographic model. 
A (small) mass term for the quarks can be introduced using a prescription suggested in \cite{AK,Hashimotomass}. 
We adopt that prescription and compute the ground state energy density of the model as well as the topological susceptibility and the pion and $\eta'$ masses as a function of $\theta$, finding agreement with the chiral Lagrangian results (for a recent holographic derivation of these observables in a bottom-up model, see \cite{kirtheta}).
Then we focus on the baryonic sector. 
Just like baryons in the large $N_c$ limit can be seen as solitons of the chiral Lagrangian, in the WSS model they are identified with instantons of the holographic Lagrangian describing the mesonic sector \cite{SS-barioni,BaryonST}.

We compute the $\theta$-corrected holographic instanton solutions, focusing on the $N_f=2$ case, finding that the baryon spectrum does not get corrections to first order in $\theta$. 
Currents are instead sensitive to CP-breaking effects. 
In particular, the dipole term in the nucleon electromagnetic form factor turns out to be different from zero and, as already pointed out in \cite{nedmshort}, exhibits complete vector meson dominance. 
We present a review and a detailed analysis of the computation of the NEDM reported in \cite{nedmshort}, complementing it with a novel study of the full momentum dependence of the dipole form factor. 
Finally, we compute the CP-breaking pion-nucleon coupling ${\bar g}_{\pi N N}$ finding that it is zero to leading order in the large $N_c$ limit. 

The paper is organized as follows. In Section \ref{WSSreview} we review the main features of the original WSS model with massless flavors. 
In Section \ref{anomaly} we recall how the $U(1)$ axial anomaly and the chiral effects on the $\theta$ term are realized in the model.
We also discuss a Horava-Witten-like solution of the anomalous Bianchi identities involved in the gravitational description of these effects.
In Section \ref{massfer} we review the inclusion of the flavor mass term and in Section \ref{masstheta} we discuss how it affects the $\theta$-dependent vacuum.
In Section \ref{holobar} we focus on the holographic description of baryons in the WSS model. 
After reviewing the soliton solution describing baryons and its quantization, we compute the shift in the baryonic Hamiltonian due to the $\theta$ angle and the flavor mass term, discovering that while the shift is of leading order in $m_{\pi}$, it is subleading (${\cal O}(\theta^2)$) in $\theta$. 
In Section \ref{sec:newsol} we compute the leading order corrections (in $\theta$ and in the quark masses) to the instantonic solutions describing baryons.
In Section \ref{nedmsec} we review and discuss how, focusing on the nucleon electromagnetic form factors, these novel solutions can be used to compute the neutron electric dipole moment. Moreover we present a novel analysis of the full electromagnetic dipole form factor.
Finally, in Section \ref{gpNNsec} we focus on the CP-violating pion-nucleon coupling.  We collect some further technical comments in the Appendices.  
\subsubsection*{Conventions}
Throughout this paper we use the conventions in \cite{SS1} for the RR forms, scaling them with respect to the standard notation as
\begin{equation}
 C_{p+1} \rightarrow \frac{k_{0}^2 \tau_{6-p}}{\pi} C_{p+1} \,,
\end{equation}
where $2k_{0}^2=(2\pi)^7 l_s^8$ gives the ten dimensional Newton's constant, $\tau_p = (2\pi)^{-p} l_s^{-(p+1)}$ is proportional to the Dp-brane tension and $l_s\equiv\sqrt{\alpha'}$ is the string length. 
\section{Witten-Sakai-Sugimoto model}
\label{WSSreview}
\setcounter{equation}{0}
The WSS model is based on a D-brane setup in type IIA string theory.  It consists of $N_c\gg1$ D4-branes wrapped on a circle $S_{x_4}$ \cite{witten} and $N_f$ D8$-{\bar {\rm D}}$8-branes placed at fixed antipodal points on the circle \cite{SS1}. Along the circle, of length $2\pi M_{KK}^{-1}$, fermions obey anti-periodic boundary conditions. In such a way, at energies $E\ll M_{KK}$ the original $(4+1)$-dimensional theory on the D4-D8 brane intersection, reduces to pure non-supersymmetric $SU(N_c)$ Yang-Mills in $3+1$ dimensions coupled to $N_f$ massless quarks. Other matter fields, transforming in the adjoint representation, get masses of the order of $M_{KK}$.  

The holographic dual description of the above large $N_c$ QCD model simplifies if the quarks are treated in the quenched approximation and (unfortunately) if the spurious adjoint matter fields are not decoupled. 
In this case, the dual picture is provided by a classical gravity background sourced by the wrapped D4-branes and probed (without backreaction) by the D8-branes.
\subsection{The background}
The relevant type IIA gravity action, in string frame, reads 
\be
S = \frac{1}{2k_0^2}\int d^{10}x\sqrt{-g}\left[e^{-2\phi}\left({\cal R}+4(\partial\phi)^2\right) -\frac12 (2\pi)^4 l_s^6 |F_4|^2 -\frac12 l_s^2 |F_2|^2\right]\,.
\label{action1}
\ee 
Here $F_4 = dC_3$ is the RR four-form which is magnetically sourced by the $N_c$ D4-branes, $\phi$ is the dilaton and $F_2=dC_1$ is the RR two-form which, as we will review in a moment, accounts for the topological $\theta$ term in the dual field theory. Neglecting its backreaction on the background amounts on working at small $\theta/N_c$ and getting only the leading order corrections in this parameter \cite{Wittentheta}. In this  paper we will work in this approximation.\footnote{See \cite{wymtheta} and references therein for a study of the physics of the model in the fully backreacted case.} 

Treating the $F_2$ form as a probe, the background has the following features \cite{witten}. The string frame metric reads
\be
\label{metricu}
ds_{10}^2 =\left(\frac{U}{R}\right)^{3/2}\left[dx_{\mu}dx^{\mu} + f(U) dx_4^2\right]+ \left(\frac{R}{U}\right)^{3/2}\left[\frac{dU^2}{f(U)} + U^2 d\Omega_4^2\right]\,,
\ee
where
\be
f(U) = 1-\frac{U_{KK}^3}{U^3}\,.
\ee
The dilaton and the four-form field strength are given by
\be\label{dilatonu}
e^{\phi} = g_s \left(\frac{U}{R}\right)^{3/4}\,,\qquad F_4 = \frac{3 R^3}{(2\pi)^2 l_s^3} \omega_4\,,
\ee
with the flux quantization condition fixing the value of $R$ as
\be
\int_{S^4} F_4 = 2\pi g_s N_c\,, \quad R= (\pi g_s N_c)^{1/3} l_s\,.
\label{f42}
\ee 
In the formulae above, $\mu = 0,1,2,3$ are the $1+3$ Minkowski directions where the Yang-Mills theory is defined, $d\Omega_4^2$ is the metric of a four-sphere $S^4$ of radius one, $U$ is the transverse radial coordinate $U\in [U_{KK},\infty)$, $x_4$ is the compact coordinate of length $2\pi M^{-1}_{KK}$ and $R$ is a curvature radius. Moreover, $g_s$ is the string coupling and $\omega_4$ is the volume form of the transverse $S^4$, of volume $V_{S^4}=8\pi^2/3$. The isometry group of $S^4$ is mapped into a global $SO(5)$ symmetry group in the dual field theory, which acts non-trivially on the adjoint Kaluza-Klein massive modes (signaling that these are, in fact, not decoupled in the limit we are considering). 

The $S_{x_4}$ circle shrinks to zero size when $U=U_{KK}$. Absence of conical singularities at $U=U_{KK}$ is guaranteed if the coordinate $x_4$ has period\begin{equation}
\delta x_4 = \frac{4\pi}{3}\frac{R^{3/2}}{U^{1/2}_{\mathrm{KK}}} = \frac{2\pi}{M_{\mathrm{KK}}}\,.\label{deftau}
\end{equation}
The resulting $(U,x_4)$ subspace has a cigar-like shape. Most of the relevant physics in the model is captured by this geometry. Regularity and the property $g_{00}(U_{KK})\neq0$ imply confinement and the formation of a mass gap in the dual field theory \cite{witten}.  

The Yang-Mills theory dual to the above background has two distinct mass scales: the Kaluza-Klein scale $M_{KK}$ (which is also the glueball mass scale) and the string tension $T_s$.
Their ratio is determined by the parameter $\lambda\sim T_s/M^2_{KK}$. Reliability of the background requires $\lambda\gg1$: this is a further indication that the spurious KK modes cannot be decoupled when the dual description sticks in the classical gravity regime.
Reliability of the background also requires $e^{\phi}$ to be small: when this condition is violated (namely, at large $U$) we should better make use of the eleven dimensional (``M-theory") completion of the model, which is an asymptotically $AdS_7\times S^4$ solution of eleven dimensional supergravity \cite{witten}.

The UV 't Hooft coupling and the Yang-Mills $\theta$ angle can be related to the gravity parameters by considering the low energy limit of the D4-brane action 
\be
S_{\rm D4} = - \tau_4 {\rm Tr} \int d^4x\, dx_4 e^{-\phi}\sqrt{-\det (G+ {\cal F})} +  \int C_5 + \frac{1}{2}\tau_4 l_s \int C_1 \wedge {\rm Tr} {\cal F}\wedge {\cal F}\,,
\ee 
where ${\cal F_{\alpha\beta}}\equiv 2\pi\alpha' F_{\alpha\beta}$ is proportional to the gauge field strength, $C_5$ is the electric five-form sourced by the branes (its field strength $F_6$ is the Hodge dual to $F_4$) and $G_{\alpha\beta}$ is the induced metric on the world-volume. 
Expanding the action to second order in derivatives, considering the UV asymptotics $U\rightarrow\infty$ and integrating over the compact $x_4$ direction one gets the Yang-Mills Lagrangian
\be
{\cal L} = \frac{N_c}{4\lambda}\left[-{\rm Tr} F^2 + \frac{\lambda}{2\pi^2}\frac{\theta}{N_c}{\rm Tr}F\wedge F\right]\,,
\label{lagthetaE}
\ee
where\footnote{The parameters $g_{YM}$ and $\lambda$ are conventionally referred to as the (UV) four dimensional gauge and 't Hooft couplings of the model, despite the fact that they differ by a factor of $2$ from the standard ones.}
\be
\lambda\equiv g_{YM}^2N_c = 2\pi g_s N_c l_s M_{KK}\,,\qquad  \theta + 2\pi k=\int_{S_{x_4}} C_1\,,
\label{hd}
\ee
and $k$ is an integer. The second relation in (\ref{hd}) defines $\theta$ mod $2\pi$ integer shifts (since the integral of $C_1$ is gauge invariant only modulo $2\pi{\mathbb Z}$). 

Solving the equation of motion for $C_1$, treated as an external field on the type IIA background given above, one finds, imposing (\ref{hd}), that
\be
C_1 = \frac{\Theta}{l_s g_s} f(U) dx_4\,,\quad {\rm where}\,\quad \Theta\equiv\frac{\lambda}{4\pi^2}\left(\frac{\theta+2\pi k}{N_c}\right)\,.
\ee
Since this parameter depends on $k$, what we actually get on the gravity side is an infinite family of solutions corresponding to possible field theory vacua. This behavior precisely reflects the expected multi-branched structure \cite{WittenLargeNChi} of the $\theta$-dependent vacuum of the theory. 
Actually, following standard holographic rules \cite{Wittentheta}, the field theory ground-state energy density $f(\theta)$ (related to the on-shell renormalized gravity action) reads, to leading order in $\Theta\ll1$,
\be
f(\Theta)= -\frac{2 N_c^2 \lambda}{3^7 \pi^2}M^4_{KK}\left(1-3\Theta^2\right)\,.
\label{en0}
\ee
Since $\Theta$ is proportional to $\theta+2k\pi$, for a given value of $\theta$ the true vacuum energy is obtained by minimizing the previous expression over $k$
\be
f(\theta) = {\rm min}_k f(\Theta)\,.
\ee
As a result, the ground state energy density turns out to be a periodic function of $\theta$, as expected \cite{WittenLargeNChi}. To any given interval, of length $2\pi$, of possible values of $\theta$, it corresponds a precise value of $k$. For example, $k=0$ when $\theta\in(-\pi,\pi)$ and so on. 

Notice that the probe approximation for $C_1$ requires that, in the $k=0$ branch,
\be
\Theta = \frac{\lambda}{4\pi^2}\frac{\theta}{N_c}\ll 1\,.
\ee
This is actually one of the limits we will work with.

When $\theta\ll1$ the energy density reads
\be
f(\theta) - f(0) 
=\frac12\chi_g \theta^2\left[1+ {\cal O}(\theta^2)\right]\,,
\label{fthetau}
\ee
with the topological susceptibility given by \cite{Wittentheta}
\be
\chi_g = \frac{\lambda^3 M_{KK}^4}{4(3\pi)^6}\,.
\label{tsusc}
\ee
See \cite{wymtheta} for an exact-in-$\Theta$ analysis of the ground state energy density and many other physical observables. 

To simplify the formulae it is sometimes convenient to set $M_{\mathrm{KK}}=1$ working in the following units 
\begin{equation}
R^3 = \frac{9}{4}\;,\quad U_{\mathrm{KK}} = M_{\mathrm{KK}} =1\;,\quad g_s\,l_s = \frac{g_{\mathrm{YM}}^2}{2\pi}\,. \label{adscftdict}
\end{equation}
\subsection{Adding probe flavor branes}
Treating the $N_f$ flavor D8-branes as probes on the background requires taking (see e.g. \cite{smearedSS})
\be\label{epsilonf}
\epsilon_f \equiv\frac{1}{12\pi^3}\lambda^2\frac{N_f}{N_c}\ll1\,,
\ee
and neglecting ${\cal O}(\epsilon_f)$ corrections on the background fields. This is another limit in which we will work.\footnote{See \cite{smearedSS} for an account of the ${\cal O}(\epsilon_f)$ corrections.} In the probe approximation, the background metric, dilaton and four-form RR field strength will be kept fixed as in (\ref{metricu}), (\ref{dilatonu}) so that the equations of motion to be solved for, arise from a string frame action of the form
\begin{equation}
\label{acti}
\begin{aligned}
S &= -\frac{1}{4\pi}\sum_{p\;\mathrm{odd}}(2\pi l_s)^{2(p-4)}\int \,F_{p+1}\wedge\hodge F_{p+1} + \int_{\rm D8} \sum_{k=1}^4 C_{9-2k} \wedge \frac{1}{k! (2\pi)^k}\Tr \mathcal{F}^{ k}\, + 
\\ & \quad -\frac{1}{(2\pi)^8 l_s^9}\int_{\rm D8} d^{9}\xi\,e^{-\phi}\Str\sqrt{|\det(\mathcal{P}[g] + 2\pi l_s^2 \mathcal{F})|}\,.
\end{aligned} 
 \end{equation}
The $F_{p+1}$ are the RR field strengths of the bulk $C_{p}$ forms while the $\mathcal{F}$ are the $U(N_f)$ field strengths of the gauge fields living on the D8-branes, $\mathcal{F} = d \mathcal{A} + i\mathcal{A}\wedge \mathcal{A}$. 
Powers of differential forms are done by means of the wedge product. The symbol $\mathcal{P}[g]$ denotes the pullback of the metric on the D8 worldvolume and the symbol ``$\Str$'' denotes the symmetrized trace on the gauge group indexes. 

The energy density for the D8-branes (corresponding to the antipodal embedding on the $S_{x_4}$ circle) is minimized by the $u$-shaped embedding $x_4(U)={\rm const}$. 
Its physical meaning is remarkable. 
The $U(N_f)\times U(N_f)$ symmetry on the antipodal D8$-{\bar {\rm D}}8$ branes, which in turn corresponds to the classical chiral symmetry group in the dual field theory, is broken to the diagonal subgroup since the two different branches actually join at the tip of the cigar in the background. 
This is how the holographic model realizes the spontaneous chiral symmetry breaking of the dual QCD-like theory.

It is often more convenient to redefine the cigar coordinates in the following way \cite{SS1}
\begin{equation}
U^3 = U_{\mathrm{KK}}^3 + U_{\mathrm{KK}} {\tilde u}^2\;,\quad \varphi = \frac{2\pi}{\delta x_4} x_4 \,,\label{defUcoord}
\end{equation}
parameterizing the $({\tilde u},\varphi)$ plane in Cartesian coordinates $(y,z)$
\begin{equation}
y=\tilde u\cos\varphi\;,\quad z=\tilde u \sin \varphi\,.
\label{yzdef}
\end{equation}
The cigar metric then reads
\begin{equation}
d s^2_{(y,z)} = \frac{4}{9}\left(\frac{R}{U}\right)^{3/2}\left[\left(1-q(\tilde u)z^2\right)d z^2 + \left(1-q(\tilde u)y^2\right)d y^2 - 2zy \, q(\tilde u)d x d y\right]\,,
\end{equation}
with $U$ given as a function of $z$ and $y$ and $q(\tilde u)$ defined by $q(\tilde u)= \frac{1}{{\tilde u}^2}\left(1-\frac{U_{\mathrm{KK}}}{U}\right)$. 

Using these coordinates, the antipodal embedding just reads $y=0$. Correspondingly, putting the $S^4$ components of the $\mathcal{F}$ field to zero, assuming that the other components do not depend on the $S^4$ angular coordinates, integrating over $S^4$ and expanding to second order in derivatives, the relevant action, from (\ref{acti}), reduces to
\begin{equation}\label{SYM}
S = -\kappa\int d^4x d z\, \left(\frac{1}{2}h(z)  \, \Tr\mathcal{F}_{\mu\nu}\mathcal{F}^{\mu\nu} +   k(z)  \Tr\mathcal{F}_{\mu z}\mathcal{F}^\mu_{\;\; z}\right)
 + \frac{N_c}{24\pi^2}\int  \omega_5(\mathcal{A}) + S_{C_7}\,,
\end{equation}
where (in units $M_{KK}=U_{KK}=1$)
\be
\kappa = \frac{N_c\lambda}{216\pi^3}\,,\quad h(z) = (1+z^2)^{-1/3}\,,\quad k(z) = (1+z^2)\,,
\ee
and 
\begin{equation}
\omega_5(\mathcal{A}) =  \Tr\left(\mathcal{A}\wedge \mathcal{F}^{ 2}- \frac{i}{2}\mathcal{A}^{ 3} \wedge\mathcal{F} - \frac{1}{10}\mathcal{A}^{ 5}\right)\;,\quad d \omega_5(\mathcal{A}) = \Tr\mathcal{F}^{  3}\,.
\label{omega5}
\end{equation}
Among all the RR forms $F_{p+1}$ in (\ref{acti}) we are keeping only $F_8$ (in $S_{C_7}$), dual to $F_2$. For a moment let us neglect the $S_{C_7}$ term (we will discuss in detail its implications in Section \ref{anomaly}) and focus on the physical meaning of the remaining part of the action (\ref{SYM}). It provides the holographic description to the mesonic sector of the model.
\subsection{Holographic mesons}\label{section:mesons}
Let us consider inserting into (\ref{SYM}) the following expansions for the gauge field 
\bea\label{expA}
&&{\mathcal A}_{z} (x^{\mu},z) = \sum_{n=0}^{\infty} \varphi^{(n)}(x^{\mu})\phi_n(z)\,,\nonumber \\ 
&&{\mathcal A}_{\mu}(x^{\mu},z) = \sum_{n=1}^{\infty} B_{\mu}^{(n)}(x^{\mu})\psi_n(z)\,.
\eea
If we choose the functions $\phi_n(z)$, $\psi_n(z)$ to form complete, suitably normalized sets, the fields $\varphi^{(n)}$ and $B_{\mu}^{(n)}$ get canonical mass and kinetic terms in four dimensions. In particular, we set
\be \label{eqforpsi}
-h(z)^{-1} \partial_z (k(z)\partial_z\psi_n(z))=\lambda_n \psi_n(z)\,, \qquad \kappa  \int dz \,h(z) \psi_n(z) \psi_m(z) = \delta_{mn}\,.
\ee
From these conditions, as we review in Appendix \ref{mesons}, it follows that the $B_{\mu}^{(n)}$ modes correspond to massive vectors (resp. axial vectors), for odd (resp. even) $n$, with masses $m_n^2 = \lambda_n M_{KK}^2$. 
For example, $B_{\mu}^{(1)}$ is identified with the $\rho$ meson and $B_{\mu}^{(2)}$ with the $a_1$ meson. 
The scalar modes $\varphi^{(n)}$ for $n\ge1$ get eaten by the $B_{\mu}^{(n)}$, while the mode $\varphi^{(0)}$ corresponds to the pion. Other massive scalar mesons are given by fluctuations of the D8-brane embedding.

Thus, a remarkable feature of the effective action (\ref{SYM}) is the fact that it includes automatically, into a unified picture, the low lying modes and the whole tower of massive mesons. 
All the parameters in the meson action are fixed in terms of the few parameters of the model, i.e. $N_c,N_f,M_{KK}$ and $\lambda$.

As we review in Appendix \ref{mesons}, the effective action for the pion precisely reduces to the chiral Lagrangian and the Skyrme model, with the pion decay constant $f_{\pi}$ and the coupling $e$ defined as
\begin{equation}
f_{\pi} = 2 \sqrt{\frac{\kappa}{\pi}}\,, \qquad e \sim -\frac{1}{2.5\kappa}\,,
\label{fpaie}
\end{equation}
and the pion matrix given by
\begin{equation}
\U(x^\mu) = \mathcal{P} \exp\left( -i \int_{-\infty}^\infty d z\, \mathcal{A}_z(x^\mu,z)\right)\,.
\label{pionm}
\end{equation} 
\section{The $U(1)_A$ anomaly and flavor effects on $\theta$}
\label{anomaly}
\setcounter{equation}{0}
In this section we describe how the presence of massless quarks in the WSS model erases any physical effect of the $\theta$ parameter. 

Before the reduction on $S^4$, the $S_{C_7}$ term in the action (\ref{SYM}) reads 
 \begin{equation}
\begin{aligned}
S_{C_7} &= -\frac{1}{4\pi}(2\pi l_s)^{6}\int d C_{7}\wedge \hodge d C_{7} + \frac{1}{2\pi}\int C_{7} \wedge \Tr\mathcal{F}\wedge \omega_y\,,
\label{sc7}
\end{aligned}
\end{equation}
where we have introduced the one-form $\omega_y = \delta(y)d y$, in order to extend the D8 integral to the whole spacetime. The equation of motion for $C_7$ reads
\begin{equation}
d \hodge d C_{7}  = \frac{1}{(2\pi l_s)^6}\Tr \mathcal{F} \wedge\delta(y) d y\,.
\label{maxwellbianchi}
\end{equation}
By using the Hodge relation\footnote{The notation $\tilde F_2$ is due to the fact that $\tilde F_2 \neq dC_1$.} 
\begin{equation}
\hodge F_8 = (2\pi l_s)^{-6} \tilde F_2 \,,
\end{equation}
we see that the equation of motion above is translated into an anomalous Bianchi identity
\be
d \tilde F_2 = \Tr \mathcal{F} \wedge\delta(y) d y\,.
\label{bianchi}
\ee
Notice that the ``anomaly" is only driven by the Abelian component of the $U(N_f)$ gauge field, i.e. the hatted field in the decomposition
\begin{equation}
\mathcal{A} = \widehat{A} \frac{\unit}{\sqrt{2N_f}} + A^a T^a\,,
\label{separ}
\end{equation}
where $T^a$ are the $SU(N_f)$ generators.

We can formally solve (\ref{bianchi}) by writing
\be
\tilde F_2 = dC_1 + \sqrt{\frac{N_f}{2}}\widehat{A} \wedge\delta(y) d y\,.
\label{intB}
\ee
Now, as it was already observed in \cite{SS1},  following the results in \cite{greenharvey}, this form is gauge invariant if we allow for the following combined gauge shifts
\begin{equation}
\delta_\Lambda dC_{1} = \sqrt{\frac{N_f}{2}}d\Lambda\wedge \delta(y)d y\;,\qquad \delta_\Lambda \widehat{A} = - d \Lambda\,.
\label{C1shift}
\end{equation}
This actually implies that when D8-branes are present, $dC_1$ is not a gauge invariant form. The correct gauge invariant combination is $\tilde F_2$.  
Moreover, with a gauge shift on the Abelian component of the gauge field on the brane, the components of $dC_1$ along the cigar directions can be gauged away. Since the integral of $dC_1$ along the cigar gives the bare $\theta$ parameter of the theory, this implies, consistently with field theory expectations, that the bare $\theta$ parameter can be rotated away by a chiral $U(1)_A$ phase shift of the fermionic fields. This is explicitly realized by considering
\be
\delta_\Lambda \widehat{A}_z = - \partial_z \Lambda\,.
\label{gaugeshift}
\ee
Integrating along the cigar, eq. (\ref{C1shift}) gives
\be
\delta_{\Lambda}\,\theta = \sqrt{\frac{N_f}{2}}(\Lambda|_{z=+\infty}-\Lambda|_{z=-\infty})\,,
\ee
which corresponds to the shift
\be
\theta \rightarrow \theta+ 2 N_f \alpha\,,
\ee
after recalling that a gauge transformation with $\Lambda|_{z=\pm \infty}=\pm\sqrt{2 N_f}\alpha$ is holographically mapped into the $U(1)_A$ rotation
\be
q^f\rightarrow e^{i\gamma_5 \alpha} q^f\,,
\ee
on the fundamental fermionic fields \cite{SS1}.

Since with a chiral rotation the $\theta$ parameter can be rotated away, it is clear that when the model contains (even just one) massless flavors its topological susceptibility as well as any $\theta$-dependence of its observables vanishes.

A non-zero $\theta$-dependence can be obtained when the quarks are massive, as in the real world.

As we recall in Appendix \ref{applu}, the action $S_{C_7}$ is equivalent to 
\be
S_{{\tilde F}_2} = -\frac{1}{4\pi (2\pi l_s)^6} \int d^{10}x |\tilde F_2|^2\,.
\ee
Considering a zero mode for ${\widehat A}_z$ such that
\be
\int dz {\widehat A}_z = \frac{2 \eta'}{f_{\pi}}\,, 
\ee
we see that using the integrated Bianchi identity for $\tilde F_2$ and its equation of motion $d\hodge \tilde F_2 = 0$, the on-shell value of the action above reduces to
\be\label{SF2}
S_{{\tilde F}_2} = -\frac{\chi_g}{2}\int d^4x\left(\theta + \frac{\sqrt{2N_f}}{f_{\pi}}\eta'\right)^2\,,
\ee
where $\chi_g$ is the topological susceptibility of the unflavored model (\ref{tsusc}). As it has been observed in \cite{SS1} this precisely gives the large $N_c$ estimate of the $\eta'$ mass predicted by the Witten-Veneziano formula
\be
m_{\eta'}^2 = m_{WV}^2\equiv\frac{2N_f}{f_{\pi}^2}\chi_g\,.
\ee
Being explicit this gives, in our model,
\be
m_{WV}^2 = \frac{1}{27\pi^2}\frac{N_f}{N_c} \lambda^2 M_{KK}^2 \sim \epsilon_f M_{KK}^2\,.
\label{mWV}
\ee
Hence, working in the probe approximation requires taking 
\be
m_{WV}\ll M_{KK}\,.
\ee
This is then another limit in which we are forced to work.  

What we just did can be understood in terms of a Stueckelberg mechanism, in which a massless vector field $\widehat{A}_M$ ``eats'' a scalar (from the D8 point of view) field $C_y$. Acquiring a new degree of freedom $\widehat{A}_M$ becomes massive, hence explaining the mass of the $\eta'$ arising from the $U(1)_A$ anomaly. 
\subsection{Horava-Witten solution of the anomalous Bianchi identity}
In general, the formal solution \eqref{intB} of the anomalous Bianchi identity \eqref{bianchi} does not solve the equation of motion $d\hodge\tilde F_2 =0$. 
The main problem is the presence of the delta function. The present setup shares many common points with the Horava-Witten one \cite{HoravaWitten}. 
As in that case, we can solve the Bianchi identity in a way which is compatible with the equations of motion by writing
\bea
\tilde{F}_{MN} &=& \Theta(y)\sqrt{\frac{N_f}{2}} \widehat{F}_{MN} + f_{MN},\qquad M,N,...= 0,1,2,3,z\,,\nonumber \\
\tilde{F}_{zy} &= &f_{zy}\,,
\label{horavawitt}
\eea
where $\Theta(y)$ is the step function, $\Theta(y) = |y|/2y$, $f_{MN}$ are regular terms vanishing at $y=0$ and $f_{zy}$ will be discussed in a moment. The extra terms $f_{MN}$ are necessary to satisfy the equation of motion $d \hodge \tilde{F}_{(2)} = 0$. 

The Bianchi identity $d \tilde F_2 = \Tr \mathcal{F} \wedge\delta(y) d y$ is satisfied provided $d f = 0$, hence one can always put
\begin{equation}
f_{AB} = \partial_A g_B -\partial_B g_A\,,\quad {A,B,...= 0,1,2,3,z,y}\,.
\label{f=dg}
\end{equation}
A solution $f_{AB}$ of the Bianchi-Maxwell system (i.e. the Bianchi identity and the equation of motion for ${\tilde F}_{AB}$), provided it exists, is not unique. In fact it is always possible to add to a given solution the zero mode
\begin{equation}
f_{zy}^{(0)} = \frac{C}{U^6}\;,\quad f^{(0)}_{AB \neq zy} = 0\,,\label{zeromode}
\end{equation}
satisfying $d f^{(0)} = 0$ and $d \hodge f^{(0)} = 0$, for any value of the constant $C$. 

Let us thus write
\begin{equation}
f_{AB} = f^{(0)}_{AB} + f^{(1)}_{AB}\;,
\end{equation}
imposing 
\begin{equation}
\lim_{|\mvec{x}|\to \infty} \int d z d y \, f_{zy}^{(1)} = 0\;, \label{vanishbc}
\end{equation}
which is a consistent boundary condition. Now, the constant $C$ acquires a physical meaning (in terms of the $\theta$ parameter) after imposing the boundary condition
\begin{equation}
\lim_{|x|\to \infty}\int d z d y \,\tilde{F}_{zy} = \theta + \lim_{|x|\to \infty}\sqrt{\frac{N_f}{2}} \int {\widehat A}_z dz \,,
\label{constraintmod}
\end{equation}
which gives
\begin{equation}
C = \frac{1}{\pi}\left(\theta + \lim_{|\mvec{x}|\to \infty} \sqrt{\frac{N_f}{2}} \int d z\,\widehat{A}_z \right)\;.\label{C}
\end{equation}
Let us now add that $\lim_{|\mvec{x}|\to \infty} \widehat{F}_{MN} = 0$: from this it follows that the limit $|\mvec{x}|\to \infty$ of the Bianchi-Maxwell system is linear in $f_{AB}^{(1)}$, hence we can find solutions that vanish at spatial infinity.\footnote{We are implicitly exchanging $\lim_{|\mvec{x}|\to \infty}$ and $\int d z d y$.}
 
Moreover, whatever the explicit form of $f^{(1)}_{AB}$ is, it does not mix with the equations of motion of the gauge fields $\widehat{A}$. This follows from the fact that the Horava--Witten solution (\ref{horavawitt}) is antisymmetric under $y\to -y$. 
Since $f^{(1)}_{AB}$ is smooth in $y$ it must be
\begin{equation}
f^{(1)}_{AB}{\big|_{y=0}} = 0\;.
\end{equation}
Recalling that the mixing between the equations is schematically given by\footnote{The constant $\gamma$ does not matter in this discussion; it is equal to $6$ for the $z$ component and to $7/2$ for the $\mu$ components. $S_{\mathrm{mass}}$ is discussed in the next Section.}
\begin{equation}
\frac{\delta S_{\mathrm{DBI+CS+mass}}}{\delta \widehat{A}} = (\mbox{\footnotesize const.})\,\delta(y) \,U^{\gamma} \tilde{F}_{2}\;,
\end{equation}
we see that $y$ has to be zero, hence only the zero mode $f^{(0)}_{AB}$ can contribute. Thus we can effectively set $f_{zy}=f_{zy}^{(0)}$ in this kind of computations.

Finally, we want to show that there is indeed an explicit solution for $f^{(1)}_{AB}$, even though we will not need it. 
According to the observations above, the solution is antisymmetric in $y$ and thus we can first solve the Bianchi-Maxwell system for $y>0$ and then continue the solution for negative values. At this point the existence of the solution is obtained by a counting: there are three independent equations, while the unknowns are the $g_A^{(1)}$ defined by
\begin{equation}
f_{AB}^{(1)} = \partial_A g_B^{(1)} -\partial_B g_A^{(1)}\;,
\end{equation}
analogously to \eqref{f=dg}. 
The independent components are three because the Lorentz symmetry relates the $\mu$ indexes. The system is solvable having the same number of components and unknowns.
\section{WSS model with massive fermions} 
\label{massfer}
\setcounter{equation}{0}
In view of the relation (\ref{pionm}), defining the pion matrix as a path ordered holonomy matrix and in analogy with the chiral Lagrangian approach, a natural term to add to the effective action (\ref{SYM}) in order to describe massive quarks is
\begin{equation}\label{Smass}
S_\mathrm{mass} = c \int d ^4x\, \Tr\mathcal{P}\left[M  \exp\left(-i\int_{-\infty}^\infty\mathcal{A}_z d z\right) + \mathrm{c.c.}\right]\,,  
\end{equation}
where $c$ is a constant and $M$ is the mass matrix. 
This term has actually a very precise meaning in string theory \cite{AK, Hashimotomass}: it is the deformation due to open string worldsheet instantons stretching between the D8-branes. 
A basic observation in \cite{AK} is that the $U(N_f)$ holonomy matrix $\U$ which is the order parameter for chiral symmetry breaking, is not gauge invariant, when embedded in the full string theory model, under gauge transformations of the NSNS $B$-field. 
A gauge invariant object can be obtained by multiplying $\U$ by $e^{i\int B}$ where the integral is done over the cigar directions of the background. 
A way to construct an operator carrying such a phase is to insert an open fundamental string (actually a worldsheet instanton) stretching between the branes. 
The string worldsheet will be extended along the cigar directions $U,x_4$ from $U=U_{KK}$ up to a cutoff $U=U_m$ which will set the quark bare mass parameter. 
Introducing such a worldsheet instanton 
corresponds to deforming the dual gauge theory by a non-local mass term for the fermions. 

The Nambu-Goto part of the open string action is put on-shell and its exponentiation contributes to the constant $c$ and the mass terms. 
What remains is just the boundary interaction of the open string with the gauge fields on the D8-branes. 
The constant, up to an irrelevant normalization factor, reads\footnote{The observables will depend on the combination $c m_q$, where $m_q$ is the quark mass. 
This combination will be fixed by the GMOR relation. 
One has thus the freedom to fix one of the two parameters at will, so we can take the normalization factor to be 1 without loss of generality.}
\begin{equation}
c = \frac{1}{3^{9/2}\pi^3}g_{\mathrm{YM}}^3N_c^{3/2}M_{\mathrm{KK}}^3 \,. \label{cdef}
\end{equation}

When the mass term (\ref{Smass}) is added to the original WSS model, in such a way that all flavor fields get masses, we should expect that the $\theta$ dependence emerges again. This is actually what happens.

As reviewed in Sections \ref{WSSreview} and \ref{anomaly}, the $\theta$ term can be introduced as an integral of $C_{1}$ and then removed (in absence of flavor mass terms) via a gauge shift (\ref{gaugeshift})
\begin{equation}
\frac{1}{\sqrt{2N_f}}\int \widehat{A}_z \;\longrightarrow\; \frac{1}{\sqrt{2N_f}}\int \widehat{A}_z -\frac{\theta}{N_f}\,.
\end{equation}
After this shift, however, $S_\mathrm{mass}$ becomes
\begin{equation}
S_\mathrm{mass} = c \int d ^4x\, \Tr\mathcal{P}\left[M e^{i\frac{\theta}{N_f}}  \exp\left(-i\int_{-\infty}^\infty\mathcal{A}_z d z\right) + \mathrm{c.c.}\right]\,, 
\label{smass}
\end{equation}
which amounts to redefining
\begin{equation}
M  \;\longrightarrow \; M e^{i\theta/N_f}\,.
\end{equation}
The $\theta$-dependence is thus not erased anymore. Moreover, as expected in QCD, the physical $\theta$ parameter is not just the coefficient of $F \wedge F$ but the combination
\begin{equation}
\bar\theta = \theta + \arg \det M\,.
\label{thetaphys}
\end{equation}
In the following we will mostly focus on the mass-degenerate case $M_{ij} = m_q \delta_{ij}$, choosing $m_q$ to be real.
\section{$\theta$ dependence of the vacuum energy} \label{masstheta}
\setcounter{equation}{0}
Let us now see how the mass deformation introduced above modifies the vacuum solution. Let us first notice that the chiral condensate satisfies the Gell-Mann--Oakes--Renner (GMOR) relation
\cite{AK,Hashimotomass}
\begin{equation}
\sum_{f=1}^{N_f}\langle \overline{\psi}_f\psi_f\rangle = -2cN_f\,,
\end{equation}
where $c$ is defined in \eqref{cdef}. In the particular mass-degenerate case $M_{ij} = m_q \delta_{ij}$ the GMOR relation
\be
f_\pi^2m_\pi^2 =  - 2 \frac{m_q}{N_f}\sum_{f=1}^{N_f}\langle\overline{\psi}_f\psi_f\rangle\,,
\ee
implies that
\be\label{gmor1}
cm_q = \frac{1}{4}f_{\pi}^2m_{\pi}^2\,.
\ee
In this case the minimum of the energy is found by setting the non Abelian component of the gauge field $A$ to zero modulo gauge transformations. The vacuum will then be described by a pure gauge solution $\widehat{F}=0$.

The only relevant part of the effective action determining the vacua is that for the $\widehat{A}_z$ Abelian component.\footnote{Notice that at $y=0$, where the gauge fields are defined, from the Horava-Witten-like solution \eqref{horavawitt} it follows that the equation of motion for $\widehat{A}_{\mu}$ does not receive any contribution from $S_{C_7}$. This is true for the two following reasons: a) the metric on the cigar directions $(y,z)$ is diagonal at $y=0$; b) we are setting ${\tilde F}_{\mu y}=0$. As a consequence, only the equation of motion for $\widehat{A}_z$ receives a contribution from $S_{C_7}$ via the zero mode components of ${\tilde F}_{yz}$. All this will apply also to the instanton solutions we will look for in the following.} 
Together with (\ref{SYM}), (\ref{SF2}), the action (\ref{smass}) gives
\be
{\cal L}_{eff\, z} = -\frac{\kappa}{2} k(z) \widehat{F}_{\mu z} \widehat{F}^{\mu}_z + c\, {\rm Tr} {\cal P}[M e^{ - \frac{i}{\sqrt{2N_f}} \int \widehat{A}_z dz} \unit + c.c. ] - \frac{\chi_g}{2} \left(\theta + \sqrt{\frac{N_f}{2}}\int \widehat{A}_z dz\right)^2\,.
\label{effett}
\ee
The vacuum solution $\widehat{F}_{\mu z}=0$ can be given in terms of
\be
\varphi \equiv -\frac{1}{\sqrt{2N_f}}\int_{-\infty}^{\infty} \widehat{A}_z dz\,.
\ee
From the equation of motion of $\widehat A_z$ we actually get the following condition in the mass-degenerate case
\be
m_{\pi}^2\sin\varphi = m_{WV}^2 \left(\frac{\theta}{N_f} - \varphi \right)\,,
\label{vacuumphi}
\ee
where, also recalling (\ref{fpaie}), we have used $\kappa m_{\pi}^2 = \pi c m_q$ and eq. (\ref{mWV}). Equation \eqref{vacuumphi} is precisely the same which follows from the chiral Lagrangian approach discussed in \cite{WittenLargeNChi} (see also \cite{DiVecchiaLargeNAdSCFT} for a review). 

The on-shell four dimensional Lagrangian density on the vacuum solution we have found is
\begin{equation}
{\mathcal{L}}^{\mathrm{on-shell}} = -\frac{\chi_g}{2}(\theta - N_f \varphi)^2 + 2c N_f m_q   \, \cos\varphi\,.
\end{equation}
We can extract the vacuum solution analytically by considering the following two extreme cases:
\begin{enumerate}[\textit{\roman{enumi})}]
\item $m_{WV}^2 \ll m_{\pi}^2$: in this case we have a multi-branched solution
\be
\varphi = 2\pi k + \frac{m_{WV}^2}{m_{\pi}^2} \left(\frac{\theta}{N_f}-2\pi k \right) + {\cal O}(m^4_{WV}m_{\pi}^{-4})\,,\quad k\in\mathbb Z\,.
\ee
This is the limiting case which arises if we take the large $N_c$ limit before the chiral one. 
In a sense, this limit is analytically connected with the limit in which the quark mass is so large that the flavors can be integrated out. Correspondingly, the vacuum energy density around $\theta=0$ goes, to leading order, like 
\be
f(\theta)-f(0) \sim \frac{\chi_g}{2}\,\theta^2\,,
\ee
which is the same behavior (\ref{fthetau}) as for the unflavored theory.
\item $ m_{WV}^2 \gg m_\pi^2$: in this limit the solution is unique
\begin{equation}
\varphi = \frac{\theta}{N_f}+ {\cal O}(m_{\pi}^2 m_{WV}^{-2})\,.
\end{equation}
This limit is actually closer to the phenomenologically acceptable case because $m_\pi \simeq 135 \,\mathrm{MeV}$ while $m_{WV}\sim m_{\eta'} \simeq 958 \,\mathrm{MeV}$. 
In this case the vacuum energy density $f(\theta)$ reads
\begin{equation}
f(\theta) - f(0)= \frac{N_f}{2} m_{\pi}^2 f_{\pi}^2\left[1- \cos\left(\frac{\theta}{N_f}\right)\right] + \mathcal{O}(m_\pi^4)\,.
\end{equation}
The topological susceptibility of the theory is thus
\begin{equation}
\chi = \frac{\partial^2 f(\theta)}{\partial\theta^2}\Big|_{\theta=0}= \frac{m_{\pi}^2 f_{\pi}^2}{2 N_f}\,,
\end{equation}
as expected from chiral perturbation theory.
\end{enumerate}

In any case, expanding the effective Lagrangian (\ref{effett}) around the vacuum solution (\ref{vacuumphi}), we can obtain the following $\theta$-dependent mass spectrum 
\be
m_{\eta'}^2 (\varphi) = m_{WV}^2 + m_{\pi}^2 (\varphi)\,,\quad  m_{\pi}^2 (\varphi) \equiv m_{\pi}^2 |\cos\varphi|\,.
\ee
In the case $ m_{WV}^2 \gg m_\pi^2$, 
\be
m_{\pi}^2(\varphi)=m_{\pi}^2(\theta) = m_{\pi}^2 |\cos{\frac{\theta}{N_f}}|\,,
\ee
which implies that the masses of the low lying mesons decrease quadratically with $\theta$ for small $\theta$. This behavior reflects the general trend already observed in \cite{wymtheta} for other mass scales in the unflavored theory. 

     
\section{Holographic baryons}
\setcounter{equation}{0}
\label{holobar}

In this Section we first review how baryons are described in the WSS model, recalling the quantization of the moduli space Hamiltonian. Then, we show that the correction to the baryon spectrum due to massive quarks and the $\theta$ term is quadratic in $\theta$. 

In the WSS model, following \cite{wittenbaryon}, a baryon vertex is identified with a D4-brane wrapped on $S^4$ and the baryon number is defined as the charge of that brane. Adding a D4-brane source to the WSS setup implies including a term
\begin{equation}
\frac{1}{8\pi^2}\int_{\rm D8} C_{5} \wedge \Tr \mathcal{F}\wedge\mathcal{F}\,,
\end{equation} 
into the action. This in turn implies that a baryon corresponds to a soliton solution $\mathcal{F}$ with non trivial instanton number
\be
n_B=\frac{1}{8\pi^2}\int_B \Tr \mathcal{F}\wedge\mathcal{F}\,,
\ee
where $B$ is the space spanned by $x^{1,2,3},\,z$. The instanton number $n_B$ is then interpreted as the baryon number \cite{SS-barioni}. To show that this is indeed the case, let us write down the original WSS action (eq. (\ref{SYM}) without the $S_{C_7}$ term) separating the Abelian and the non Abelian components (see \eqref{separ}) 
\begin{equation}
\begin{aligned}
S_{\mathrm{D8}} &= -\kappa\int d^4x d z\, \left(\frac{1}{2}h(z)  \, \Tr F_{\mu\nu}F^{\mu\nu} +   k(z)  \Tr F_{\mu z}F^\mu_{\;\; z}\right)
 +\\&\quad  
 -\frac{\kappa}{2}\int d^4x d z\, \left(\frac{1}{2}h(z)  \,  \widehat{F}_{\mu\nu}\widehat{F}^{\mu\nu} +   k(z)   \widehat{F}_{\mu z}\widehat{F}^\mu_{\;\; z}\right)
 +\\&\quad  
+\frac{N_c}{24\pi^2}\int  \biggl[\omega_5^{SU(N_f)}(A)+ \frac{3}{\sqrt{2N_f}}\widehat{A}\,\Tr F^2 + \frac{1}{2\sqrt{2N_f}}\widehat{A}\,\widehat{F}^2\biggr]\,.
\label{clac}
\end{aligned}
\end{equation}
Here $\omega_5^{SU(N_f)}$ is defined as in (\ref{omega5}), written in terms of just the non Abelian components. It is worth noticing that it is identically zero for $N_f=2$. 

Defining $a(t) = \widehat{A}/\sqrt{2N_f}$ and treating it as a time dependent perturbation over the soliton solution with instanton number $n_B$, we obtain in the action a term
\begin{equation}
\frac{N_c}{8\pi^2}\int d x^0\, a \int_B \Tr F^2 = n_B N_c\int d x^0\,a\,.
\end{equation}
This describes a point-like particle with $U(1)_V$ charge equal to $N_c n_B$: precisely that of a baryon (a bound state of $N_c$ quarks) with baryon number $n_B$. 

The above holographic picture resembles the Skyrme one, where baryons at large $N_c$ are seen as solitons in the chiral Lagrangian \cite{skyrme,ANWSkyrme}. The similarity becomes more evident at low energies, since, integrating out the massive vector modes, the effective WSS action reduces to the Skyrme model with the WZW term. 

The equations of motion following from (\ref{clac}) are not easy to solve analytically. A simple static instanton solution, for $N_f=2$, can be given focusing in a tiny region around $z=0$ where one can neglect the curvature of the background setting $k(z)\approx h(z)\approx1$. In this case the solution is given by a charged BPST instanton \cite{SS-barioni,BPSTInst}
\begin{equation}
A_M^{\mathrm{cl}} = -i f(\xi) g \partial_M g^{-1}\,,\quad \widehat{A}^{\mathrm{cl}}_0 = \frac{N_c}{8\pi^2\kappa}\frac{1}{\xi^2}\left[1-\frac{\rho^4}{(\rho^2+\xi^2)^2}\right]\;,\quad A_0^{\mathrm{cl}}=\widehat{A}_M^{\rm cl}=0\,,
\label{Ainst}
\end{equation}
where 
\begin{equation}
f(\xi) = \frac{\xi^2}{\xi^2+\rho^2}\;,\quad g(x) = \frac{(z-Z)\unit - i (\mvec{x}-\mvec{X})\cdot \mvec{\tau}}{\xi}\label{eqg}\;, \quad \xi^2 \equiv (\vec{x}-\vec{X})^2+(z-Z)^2\,, 
\end{equation}
$\tau^a$ are the Pauli matrices and the index $M$ runs over the four directions $x^{1,2,3}, z$. 

The instanton solution written above depends on eight parameters: the instanton size $\rho$, the instanton center of mass position $X^{M}=(\vec X, Z)$ in the four dimensional Euclidean space, and three $SU(2)$  ``angles" related to the fact that the solution can be rotated by means of a global gauge transformation. 

Substituting the solution (\ref{Ainst}) into the action (\ref{clac}) on finds $S_{\rm{on\,shell}} = -\int d t M_B$, where, up to $\mathcal{O}(\lambda^{-2})$ corrections
\begin{equation}\label{MB}
M_B(\rho,Z) =  M_0\left[1 +  \left(\frac{\rho^2}{6}+ \frac{N_c^2}{320 \pi^4 \kappa^2}\frac{1}{\rho^2} + \frac{Z^2}{3}\right)\right]\,,\quad M_0\equiv8\pi^2 \kappa\,,
\end{equation}
with $M_0$ giving the baryon mass in the $\lambda\rightarrow\infty$, $N_c\rightarrow\infty$ limit. 

This implies that, while $\mvec{X}$ and the gauge group orientations are genuine moduli of the instanton solution, $\rho$ and $Z$ are not; in fact they are classically fixed by minimizing $M_B$ as
\begin{equation}
\rho_\mathrm{cl}^2 = \frac{N_c}{8\pi^2 \kappa}\sqrt{\frac{6}{5}}= \frac{27\pi}{\lambda}\sqrt{\frac{6}{5}}\;,\qquad
Z^\mathrm{cl} = 0\,.
\end{equation}
These relations imply that the center of the instanton is classically localized at $Z=0$ and its size $\rho \sim 1/\sqrt{\lambda}$ is very small in the $\lambda\gg1$ regime. This is perfectly consistent with the approximation we have taken to get the above instanton solution. In particular, the latter is obtained by a systematic expansion of the equations of motion in $1/\lambda$, considering a scaling  $\vec x, z \sim \mathcal{O}(\lambda^{-1/2})$, $x^0 \sim \mathcal{O}(1)$ for the space-time variables, and the following scalings for the gauge fields 
\begin{equation}\label{scalings}
\begin{aligned}
&\mathcal{A}_M \in \mathcal{O}(\lambda^{1/2})\;,\qquad \mathcal{A}_0 \in \mathcal{O}(\lambda^{0}) \,,
\\
&\mathcal{F}_{MN} \in \mathcal{O}(\lambda)\;,\qquad \mathcal{F}_{M0} \in \mathcal{O}(\lambda^{1/2})\,.
\end{aligned}
\end{equation}
In the following discussion we will treat $\rho$ and $Z$ as approximate moduli, allowing them to fluctuate quantistically around their classical values. This is not completely correct because they modify the potential energy, but it remains a good approximate description if the fluctuations are small.

\subsection{Quantization}\label{barquantization}
The quantization of the WSS soliton proceeds following the moduli space approximation method as described in \cite{SS-barioni} and takes inspiration from the Skyrmion quantization \cite{ANWSkyrme}. Since $M_0=8\pi^2 \kappa \propto \lambda N_c\gg1$, the baryon is very heavy and the system reduces to a quantum mechanical model for the instanton (pseudo) moduli. In the $SU(2)$ case the one-instanton moduli space, topologically equivalent to ${\mathbb R}^4\times {\mathbb R}^4/{\mathbb Z}_2$, is parameterized by $X^{M}$ and $y^I$, ($I=1,2,3,4$) with the ${\mathbb Z}_2$ action $y^{I}\rightarrow-y^{I}$.  The instanton size $\rho$ is given by $\rho^2 = y^Iy^I$ and $a^I=y^I\rho^{-1}$ are the $SU(2)$ directions.

Technically, the above parameters are promoted to time dependent variables and the $SU(2)$ field describing the slowly moving soliton is defined through a ``wrong'' gauge transformation \begin{equation}
\begin{aligned}
A_0^{\rm cl} &\;\longmapsto\;& A_0' &= 0 \,,\\
A_M^{\rm cl} &\;\longmapsto\;& A_M' &= V A_M^{\rm cl} V^{-1} - i V \partial_M V^{-1} \,.
\label{modrot}
\end{aligned}
\end{equation}
The $SU(2)$ matrix $V(t, \vec x, z)$ is necessary for ensuring that the new time-dependent soliton still solves the equations of motion following from the action (\ref{clac}). The only non trivial condition comes from the Gauss's law constraint 
\begin{equation}
D_M F_{0M} + \frac{N_c}{64\pi^2 \kappa} \varepsilon_{MNPQ} \widehat{F}_{MN} F_{PQ}= 0\,,
\end{equation}
which actually reduces to $D_M F_{0M} = 0$ on the solution. This equation can be solved by
\be
\Phi \equiv -i V^{-1}\dot V= - \dot{X}^N A^{\rm cl}_N -i f(\xi) g (\bvec{a}^{-1} \dot{\bvec{a}}) g\,,
\ee
where the dot is a time derivative, $f(\xi)$ and $g$ are defined in \eqref{eqg}, $\bvec{a}(t)=a_4(t) + i a_a(t)\tau^a$ contains the gauge group orientation moduli and  the boundary condition
\begin{equation}
\lim_{z\to \pm \infty} V(t, \vec x, z) = \bvec{a}(t)\,,
\end{equation}
has been imposed.

Inserting the slowly moving soliton solution into the action (\ref{clac}) one gets the quantum mechanical Lagrangian 
\begin{equation}
\begin{aligned}
\mathcal{L} 
&=\frac{M_0}{2}\left(\dot{\mvec{X}}^2 + \dot{Z}^2 + 2(\dot{y}^I)^2\right) - M_0\left[1+\frac{\rho^2}{6} + \frac{N_c^2}{320\pi^4\kappa^2} \frac{1}{\rho^2} + \frac{Z^2}{3}\right]\,,
\end{aligned}
\end{equation}
and thus the Hamiltonian
\begin{equation}
\mathcal{H} = M_0 + \mathcal{H}_X +\mathcal{H}_Z + \mathcal{H}_y\,,
\end{equation}
where
\begin{equation}
\begin{aligned}
\mathcal{H}_X &= - \frac{1}{2M_0}\frac{\partial^2}{\partial {X^i}^2}\,,\\
\mathcal{H}_Z &= - \frac{1}{2M_0}\frac{\partial^2}{\partial {Z}^2} + \frac{M_0}{3} Z^2 \,,\\
\mathcal{H}_y &= - \frac{1}{4M_0}\sum_{I=1}^4\frac{\partial^2}{\partial {y^I}^2} + \frac{M_0}{6} \rho^2 + \frac{N_c^2 M_0}{320 \pi^4 \kappa^2} \frac{1}{\rho^2}\,.
\end{aligned}
\end{equation}
They respectively describe a free particle in three dimensions, a harmonic oscillator in one dimension and a harmonic oscillator in four dimensions with an extra centrifugal energy. 

The eigenfunctions and eigenvalues for the first two pieces are \cite{SS-barioni}
\begin{equation}
\Psi_1(\mvec{X}) = \frac{1}{(2\pi)^{3/2}} e^{i \mvec{P} \cdot \mvec{X}}\;,\qquad E_X = \frac{\mvec{P}^2}{2M_0}\,,
\end{equation} 
\begin{equation}
\Psi_2(Z) = H^{(n)}((\sqrt{2/3}M_0)^{1/2}\,Z) e^{-\frac{M_0}{\sqrt{6}}Z^2}\;,\qquad 
E_Z = \frac{2 n_Z + 1}{\sqrt{6}}\,,
\end{equation}
where $H^{(n)}$ are Hermite polynomials. Concerning the third one, switching to spherical coordinates in $\mathbb{R}^4$, the Laplacian decomposes as 
\begin{equation}
\sum_{I=1}^4\frac{\partial^2}{\partial {y^I}^2} = \frac{1}{\rho^2}\partial_\rho \left( \rho^3 \partial_\rho \cdot \right) + \frac{1}{\rho^2}\nabla_{S^3}^2\,,
\end{equation}
and the obvious ansatz for $\Psi_3(y^I)$ is
\begin{equation}
\Psi_3(y^I) = R(\rho) Y^{(\ell)}(a^I)\,,
\end{equation}
where $Y^{(\ell)}$ are the scalar spherical harmonics on $S^3$ with eigenvalue $\ell (\ell+2)$. Such a wave function has spin and isospin equal to $\ell/2$ where the spin and isospin operators are identified with the generators of the $SO(4)$ symmetry group acting on the $y^I$
\bea
J_k &=&  \frac{i}{2}\left(-y_4\frac{\partial}{\partial y_k}+ y_k\frac{\partial}{\partial y_4}-\epsilon_{klm}y_l\frac{\partial}{\partial y_m}\right)\,,\nonumber \\
I_k &=& \frac{i}{2}\left(y_4\frac{\partial}{\partial y_k}- y_k\frac{\partial}{\partial y_4}-\epsilon_{klm}y_l\frac{\partial}{\partial y_m}\right) \,.
\label{spiniso}
\eea
These relations imply that only states with $I=J$ appear in the spectrum. A crucial observation is that $a^I$ and $-a^I$ are identified on the instanton moduli space. 
If we want to quantize the solitons as fermions we have to require the wave function to be antiperiodic $\psi(a^I) = -\psi(-a^I)$. This selects $\ell=1,3,5,\cdots$ to be positive odd integers. The related states have $I=J=\ell/2$.

The solution for $R(\rho)$ can be found by noticing that the centrifugal term in $\mathcal{H}_y$ modifies the angular momentum as
\begin{equation}
\ell(\ell + 2) \to \ell(\ell + 2 ) + \frac{N_c^2 M_0^2}{80\pi^4 \kappa^2} \equiv \tilde{\ell}(\tilde{\ell} + 2)\,.
\end{equation}
Thus, upon substituting $\ell \to \tilde{\ell}$ we end up with a regular harmonic oscillator in four dimensions in spherical coordinates. The solution is
\begin{equation}
R(\rho) = e^{-\frac{M_0}{\sqrt{6}}\rho^2}\rho^{\tilde{\ell}} \ _1F_1\left(n_\rho,\tilde{\ell}+2,\sqrt{2/3}M_0 \,\rho^2\right)\;,\quad n_\rho \in \mathbb{N}\,,
\end{equation}
where $_1F_1\left(a,b,\rho\right)$ is the Confluent Hypergeometric Function. The corresponding eigenvalues are
\begin{equation}
E_\rho = \frac{1}{\sqrt{6}}(2n_\rho + \tilde{\ell} + 2) = \frac{2n_\rho + 1}{\sqrt{6}} + \sqrt{\frac{(\ell + 1)^2}{6} + \frac{2 N_c^2}{15}}\,.
\end{equation}

A baryon is a state $|B, s\rangle$ in the Hilbert space defined by the Hamiltonian ${\cal H}$, where $s$ is the (iso)spin of the baryon. The quantum numbers $n_\rho$ and $n_Z$ describe excited baryons and/or resonances; the case $\ell=1$, $n_\rho = n_Z = 0$ corresponds to the neutron (with isospin component $I_3=-1/2$) and the proton ($I_3=1/2$) and the corresponding wavefunctions are
\bea
&& |p\uparrow\rangle \propto R(\rho) \psi_Z(Z) (a_1 + i a_2)\,,\quad  |p\downarrow\rangle \propto R(\rho) \psi_Z(Z) (a_4 - i a_3)\,,\nonumber\\
&& |n\uparrow\rangle\propto R(\rho) \psi_Z(Z) (a_4+ i a_3)\,,\quad  |n\downarrow\rangle \propto R(\rho) \psi_Z(Z)(a_1 - i a_2)\,,
\label{pnwave}
\eea
with
\be
R(\rho)=\rho^{-1+2\sqrt{1+N_c^2/5}}e^{-\frac{M_0}{\sqrt{6}}\rho^2} \, ,\qquad \psi_Z(Z)=e^{-\frac{M_0}{\sqrt{6}}Z^2} \,.
\end{equation}

\subsection{Baryon Hamiltonian with quark mass and $\theta$}
Let us now consider adding to the action (\ref{clac}) the mass term for the flavors introduced in (\ref{Smass}) at $\theta\neq0$. This term gives a novel contribution to the baryon Hamiltonian and modifies the WSS soliton solution. At leading order in the small $m_q$ limit (let us focus on the simpler case of degenerate quark masses), the contribution can be computed, along the same lines as in \cite{hadronmass}, from the on-shell value of 
\begin{equation}
S_{\mathrm{mass}} = c \int d^4 x\,\Tr \mathcal{P}\left[Me^{i\varphi}\left(e^{-i\int_{-\infty}^\infty dz {\cal A}_z}- \unit\right)+ \mathrm{c.c.}\right]\,,
\end{equation}
on the WSS instanton soliton solution (\ref{Ainst}). Here the $\unit$ subtraction corresponds to the subtraction of the vacuum energy (in the case of degenerate masses the minimum is for $\U=\unit$), while $e^{i\varphi}$ comes from the vacuum $\theta$-dependent contribution discussed in Section \ref{masstheta}. 

Let us work in singular gauge, where the $A_z^{\rm cl}$ field is given by
\begin{equation}
A_z^{\rm cl} = \left[\frac{1}{\xi^2}-\frac{1}{\xi^2+\rho^2}\right](\mvec{x}-\mvec{X})\cdot \mvec{\tau}\,,
\end{equation}
which is obtained from (\ref{Ainst}) after implementing a gauge transformation $A^{\rm cl}_z \to g^{-1}A_z^{\rm cl} g - ig^{-1}\partial_z g$, where $g$ is given by \eqref{eqg}. 

The pion matrix is easily computed (we also set $\mvec{X}=0$ without loss of generality) as
\begin{equation}
\U = \exp\left[-i\pi\frac{\mvec{\tau}\cdot\mvec{x}}{|\mvec{x}|}\left(1-\frac{1}{\sqrt{1+\rho^2/|\mvec{x}|^2}}\right)\right] \equiv \exp\left[-i \frac{\mvec{\tau}\cdot\mvec{x}}{|\mvec{x}|}\,\hat\alpha(|\mvec{x}|)\right]\,.
\end{equation}
The shift in the baryon mass $\delta M_B$ is given by $-\int d^3 x\,\mathcal{L}_{\mathrm{mass}}$, where $S_\mathrm{mass} = \int d^4 x\,\mathcal{L}_{\mathrm{mass}}$. We have
\begin{equation}
\delta M_B = -2c\int d^3 x\,\Tr\left[M\cos(\varphi)(\cos\hat\alpha-1) \right]\,.
\end{equation}

Let us now focus on the $N_f=2$ degenerate case in the physical mass regime $m_{\pi}\ll m_{WV}$, so that, as we found in Section \ref{masstheta}, we can set $\varphi=\theta/2$ up to subleading corrections in the mass ratio. 
We define the integration variable $y = |\mvec{x}|/\rho$ and get
\begin{equation}
\delta M_B = 16\pi\rho^3c\,m_q\cos(\theta/2)\int_0^\infty d y\,y^2\left[ 1+\cos\left(\frac{\pi}{\sqrt{1+y^{-2}}}\right)\right] \label{int1}\,.
\end{equation}
The integral is evaluated numerically and the final result is
\begin{equation}
\delta M_B = 16\pi \rho^3 c\, m_q\cos(\theta/2)\cdot 1.104 \,.
\end{equation}
The quantum contribution to this mass splitting, that differentiates the various species of baryons, follows in the same way as in \cite{hadronmass}, so we will skip it.

A relevant result of this Section is that the baryon Hamiltonian, hence the spectrum, through the mass term piece $\delta M_B$ computed above, gets second order ${\cal O}(\theta^2)$ corrections at small $\theta$. The mass splitting $\delta M_B$ at $\theta=0$ will anyway perturb some of the baryonic properties. In the semiclassical limit it will in fact affect the size of the baryon $\rho$ which will get an ${\cal O}(m_q)$ correction. 

When two different quark masses $m_u, m_d$ are considered the result is modified. 
First of all we should impose that the pion matrix $\hat{\mathcal{U}}=e^{i\frac{\theta}{2}}\U$ approaches $\hat{\mathcal{U}}_0$ when $|\mvec{x}|\to\infty$ (the vacuum configuration). The matrix $\hat{\mathcal{U}}_0$ turns out to be:
\begin{equation}
\hat{\mathcal{U}}_0 = e^{i\frac{\theta}{2}}\left(\begin{array}{cc} e^{i\Phi} & 0\\ 0 & e^{-i\Phi}\end{array}\right) = e^{i\frac{\theta}{2}} \U_0\,,
\end{equation}
\begin{equation}
\cos \Phi = \frac{\cos \frac{\theta}{2}}{\sqrt{\cos^2\frac{\theta}{2}+\left(\frac{m_d-m_u}{m_u+m_d}\right)^2\sin^2\frac{\theta}{2}}}\;,\quad\sin \Phi = \frac{\frac{m_d-m_u}{m_u+m_d}\sin \frac{\theta}{2}}{\sqrt{\cos^2\frac{\theta}{2}+\left(\frac{m_d-m_u}{m_u+m_d}\right)^2\sin^2\frac{\theta}{2}}}\,.
\end{equation}
The classical action has to be modified as
\begin{equation}
S_{\mathrm{mass}} = c \int d^4 x\,\Tr \mathcal{P}\left[Me^{i\frac{\theta}{2}}\left(e^{-i\int_{-\infty}^\infty dz {\cal A}_z}- \U_0\right)+ \mathrm{c.c.}\right]\,,
\end{equation}
and the solution ${\cal A}^{\rm cl}_z$ must be computed after a global gauge rotation that satisfies $\lim_{|x|\to \infty} \U = \U_0$ (we could take for instance $g(\infty) = \U_0$ and $g(-\infty)=\unit$). The result follows easily:
\begin{equation}
\delta M_B = 8\pi \rho^3 c \Tr(M \U_0)\cos(\theta/2)\cdot 1.104 \,.
\label{massformula}
\end{equation}
An interesting feature of the non-degenerate mass case is that the $SU(2)$ modulus $\bvec{a}$ gets  a potential term
\begin{equation}
\delta M_B \propto \Tr(M \bvec{a} \U_0 \bvec{a}^{-1})\,,
\end{equation}
thus giving a mass splitting between states with different isospin. For the case of the proton and the neutron this splitting would be too small compared to the electromagnetic splitting (not included in this analysis), so we ignore this computation.
\section{Mass and $\theta$ perturbations to holographic baryons}\label{sec:newsol}
\setcounter{equation}{0}
Let us now show how the original WSS instanton solution holographically describing a baryon gets modified by the mass and the $\theta$ term. This is done at leading order both in $m_q$ and $\theta$. For simplicity we mostly focus on the case of two degenerate masses $m_u = m_d$. 

The equations following from the action given by the sum of (\ref{SYM}) (with $S_{C_7}$ given in (\ref{sc7})) and (\ref{Smass}) are
\begin{align}
&-\kappa \left(h(z)\partial_\nu \widehat{F}^{\mu\nu} + \partial_z(k(z)\widehat{F}^{\mu z})\right) + \frac{N_c}{128\pi^2}\varepsilon^{\mu\alpha\beta\gamma\delta}
\left(F_{\alpha\beta}^aF_{\gamma\delta}^a + \widehat{F}_{\alpha\beta}\widehat{F}_{\gamma\delta}\right)=0\,,
\label{uno}\\
&-\kappa \Bigl( h(z)D_\nu F^{\mu\nu} + D_z(k(z)F^{\mu z})\Bigr)^a + \frac{N_c}{64\pi^2}\varepsilon^{\mu\alpha\beta\gamma\delta}
 F_{\alpha\beta}^a\widehat{F}_{\gamma\delta}=0\,,
 \label{due}\\
&-\kappa\, k(z)\partial_\nu\widehat{F}^{z\nu} + \frac{N_c}{128\pi^2}\varepsilon^{z\mu\nu\rho\sigma}
\left(F_{\mu\nu}^aF_{\rho\sigma}^a + \widehat{F}_{\mu\nu}\widehat{F}_{\rho\sigma}\right)=\nonumber\\&
\qquad \qquad \qquad \  = -\frac{4\pi}{3}\sqrt{\frac{N_f}{2}}[d C_{7}]_{0123} - i c \,\Tr \left[\frac{M}{\sqrt{2N_f}}\left(\mathcal{P}e^{-i\int_{-\infty}^\infty \mathcal{A}_zd z}- \mathrm{c.c.}\right)\right]\,,\label{tre}\\
&-\kappa\,k(z) (D_\nu F^{z\nu})^a + \frac{N_c}{64\pi^2}\varepsilon^{z\mu\nu\rho\sigma}F^a_{\mu\nu} \widehat{F}_{\rho\sigma} = 
- i c \,\Tr \mathcal{P}\left[M\frac{\tau^a}{2}\left(e^{-i\int_{-\infty}^\infty \mathcal{A}_zd z}- \mathrm{c.c.}\right)\right]\,.\label{quattro}
\end{align}
The factors $N_f$ are displayed explicitly but will soon be substituted by ``2''. 

The solution will be decomposed in three different contributions: $\mathcal{A}^\mathrm{vac}$, $\mathcal{A}^\mathrm{inst}$ and $\mathcal{A}^\mathrm{mass}$. The first one is the vacuum solution found in Section \ref{masstheta}
\begin{equation}
\begin{aligned}
A_z^\mathrm{vac} &= \mathcal{A}^\mathrm{vac}_\mu = 0\;,\qquad &\int \widehat{A}_z^\mathrm{vac} dz& = -\sqrt{2N_f}\varphi \ .
\end{aligned}
\end{equation}
The second one, in the $\xi\ll1$ region, is the WSS instantonic solution (\ref{Ainst}) in singular gauge
\begin{equation}\label{wssinst}
\begin{aligned}
A_M^{\mathrm{inst}} &= -i(1-f(\xi)) g^{-1} \partial_M g\;, \quad &  \widehat{A}^{\mathrm{inst}}_0 &= \frac{N_c}{8\pi^2\kappa}\frac{1}{\xi^2}\left[1-\frac{\rho^4}{(\rho^2+\xi^2)^2}\right]\,,
\\ A_0^{\mathrm{inst}}&=\widehat{A}_M^{\mathrm{inst}}=0\,.
\end{aligned}
\end{equation}
The solution in the remaining range of $\xi$ values will be presented in a moment.

The last piece, $\mathcal{A}^\mathrm{mass}$, is the perturbation due to the presence of the mass term that we wish to compute.
Since we are looking for solutions with non trivial field strength $\widehat{F}$, the components of the RR two-form $[\tilde{F}_{2}]_{AB}$ with $A,B \neq y$ are given by the Horava--Witten solution \eqref{horavawitt}. 
The component $zy$, instead, will be kept to be the same as in the vacuum 
\begin{equation}
[{\tilde F}_{2}]_{zy} = \frac{1}{\pi U^6}\frac{2 m_\pi^2}{m^2_{\eta'}} \sin \varphi\,.
\end{equation}
This zero mode leaves all equations untouched apart from the one for $\widehat{A}_z^\mathrm{mass}$ (\ref{tre}). In terms of $C_{7}$ it reads
\begin{equation}
[d C_{7}]_{0123, S^4} = -\frac{3 c m_q}{2\pi} \sin \varphi \ .
\end{equation}
To determine the perturbation $\mathcal{A}^\mathrm{mass}$, we expand the equations of motion to first order in $m_q$ (with $\mathcal{A}^\mathrm{mass}$ being of  ${\cal O}(m_q)$). The resulting equations for the mass perturbation will be mixed by the presence of the Chern--Simons terms, making it very difficult to find a solution. The following arguments will enable us to simplify the problem. 

There are three different regions in which we can divide the space: $\xi \ll 1$, $\rho \ll \xi \ll 1$ and $ \rho \ll \xi$.\footnote{There is also another  ``large scale'' region $\xi > \log{\lambda}/M_{KK}$, beyond the asymptotic one,  where non-linear effects  become important, for example for the computation of the  baryon charge form factor at large distance \cite{Bolognesi:2013nja}. The existence of this large scale is ignored in the present computation and it does not affect the neutron electric dipole moment computation in the following Section, at least for large $\lambda$.}
 We will call them respectively the flat, the overlapping and the asymptotic  region. 
The  flat  region is where the curvature of the metric can be neglected. This is where the WSS BPST-like instanton solution (\ref{Ainst}) has been obtained. This solution has the scaling with $\lambda$ reported in (\ref{scalings}). 

In the asymptotic region the original WSS instanton solution gets modified. Far from the origin the warp factors $k(z)$ and $h(z)$ cannot be neglected anymore and the asymptotic solution,\footnote{This is a little bit different from the one in \cite{SS-form} because we have not considered the gauge group orientation moduli yet (this will be done in the following section); moreover here all the moduli of the solution are time independent.} in singular gauge, that replaces ${\cal A}^{\mathrm{inst}}$ in (\ref{wssinst}) reads
\begin{equation}
\begin{aligned}
\widehat{A}_0 &=-\frac{N_c}{2\kappa \lambda}G(\mvec{x},z,\mvec{X},Z)\,,\\
\widehat{A}_i &= \widehat{A}_z = 0\,, \phantom{\frac12}\\
A_0 &= 0 \,, \phantom{\frac12}\\
A_i &= -2\pi^2\rho^2 \tau^a\left(\varepsilon^{iaj}\frac{\partial}{\partial X^j}-\delta^{ia}\frac{\partial}{\partial Z}\right)G(\mvec{x},z,\mvec{X},Z)\,, \\
A_z &= -2\pi^2 \rho^2 \tau^a \frac{\partial}{\partial X^a}H(\mvec{x},z,\mvec{X},Z)\,,
\label{asysol}
\end{aligned}
\end{equation}
where
\begin{equation}
\begin{aligned}
G(\mvec{x},z,\mvec{X},Z) &= -\kappa \sum_{n=1}^\infty \psi_n(z)\psi_n(Z)\frac{e^{-\sqrt{\lambda_n}r}}{4\pi \,r}\;,\quad r=|\mvec{x}-\mvec{X}|\,,\\
H(\mvec{x},z,\mvec{X},Z) &= -\kappa \sum_{n=0}^\infty \phi_n(z)\phi_n(Z)\frac{e^{-\sqrt{\lambda_n}r}}{4\pi \,r}\,,
\label{GHgreen}
\end{aligned}
\end{equation}
and the functions $\psi_n, \phi_n$ are the same that have been introduced in the meson sector in Section \ref{section:mesons} (see also Appendix \ref{mesons}) and $\lambda_0\equiv0$. Actually, since the asymptotic expansions above contribute to the currents in the WSS model \cite{SS-form}, they account for the meson contributions to e.g. the form factors.

From (\ref{asysol}) we see that there is a suppression of an overall $\lambda$ factor for each field; moreover the functions $G(\mvec{x},z,\mvec{X},Z)$ and $H(\mvec{x},z,\mvec{X},Z)$ are of order $\sim e^{-r}$ in $r$, $\sim 1/z$ in $z$ and $\sim 1/r$ in $r$, $\sim 1/z^2$ in $z$ respectively. 
In the overlapping region the solution is again \eqref{asysol} but with the functions $G$ and $H$ replaced by the flat Green's function $G^\mathrm{flat} = -1/4\pi^2\xi^2$; the maximum value of the fields is reached when $\xi$ approaches $\rho$, so the scaling is precisely \eqref{scalings}, but here this behavior is reached as an upper limit (see \tablenamev \ref{tabscaling2}).
\begin{table}[h]
\centering
\begin{tabular}{l|ccc}
\hline  
 &Flat & Overlapping & Asymptotic \\
\hline
Region &$\xi \ll 1$ & $\rho \ll \xi \ll 1$ & $\rho\ll\xi$ \\
Solution & BPST instanton & function $G^\mathrm{flat}$ & functions $G$ and $H$\\
Scaling & $\lambda$ scaling & $\lambda$ scaling (limit) & $z$ and $r$ scaling \\
\hline
\end{tabular}
\caption{\label{tabscaling2} Scalings of the solution in the different regions.}
\end{table}

With this in mind let us look at the Chern--Simons terms in equations (\ref{uno})-(\ref{quattro}); in the asymptotic region all of them will be negligible as they are quadratic in the fields, in the other two regions however some of them have to be considered. 
If we look at \eqref{scalings} we conclude that, whenever an $\mathcal{A}_0$ is present in a Chern--Simons term, its $\lambda$ scaling is lowered, so the leading terms will be those with $\mu=0$. 
In fact in the equations for the $\mu = 0$ components all terms are of the same order in $\lambda$, while in those for the $\mu = i$ or $z$ components, the Chern-Simons terms happen to be suppressed as $1/\lambda$ with respect to the Yang-Mills terms, hence we will drop them in the following.

Now we are ready to write down the equations for the mass perturbation (gauge fields without superscript are $\mathcal{A}^{\mathrm{inst}}$ or the ones in (\ref{asysol}), our convention is $\varepsilon_{0123z}=- \varepsilon^{0123z}=1$). Up to subleading terms they read
\begin{align}
&-\kappa \left(h(z)\partial_\nu \widehat{F}^{0\nu}_{\mathrm{mass}} + \partial_z(k(z)\widehat{F}^{0 z}_{\mathrm{mass}})\right) - \frac{N_c}{128\pi^2}\varepsilon^{ijk}
\left(4F_{ij}^aF_{kz}^{a,\,\mathrm{mass}} + 4F_{iz}^aF_{jk}^{a,\,\mathrm{mass}}\right)=0 \label{unoz}\,, \\
&-\kappa \left(h(z)\partial_\nu \widehat{F}^{i\nu}_\mathrm{mass} + \partial_z(k(z)\widehat{F}^{i z}_\mathrm{mass})\right) =0 \label{unon}\,, \\
&-\kappa \Bigl(h(z)D_\nu {F^{0\nu}} + D_z(k(z){F^{0 z}})\Bigr)^a{\big|_\mathrm{mass}} - \frac{N_c}{64\pi^2}\varepsilon^{ijk}\left(
 2 F_{ij}^a\widehat{F}^{\mathrm{mass}}_{kz} +  2 F_{iz}^a\widehat{F}^{\mathrm{mass}}_{jk}\right)=0 \label{duez}\,, \\
&-\kappa \Bigl(h(z)D_\nu F^{i\nu} + D_z(k(z)F^{i z})\Bigr)^a{\big|_\mathrm{mass}}=0 \phantom{\frac{1}{2}} \label{duen}\,, \\
&-\kappa\, k(z)\partial_\nu\widehat{F}^{z\nu}_\mathrm{mass} = -\chi_g\left(\theta + \sqrt{\frac{N_f}{2}}\int \widehat{A}_z dz\right)^2-ic\, {\rm Tr} {\cal P}\left[M e^{ - \frac{i}{\sqrt{2N_f}} \int \widehat{A}_z dz}\U_0 - {\rm c.c.}\right] \label{tren}\,, \\
&-\kappa\,k(z) (D_\nu F^{z\nu})^a{\big|_\mathrm{mass}}= -ic\, {\rm Tr}  {\cal P}\left[M \tau^a e^{ - \frac{i}{\sqrt{2N_f}} \int \widehat{A}_z dz}\U_0 - {\rm c.c.}\right]\,,
\label{quattron}
\end{align}
where 
\be
\U_0 = - \cos\alpha \unit + \cdots \,,\quad \alpha \equiv \pi/\sqrt{1+\rho^2/r^2}\,,
\ee
and $\cdots$ denote terms which do not contribute to the trace. 

The notation ${\big|_\mathrm{mass}}$ means ``pick up the linear contribution in $m_q$". For now we work in the static gauge and we admit no time dependence for $\mathcal{A}_\mathrm{mass}$ (so the indexes ``$\nu$" in the equations above become ``$j$").

The above system of equations can be divided into four parts:
\begin{enumerate}[\textit{\roman{enumi})}]
\item Abelian space component equations \eqref{unon}, \eqref{tren}.
\item Non Abelian time component equation \eqref{duez}.
\item Non Abelian space component equations \eqref{duen}, \eqref{quattron}.
\item Abelian time component equation \eqref{unoz}.
\end{enumerate}
\subsection{Abelian field: space components}\label{sec:abfield} 
A consistent solution to the set \textit{i)} can be found with the ansatz $\widehat{A}_i^\mathrm{mass} = 0$. We will verify in the end this assumption.

Let us first notice that \eqref{tren} can be rederived starting from the effective action for the Abelian component $\widehat A_z$, which, to first order in the mass deformation, reads
\be
{\cal L}_{eff\, z} = -\frac{\kappa}{2} k(z) \widehat{F}_{\mu z} \widehat{F}^{\mu}_z + c\, {\rm Tr} {\cal P}\left[M e^{ - \frac{i}{\sqrt{2N_f}} \int \widehat{A}_z dz}\U_0 + {\rm c.c.}\right] - \frac{\chi_g}{2} \left(\theta + \sqrt{\frac{N_f}{2}}\int \widehat{A}_z dz\right)^2\,.
\label{effaz}
\ee
Focusing on the $N_f = 2$ mass degenerate case and using the condition \eqref{vacuumphi}, we see that the equation of motion \eqref{tren} reads
\begin{align} 
& \kappa\, k(z)\partial_i\partial^i\widehat{A}_z^\mathrm{mass}  = 2 c m_q \sin \varphi(\cos \alpha +  1) \,.
\label{treu}
\end{align}
Writing the equation as above, we have neglected the mass term for $\widehat{A}_z^\mathrm{mass}$, which would arise from the effective Lagrangian (\ref{effaz}). Recalling that $\int\widehat{A}_z dz$ is holographically related to the $\eta'$ field, we see that this term actually corresponds to the $\eta'$ mass. To leading order in the small quark mass limit, the latter is given by the Witten-Veneziano relation (\ref{mWV}), which shows, in turn, that the squared $\eta'$ mass is a parameter of ${\cal O}(\epsilon_f)$. Since we are working in the probe approximation, the $\eta'$ mass term is thus subleading. We will return to this point in Section \ref{sec:FF} where we will see that the $\eta'$ mass term can be used to regularize the integral which defines the full electromagnetic dipole form factor.  

The $z$ dependence of equation (\ref{treu}) can be factorized by setting
\begin{equation}
\widehat{A}^\mathrm{mass}_z = \frac{u(r)}{1+z^2}\,,
\label{azab}
\end{equation}
yielding
\begin{equation}
\frac{1}{r^2}\partial_r(r^2 \partial_r u(r)) = \frac{2 c m_q }{\kappa}  \sin \varphi (\cos \alpha +  1) \label{eqforu} \ .
\end{equation}
When $r\to \infty$ the function $\alpha$ approaches a constant $\alpha\to \pi$, so the source term vanishes. The standard way to solve this equation is to use the Green's function
\begin{equation}
u_G(r,r') = \left\lbrace \begin{aligned}
& - r' & \qquad & r<r' \,,\\
& - r'\left(\frac{r'}{r}\right)  & \qquad & r>r' \,.
\end{aligned}\right.
\label{ugf}
\end{equation}
The solution is given by the following integral
\begin{equation}
u(r) = \frac{2 cm_q}{\kappa}\sin\varphi \int_0^\infty d r'\,u_G(r,r')\left(1+\cos \frac{\pi}{\sqrt{1+\rho^2/{r'}^2}}\right) \ . \label{defupiccolo}
\end{equation}
The above solution is sufficient to identically solve equation \eqref{unon}, hence we can put $\widehat{A}^\mathrm{mass}_i$ to zero: the ansatz claimed at the beginning was correct. 

It may be interesting to see the asymptotic solution for large $\lambda$. Changing variables $r' = \rho y$, since $\rho$ tends to zero, from \eqref{defupiccolo} we get that far away from $r=0$ the solution can be approximated by
\begin{equation}
u(r) \simeq -\frac{2cm_q \sin\varphi}{\kappa}\frac{\rho^3}{r} \gamma\,,\quad \gamma\equiv\int_0^\infty d y\, y^2\left(1+\cos \frac{\pi}{\sqrt{1+1/y^2}}\right)\sim 1.104 \,.
\label{ularger}
\end{equation}
In the following we will focus on the phenomenologically acceptable regime $m_{\pi}\ll m_{\eta'}$ where (for $N_f=2$) $\varphi\approx \theta/2$.
  
\subsection{Non Abelian field: time component} 
\label{azeromass}
Let us now look at equation \eqref{duez}. To first order in $m_q$ the equation for the perturbation is the following
\begin{equation}
\begin{aligned}
&  h(z) D_\nu\left(-\partial^\nu A^0_\mathrm{mass} + i [A^0_\mathrm{mass},A^\nu]\right) + D_z\left(-k(z) \partial^z A^0_\mathrm{mass} +  ik(z) [A^0_\mathrm{mass},A^z]\right)= \\& \qquad \qquad  \qquad  \qquad   \qquad  \qquad \qquad  \qquad \qquad\!\!= - \frac{N_c}{8\pi^2\kappa }\frac{\rho^2}{(\xi^2+\rho^2)^2}\frac{u'(r)}{1+z^2}\frac{(\mvec{x}-\mvec{X})\cdot\mvec{\tau}}{r}\,.
\end{aligned}\label{eqperazero}
\end{equation}
In static gauge the only field excited by this perturbation is $A_0^\mathrm{mass}$. Let us consider the following ansatz
\begin{equation}
A_0^\mathrm{mass} = W(r,z)(\mvec{x}-\mvec{X})\cdot\mvec{\tau}\,.
\end{equation}
When plugging this ansatz into the equations, the $(\mvec{x}-\mvec{X})\cdot\mvec{\tau}$ piece factorizes and we are left with a partial differential equation for $W$
\begin{equation}
\begin{aligned}
& h(z) \left(\partial_r^2 W(r,z) + \frac{4}{r}\partial_r W(r,z) + \frac{8\rho^2}{(\xi^2+\rho^2)^2}W(r,z)\right) + \partial_z(k(z) \partial_zW(r,z)) = \\ & \qquad \qquad  \qquad  \qquad \qquad  \qquad \qquad  \qquad \quad= \frac{27\pi}{\lambda} \frac{\rho^2}{(\xi^2+\rho^2)^2}\frac{1}{r}\frac{u'(r)}{1+z^2} \equiv {\cal G}(r,z) \,. \label{eqW}
\end{aligned}
\end{equation}
It is worth noting that this equation has been derived using as background the BPST-like instanton solution (\ref{Ainst}), valid in the ``flat'' part of the geometry.
Nevertheless, one can check that in the ``asymptotic'' region one would obtain precisely the expansion of equation (\ref{eqW}) for large $z$.
Thus, this equation is correct in the whole range of the radial variable.

There are two possible approaches that can be used to solve equation (\ref{eqW}): a) numerical PDE analysis; b) expansion in the eigenfunctions $\psi_n$. 
The latter, which we are going to describe here, provides interesting insights about the physical content of our results \cite{nedmshort}. The direct numerical analysis will be used later in the review of the calculation of the NEDM.

The last term in the $l.h.s$ of eq. \eqref{eqW}, being essentially the $l.h.s$ of the eigenvalue equation for the $\psi_n$ (\ref{eqforpsi}), suggests an expansion of the form
\begin{equation}
W(r,z) = \sum_{n=1}^\infty R_n(r) \psi_n(z)\,.
\label{espa}
\end{equation}
Inserting the expansion into the equation, using the eigenvalue equation (\ref{eqforpsi}) and the orthonormality conditions on the $\psi_n$ we find \cite{nedmshort}
\begin{eqnarray}
& \partial_r^2 R_m(r) + \frac{4}{r}\partial_r R_m(r) - \lambda_m R_m(r) + \\
&\sum_{n=1}^\infty \left\langle m \left| \frac{8\rho^2}{(\xi^2+\rho^2)^2}\right|n\right\rangle R_n(r) = \langle m | h^{-1}\, {\cal G} \rangle  \label{eqproj}\,, \nonumber
\end{eqnarray}
where
\begin{eqnarray}\label{matrices}
& \left\langle m \left| \frac{8\rho^2}{(\xi^2+\rho^2)^2}\right|n\right\rangle \equiv \kappa \int dz\, h(z) \psi_n(z)\psi_m(z) \frac{8\rho^2}{(\xi^2+\rho^2)^2}\,, \nonumber \\
& \langle m | h^{-1}\,{\cal G}\rangle \equiv \kappa\int dz\, \psi_m(z) {\cal G}(r,z)\,.
\end{eqnarray}
With the solution of (\ref{eqforpsi}) and (\ref{eqforu}) in hand, one can obtain an approximate solution of the above system by truncating it at some level $m$.
\subsubsection{The solution in the ``flat region"}
In order to gain intuition on the physical meaning of the solution, let us consider the flat region around $z=0$, where we can neglect the curvature effects driven by the functions $h(z), k(z)$. In this limit the equation (\ref{eqW}) reads
\begin{equation}
\partial_r^2 W(r,z) + \frac{4}{r}\partial_r W(r,z) + \frac{8\rho^2}{(\xi^2+\rho^2)^2}W(r,z) + \partial^2_z W(r,z) \approx \frac{27\pi}{\lambda} \frac{\rho^2}{(\xi^2+\rho^2)^2}\frac{u'(r)}{r}\,. \label{eqWz0}
\end{equation}
Let us also consider the $r\gg0$ limit, where the function $u(r)$ is given by eq. (\ref{ularger}). In this limit a solution of the above equation is simply
\be
W\approx \frac{27\pi}{8\lambda}\frac{u'(r)}{r} \approx \frac{27\pi}{8\lambda}\frac{c m_q}{\kappa}\gamma\, \theta\, \frac{\rho^3}{r^3}\,.
\ee  
As a result we can write (setting $\mvec{X}=0$, which we can do without loss of generality)
\begin{equation}
A_0^\mathrm{mass}= W \mvec{x}\cdot\mvec{\tau}\approx \frac{\mvec{D}\cdot \mvec{x}}{|\mvec{x}|^3}\,,
\end{equation}
where
\be
\mvec{D} =  \frac{27\pi}{8\lambda}\frac{c m_q}{\kappa}\gamma\, \theta\, \rho^3 \mvec{\tau}\,.
\ee
The above expression recalls that of an electric dipole term in the five dimensional space (at $z=0$) induced by the $\theta$ parameter. 
As we will see in Section \ref{nedmsec}, this is precisely what contributes to the electric dipole term in the dual four dimensional gauge theory.
\subsection{Non Abelian field: space components}
\label{solnonabelianApp}
The solutions we have discussed above exhaust the list of leading ${\cal O}(\theta)$ corrections to the original WSS instanton solution. At first order in $m_q$, however, we have also to consider the corrections coming from solutions to the non Abelian equations (item \textit{ii)} in the list given above). Since in the present work we are mainly interested just in the ${\cal O}(\theta)$ corrections, we present here the formal solutions to those equations discussing only their algebraic structure. 

Before expanding in $m_q$, the equations we have to consider read
\begin{equation}
\begin{aligned}
D_M F_{Mi}^a &= 0\,,\\
D_M F_{Mz}^a &= \frac{2c m_q}{\kappa} \cos\frac{\theta}{2} \frac{(x-X)^a}{r}\sin \left(\frac{\pi}{\sqrt{1+\rho^2/r^2}}\right)\,.
\end{aligned}\label{eqnonabelian}
\end{equation}
Let us first rewrite the background instanton fields as
\begin{equation}
A_M^{a,\,\mathrm{inst}}  = -\eta^a_{MN} \partial_N \log f_0(\xi)\;,\quad f_0(\xi) = 1+\frac{\xi^2}{\rho^2}\,, \label{thooftsolution}
\end{equation}
where the $\eta^a_{MN}$ are the 't Hooft symbols, which constitute a basis for the self dual tensors. The above solution represents an instanton with instanton number +1. The anti--instanton is given by the same expression with $\eta$ replaced by $\overline{\eta}$, where
\begin{equation}
\begin{aligned}
\eta^a_{MN} &= \varepsilon_{aMNz} + \delta_{aM}\delta_{Nz} - \delta_{aN}\delta_{Mz}\,, \\
\overline{\eta}^a_{MN} &= \varepsilon_{aMNz} - \delta_{aM}\delta_{Nz} + \delta_{aN}\delta_{Mz}\,.
\end{aligned} \label{defthooftsymb}
\end{equation}
Our ansatz will be composed by two functions, one modifies the $f_0$ and the other will be an extra contribution to $A_z$
\begin{equation}
A_M^a = -\eta^a_{MN} \partial_N (\log f_0(\xi) + \phi(r,z)) + \delta_{Mz} \partial_a \psi(r)\,.
\end{equation} 
Notice the different arguments in $\phi(r,z)$ and $\psi(r)$: we will see later that this is the correct assumption. These two functions have to be regarded as $\mathcal{O}(m_q)$, 
so the resulting equations will be linear in them (of course the zeroth order is already satisfied by $f_0$).

The most lengthy part now consists in putting the ansatz above into equations \eqref{eqnonabelian} and write down the equations for $\phi$ and $\psi$. Let us first focus on the tensor structure
\begin{equation}
\begin{aligned}
& \mbox{\hspace{-1cm}\footnotesize With the ansatz $\phi(r,z)$} &\hspace{2cm} & \mbox{\hspace{-1cm}\footnotesize With the ansatz $\psi(r)$} \\
D_M F_{Mi}^a &= - \varepsilon_{aij} x^j \bigl(\mbox{\footnotesize $\phi$ eqn.}\bigr)\,, &
D_M F_{Mi}^a &=  \varepsilon_{aij} x^j \bigl(\mbox{\footnotesize $\psi$ radial eqn.}\bigr)\,,  \\
D_M F_{Mz}^a &=  x^a \bigl(\mbox{\footnotesize $\phi$ eqn.}\bigr)\,,  &
D_M F_{Mz}^a &=  x^a \bigl(\mbox{\footnotesize $\psi$ zeta eqn.}\bigr)\,.
\end{aligned}\label{perturbansatz}
\end{equation}
As we can see the structure is very simple; moreover we have three different parenthesis, the ones with $\phi$ (they are identical) and the two different ones with $\psi$. The latter will be written down here in a simpler case where the function $\phi$ depends only on $\xi$ (the general solution will be given in the following)
\begin{equation}
\begin{aligned}
\bigl(\mbox{\footnotesize $\psi$ radial eqn.}\bigr) &= -\frac{8\rho^2}{r(\xi^2+\rho^2)^2}\psi'(r)\,,\\
\bigl(\mbox{\footnotesize $\psi$ zeta eqn.}\bigr) &= -\frac{2(\rho^4+\xi^4 -2\rho^2 r^2 + 2\rho^2z^2)}{r^3(\xi^2+\rho^2)^2}\psi'(r) + \frac{2}{r^2}\psi''(r) + \frac{1}{r}\psi'''(r)\,.
\end{aligned}
\end{equation}
The third derivative comes from the fact that in our definition of $A_M$ only the derivatives of $\phi$ and $\psi$ enter. The actual variables thus are $\Phi \equiv \phi'(\xi)$ and $\Psi\equiv \psi'(r)$. The equations we were looking for finally read
\begin{equation}
\begin{aligned}
-&\bigl(\mbox{\footnotesize $\phi$ eqn.}\bigr) + \bigl(\mbox{\footnotesize $\psi$ radial eqn.}\bigr) = 0\,, \\
&\bigl(\mbox{\footnotesize $\phi$ eqn.}\bigr) + \bigl(\mbox{\footnotesize $\psi$ zeta eqn.}\bigr)
=\frac{2c m_q}{\kappa} \cos\frac{\theta}{2} \frac{1}{r}\sin \left(\frac{\pi}{\sqrt{1+\rho^2/r^2}}\right)\,.
\end{aligned}
\end{equation}
Combining these equations one gets 
\begin{equation}
\begin{aligned}
&-\frac{2}{r^2}\Psi(r) + \frac{2}{r}\Psi'(r) + \Psi''(r) =
\frac{2c m_q}{\kappa} \cos\frac{\theta}{2}\sin \left(\frac{\pi}{\sqrt{1+\rho^2/r^2}}\right)\,, \\ 
&-\frac{3(\rho^4 + \xi^4 - 6\rho^2\xi^2)}{\xi^3(\xi^2+\rho^2)^2}\Phi(\xi) + \frac{3}{\xi^2}\Phi'(\xi) + \frac{1}{\xi} \Phi''(\xi) =  -\frac{8\rho^2}{r(\xi^2+\rho^2)^2}\Psi(r)\,.
\end{aligned}\label{eqforPhiPsi}
\end{equation}
Notice that in the first one  the $\xi$ dependence completely disappears. It is an ODE that can be easily integrated numerically.

In the general case $\phi$ has to be regarded as a two-variable function $\phi(r,z)$. Remarkably, as stated above in \eqref{perturbansatz}, also in this case we have a very simple tensor structure and a dependence on only one parenthesis $\bigl( \phi \mbox{ eqn.}\bigr)$, so all the manipulation made above are still valid. In this case however the equation is far more complicated
\begin{equation}
\begin{aligned}
\bigl(\mbox{\footnotesize $\phi$ eqn.}\bigr)  &= \frac{\phi^{(1,2)}+\phi^{(3,0)}}{r} + \frac{2}{r^2}\phi^{(2,0)} +\frac{\phi^{(1,0)}\left(4r^2(z^2-r^2+5\rho^2)-2(\xi^2+\rho^2)^2\right)}{r^3
   \left(\xi^2 +\rho ^2\right)^2} +\\&+  \frac{4\left(r\phi^{(0,2)} - z\phi^{(1,1)}\right)}{r(\xi^2+\rho^2)}\ - \frac{8z\phi^{(0,1)}}{(\xi^2+\rho^2)^2}\,,
\end{aligned}   
\end{equation}
where for $\phi^{(i,j)}$ we mean $\partial^i_r\partial_z^j\phi(r,z)$.

The final equation to be solved is
\begin{equation}
\bigl(\mbox{\footnotesize $\phi$ eqn.}\bigr)  = - \frac{8\rho^2}{(\xi^2+\rho^2)^2}\frac{\Psi(r)}{r}\,,
\end{equation}
where $\Psi(r)$ is substituted by the solution found above. This equation can be integrated via numerical methods, even though now we are dealing with a PDE which is certainly more challenging. 
We will not show the numerical results here because the only purpose of this Section is to show what is the correct tensor structure of the solution and how to get it.
\subsection{Abelian field: time component}
Let us finally consider equation (\ref{unoz}). In the static case, on the ${\widehat A}^i_{\rm{mass}}=0$ solution, it reduces to an equation for ${\widehat A}^0_{\rm mass}$
\be
h(z)\partial_i\partial_i \widehat{A}^{0}_{\mathrm{mass}} + \partial_z(k(z)\partial_z\widehat{A}^{0}_{\mathrm{mass}}) = \frac{27\pi}{4\lambda}\varepsilon^{ijk}
\left(F_{ij}^aF_{kz}^{a,\,\mathrm{mass}} + F_{iz}^aF_{jk}^{a,\,\mathrm{mass}}\right) \label{ahatzero}\,.
\ee 
The solution, of the form
\be
{\widehat A}^{0}_{\rm{mass}} = f(r,z)\,,
\label{hatazero}
\ee
can be obtained after the equations for the spatial components of the non-Abelian field are solved, in the way we have described in the previous Subsection. Precisely as those components, the field ${\widehat A}^{0}_{\rm{mass}}$ will be of ${\cal O}(\theta^2)$ in the small $\theta$ regime. 
\section{The neutron electric dipole moment}
\setcounter{equation}{0}
\label{nedmsec}
In a theory with spin $1/2$ particles where parity, time reversal and/or charge conjugation symmetries are not preserved, the form factors acquire novel contributions w.r.t. the cases with unbroken discrete symmetries. 
For example, the matrix element of the electromagnetic current between nucleon states of mass $M_N$ in the generic case reads (see e.g. \cite{VicariQCD} and references therein)
\be
\langle p', s' | J^{\mu}_{\rm{em}}|p, s\rangle = {\bar u}_{s'} (p') \Gamma^{\mu}(k^2) u_s(p)\,,
\ee
where $k=p'-p$, $u_s$ is a Dirac spinor with spin component $s$ and
\bea
\Gamma^{\mu}(k^2) &=& F_1(k^2) \gamma^{\mu} + \frac{1}{2M_N} F_2(k^2)i \sigma^{\mu\nu}k_{\nu}+\nonumber \\
&&+ \frac{1}{2M_N} F_3(k^2) \sigma^{\mu\nu}\gamma_5 k_{\nu} + F_A(k^2)( \gamma^{\mu}\gamma^5 k^2-2M_N\gamma^5 k^{\mu})\,.
\label{defFF}
\eea 
Here, $F_1$ and $F_2$ are the standard (C,P,T even) Dirac and Pauli form factors: when $k^2\rightarrow0$ $F_1(0)$ gives the electric charge of the fermion and $F_2(0)$ gives the anomalous part of the magnetic moment. 

The novel contributions are the dipole ($F_3$) and the anapole ($F_A$) form factors.  When $k^2\rightarrow0$, $F_A(0)$ gives the (T- and C-breaking) anapole moment and $F_3(0)$ gives the (T and P-breaking) electric dipole moment (EDM). In particular, the nucleon EDM reads
\be
d_N = \frac{F_3(0)}{2M_N}\,.
\ee
The QCD Lagrangian with non zero $\theta$ parameter is invariant under charge conjugation and thus the corresponding anapole term vanishes (anapole moments can be induced by electroweak effects). The dipole form factor, instead, is expected to be proportional to $\theta$, in the $\theta\rightarrow0$ limit.

As an example of application of (some of) the instantonic solutions found in Section \ref{sec:newsol}, in this Section we review and discuss in details the holographic computation of the neutron electric dipole moment (NEDM) performed in \cite{nedmshort}. 
Moreover, in Section \ref{sec:FF} we report the computation of the whole form factor $F_3(k)$.

\subsection{NEDM state of the art}
Permanent electric dipole moments of composite or fundamental particles with spin are sensitive observables of CP-violating effects in nature. The electric dipole couples to the electric field in the standard way $\mvec{E}\cdot \mvec{d}$. For a neutral particle, like the neutron, the dipole has to be proportional to the spin, which is a pseudovector, so that $\mvec{E}\cdot \mvec{d}$ is odd under parity and time reversal. 

Experimentally the electric dipole moment of a particle can be obtained by exposing it to an electro-magnetic field and measuring the Larmor frequency shift as the directions of the electric and magnetic fields are flipped. For neutral particles the measurement is much easier, since charged ones are accelerated by the electric field and would better require storage ring experiments. 

The history of the measurement of the neutron electric dipole moment finds its roots in the work by Purcell and Ramsey in 1950 \cite{pr}; since then many experiments followed, but no evidence for the NEDM has been found so far and the latest experimental upper bound is tiny, 
$|d_n|\le 2.9\times10^{-26} e\cdot \mathrm{cm}\, (90\%\, \rm{CL})$ \cite{pendlebury,NEDMexp}. 

This bound on the NEDM is a relevant constraint to take into account when formulating theories beyond the Standard Model (bSM). This is because in most bSM scenarios many new CP-violating effects can arise providing possibly larger NEDM than the tiny Standard Model predictions. Hence any limit on the NEDM leads to bounds on the scales of new physics. 

In principle, the NEDM can be computed by
\begin{equation}
\mvec{D}_{n,s} = \int d^3x\,\mvec{x}\, \langle n,s | J^0_\mathrm{em} |n,s\rangle\,, \label{nedmdef}
\end{equation}
where $|n,s\rangle$ is neutron state with spin $s$ and $J_\mathrm{em}$ the electromagnetic current.
In practice, computing the above matrix element requires using non perturbative tools. 

As we have recalled in the Introduction, the first order-of-magnitude theoretical estimate for the $\theta$ angle contribution to the NEDM, $|d_n|\approx 10^{-16} \, |\theta| e\cdot \mathrm{cm}$ can be found in \cite{baluni,weinberg}. In order to refine this result, various strategies have been adopted.

In lattice QCD there are essentially three possible ways for computing the NEDM (see e.g. \cite{NEDMLattice} for a recent account). A first approach consists in computing the energy difference of neutrons with spin up and spin down in a constant external electric field (see e.g. \cite{latticext}). Another one consists in taking the non-relativistic limit of the CP violating part of the matrix element of the electromagnetic current in the ground state of the neutron. Within this method, the NEDM is obtained from the electromagnetic form factor at zero momentum transfer. Finally, the NEDM can be computed by using an imaginary $\theta$ angle - to overcome the sign problem arising from the fact that the topological term is imaginary in the Euclidean Lagrangian - and then continuing back to real values. 

Lattice studies require a careful analysis of the quark mass dependence of the NEDM. Despite the fact that statistical errors are being reduced in recent lattice QCD computations (with $N_f=2$ or $N_f=2+1$ flavors) with unphysical (e.g. $m_{\pi}\ge 0.5$ GeV) pion masses, quite large systematic and statistical errors arise when pushing the pion mass to the smaller physical value. Most of the recent lattice results (see e.g. \cite{NEDMLattice}) for both $N_f=2+1$ and $N_f=2$ point towards a negative value of $d_n$, modulo the proviso above.

In chiral perturbation theory \cite{CdVVW-NEDM} the strength of the NEDM, to which just the pion cloud contributes, turns out to be proportional to the non-derivative CP-violating pion-nucleon coupling ${\bar g}_{\pi\,N\,N}$. To leading order in the chiral $m_{\pi}\rightarrow0$ limit,
\be
d_n = \frac{g_{\pi N N}\,\bar{g}_{\pi N N}}{4\pi^2 M_N} \log(M_N/m_{\pi})\approx 3.6\times 10^{-16}\, \theta\,  e\cdot \mathrm{cm}\, ,
\label{nedmchiral}
\ee 
where $g_{\pi N N}$ is the CP-preserving pseudoscalar pion-nucleon coupling. Recent computations with $N_f=3$ at next to leading chiral order, actually give $d_n = - (2.9 \pm 0.9)\times 10^{-16}\,\theta\,e\cdot \mathrm{cm}$ \cite{meissner} at the physical pion mass, after second order low energy parameters have been fitted with lattice data. 
 
In the large $N_c$ limit, the NEDM has been computed using the Skyrme model, both with $N_f=2+1$ massive flavors \cite{Dixon}, yielding $d_n = 2\times 10^{-16}\,\theta\,e\cdot \mathrm{cm}$ and in the $N_f=2$ mass degenerate case \cite{Salomonson}, where a slightly smaller value $d_n = 1.4 \times 10^{-16}\,\theta\,e\cdot \mathrm{cm}$ has been obtained. Notice that in both cases the sign of the NEDM is found to be positive. 

As it was pointed out in \cite{Dixon}, the large $N_c$ Skyrme approach gives a scaling $d_n\sim  N_c m_{\pi}^2 \theta$ when the $m_{\pi}\rightarrow0$ limit is taken (after the large $N_c$ one). 
Comparing this with the expression found in chiral perturbation theory \eqref{nedmchiral} we see explicitly how the non-commutativity of the large $N_c$ and the chiral limit show up. 
In particular no logarithmic terms are found in the Skyrme approach. The reason, as it was pointed out in \cite{Dixon}, has to be found in the different mechanisms which give rise to the NEDM in the two cases. In the chiral limit the dominant term comes from a diagram where a neutron first dissociates in a proton and a $\pi^{-}$. In the Skyrme approach, instead, virtual pion contributions are subleading in $1/N_c$.

Actually, at large $N_c$, $g_{\pi N N}\sim N_c^{3/2}$ 
 while ${\bar g}_{\pi N N}\sim m_{\pi}^2 N_c^{x}\,\theta$ where the  precise scaling factor $x$ is not known. 
 Although a first estimate gave $x=1/2$ \cite{schnitzer}, a more careful analysis pointed out that $x\le - 1/2$ \cite{gpiNNSkyrme}. The latter result would imply that at large $N_c$ the virtual pion contribution to the dipole moment (from \eqref{nedmchiral}) would scale at most like $d_n\sim m_{\pi}^2 \log(m_{\pi})\theta$ and would thus be subdominant w.r.t. the ``direct" Skyrme contribution $d_n\sim N_c m_{\pi}^2\theta$. The Skyrme computation is actually similar to the one we are going to perform for the WSS model: we can almost make a ``dictionary'' to translate our quantities with the ones in the Skyrme model. 
For instance the Skyrmion solution corresponding to a baryon here is the instanton $A^\mathrm{inst}$. The holographic model naturally extends the Skyrme one by including the contribution of the whole tower of vector mesons.

In \tablenamev \ref{tabNEDM} we summarize the estimates of the NEDM coming from different approaches, including the one in the WSS model which has already been presented in \cite{nedmshort}. In the following we are going to review that result in detail, adding further comments. Notice that in the list a previous holographic estimate \cite{hongNEDM} appears too. That result has been obtained in a simpler and less controllable bottom-up model (hard-wall) with no string theory embedding. 
\begin{table}[h!]
\centering
\begin{tabular}{llr@{\hskip -.001cm}l}
\hline
Year & Approach/model &   & $c_n= d_n/(\theta\cdot 10^{-16} e\cdot\mathrm{cm})$ \\
\hline
1979   \cite{baluni}           & bag model                       &  &      \phantom{$-1$}2.7 \\
1980  \cite{CdVVW-NEDM} & ChPT                              &  &     \phantom{$-1$}3.6 \\
1981 & ChPT                                                             &  &      \phantom{$-1$}1   \\
1981 & ChPT                                                             & &       \phantom{$-1$}5.5 \\
1982 & ChPT                                                             &  &               \phantom{$-$}20   \\
1984 & chiral bag model                                             &   & \phantom{$-1$}3.0 \\        
1984 & soft pion Skyrme model                                   &   & \phantom{$-1$}1.2 \\
1984 & single nucleon contribution                                 &   &  \phantom{$-$}11 \\
1990 \cite{Dixon}
     & Skyrme model $N_f=3$                                        &   &  \phantom{$-1$}2 \\
1991 \cite{Salomonson}
     & Skyrme model $N_f=2$                                      &   & \phantom{$-1$}1.4 \\
1991 & ChPT                                                                  &   &  \phantom{$-1$}3.3(1.8)       \\
1991 & ChPT                                                                        & &  \phantom{$-1$}4.8            \\
1992 & ChPT                                                                  & &  \phantom{$1$}$-7$.2,\,$-3$.9     \\
1999 & sum rules                                                             & &  \phantom{$-1$}2.4(1.0)  \\
2000 & heavy baryon ChPT                                                     & &  \phantom{$-1$}7.5(3.2) \\
2004 & instanton liquid                                                                    &    &  \phantom{$-$}10(4) \\
2007 \cite{hongNEDM} & holographic ``hard--wall''                      & &  \phantom{$-1$}1.08 \\
2015 \cite{NEDMLattice}   & Lattice QCD                                     &  &  \phantom{$1$}$-3$.9(2)(9) \\
2016 \cite{nedmshort}&WSS model & &      \phantom{$-1$}$1.8$ \\
\hline
\end{tabular}
\caption{An account of theoretical values for $d_n  = c_n\,  \theta \,   10^{-16} \, e\cdot \mathrm{cm}$; the table is partially taken from \cite{VicariQCD}, where all the original references are indicated. ChPT means Chiral Perturbation Theory. }
\label{tabNEDM}
\end{table}
\subsection{The currents}
In order to compute the NEDM using (\ref{nedmdef}), we need to recall, from \cite{SS-form}, how currents are holographically defined in the WSS model. 

Let us first introduce an external field in the theory by switching on non--normalizable modes for the gauge field $\mathcal{A}_\mu$, so that
\begin{equation}
\lim_{z\to\pm\infty} \mathcal{A}_\mu(x^\mu,z) = \mathcal{A}_{\mu\,L(R)}(x^\mu)\,.
\end{equation}
These modes can be seen as perturbations over the background (that approach zero at infinity), whose boundary values are kept fixed. The theory is now modified and we expect an additional term in the action
\begin{equation}
S{\big|_{\mathcal{A}_{L(R)}}} = -2\int d^4x\,\Tr\left(\mathcal{A}_{\mu\,L}\mathcal{J}^\mu_L + \mathcal{A}_{\mu\,R}\mathcal{J}^\mu_R\right)\,, 
\end{equation}
which is a source--current coupling. This term defines the chiral currents $\mathcal{J}^\mu_{L(R)}$ which turn out to be given by
\begin{equation}
\begin{aligned}
\mathcal{J}_{\mu\,L}&=-\kappa[k(z)\mathcal{F}_{\mu z}]_{z\to\infty}\,,\\
\mathcal{J}_{\mu\,R}&= +\kappa[k(z)\mathcal{F}_{\mu z}]_{z\to-\infty}\,.
\end{aligned}
\end{equation}
The axial and vector currents, associated to the vector ($+$) and axial ($-$) fields
\begin{equation}
\mathcal{V}^{(+)}_{\mu} =\frac{1}{2}(\mathcal{A}_{\mu\,L}+\mathcal{A}_{\mu\,R}) \;,\quad \mathcal{V}^{(-)}_{\mu} =\frac{1}{2}(\mathcal{A}_{\mu\,L}-\mathcal{A}_{\mu\,R})\,,
\end{equation}
are thus given by
\begin{equation}
\begin{aligned}
\mathcal{J}_{\mu\,V}&=-\kappa[k(z)\mathcal{F}_{\mu z}]^{z\to\infty}_{z\to-\infty}\,,\\
\mathcal{J}_{\mu\,A}&= -\kappa[\psi_0(z) k(z)\mathcal{F}_{\mu z}]^{z\to\infty}_{z\to-\infty}\,,
\end{aligned}\label{JVA}
\end{equation}
where $\psi_0 =\frac{2}{\pi}\arctan (z)$. 

Working in the $\theta=0$ case, in \cite{SS-form} it has been noticed that the above expressions are consistent with the source--current term in the four-dimensional action for the mesonic modes
\begin{equation}
S{\big|_{\mathcal{A}_{L(R)}}} = 2\int d^4x\,\Tr\left[\mathcal{V}^{(+)}_{\mu}\sum_{n=1}g_{v^n}v^n_\mu + \mathcal{V}^{(-)}_{\mu}\left(\sum_{n=1}^\infty g_{a^n}a^n_\mu + f_\pi \partial_\mu\Pi\right)\right]\,, 
\label{SJ}
\end{equation}
where $a_\mu^n$ and $v_\mu^n$ are, respectively, the axial-vector and vector mesons while $\Pi$ contains the pion (non Abelian part) and the $\eta'$ singlet (Abelian part). The decay constants $g_{v^n}$ and $g_{a^n}$ are given in terms of boundary values of the eigenfunctions $\psi_n$
\begin{equation}
g_{v^n} = -\kappa[k(z) \partial_z\psi_{2n-1}(z)]^{z\to\infty}_{z\to-\infty}\;,\quad
g_{a^n} = -\kappa[k(z) \partial_z\psi_{2n}(z)\psi_0(z)]^{z\to\infty}_{z\to-\infty}\,.
\label{gavn}
\end{equation}
The fact that the vector current ${\cal J}_{V\,\mu}$, as it can be read from (\ref{SJ}), is expressed as a sum over the vector meson modes $v_{\mu}^n$, reflects the complete vector meson dominance of the model. 

Splitting the Abelian and non Abelian parts of the currents as in \eqref{separ} we get the isoscalar and isovector contributions. In particular, in the case with $N_f=2$ flavors, the electromagnetic current is given by
\begin{equation}\label{currentem}
J_{\mu\,em}=-\kappa\left[k(z)\Tr(F_{\mu z}\tau^3)+\frac{k(z)}{N_c}{\widehat F}_{\mu z}\right]^{z\to\infty}_{z\to-\infty}\,.
\end{equation}
Inserting in the above definitions the instantonic solutions found in Section \ref{sec:newsol}, the result, after quantization, gives the currents in the presence of baryons. This allows to compute the expectation values $\langle B',s' | \mathcal{J}^{A,V}_\mu|B,s\rangle$, where $|B,s\rangle$ are the baryonic states with spin $s$ which are eigenstates of the baryon Hamiltonian. All the interesting static properties of baryons can be derived with this formalism. 
 
Notice that the ${\cal O}(\theta)$ term $\widehat{A}_z^\mathrm{mass}$ (\ref{azab}), modifies only the axial current $\mathcal{J}_A$ and leaves untouched the vector current $\mathcal{J}_V$. We will return to the axial form factor in Section \ref{gpNNsec}, focusing for the moment on just the electric dipole term.  
\subsection{Quantization reloaded}
The classical soliton solution we have found in Section \ref{sec:newsol} has to be quantized. Both the mass term and the $\theta$ parameter could in principle give corrections to the moduli space Hamiltonian. If this is so, the eigenstates found in Section \ref{barquantization} have to be modified accordingly. 

Crucially, however, we have found that the corrections to the Hamiltonian (i.e. those to the baryon mass formula \eqref{massformula}) are of order $\theta^2$ for small $\theta$: thus, at first order in $\theta$ we can forget about this issue and keep using the baryon eigenstates already found at $\theta=0$. Moreover, the mass term just gives rise to a ${\cal O}(m_q)$ correction to the instanton size $\rho$. We will neglect this correction since it will give rise to a subleading (in $m_q$) contribution to the NEDM.

In order to compute the electromagnetic current we need to switch on the moduli of the gauge group orientations. We would also have to consider the time dependence of  $X^I =\{\mvec{X},Z,\rho\}$, but this gives a subleading ($1/N_c$) effect and we neglect it for the moment. Using translational invariance, we also put $\mvec{X}=0$.\footnote{$\dot{\mvec{X}}\sim \mvec{P}$, the momentum of the baryon, is classically zero since we work in the baryon rest frame. Clearly for $\mvec{P} \neq 0$ we have a non zero electric dipole moment, but it would be just a magnetic moment observed from a boosted frame.} 

Since we now want to maintain $A_0 \neq 0$, we work out a moduli space quantization in a different gauge w.r.t. the one used in (\ref{modrot}). In particular, we use the following transformations\footnote{Another possible choice, which is gauge equivalent to ours, is \cite{SS-3flav}
$A_0\longmapsto A_0' = W(t) A_0 W(t)^{-1} + \Delta(x,t)$, $A_M \longmapsto A_M' = W(t) A_M W(t)^{-1}$ where the function $\Delta(x,t)$ is necessary to solve the equations of motion also in the non stationary case. Defining $Y$ so that $-iY^{-1}\dot{Y}=\Delta(x,t)$ and making the gauge transformation with parameter $Y$ allows us to find exactly \eqref{gaugerot} with $V(x,t) = W(t)Y(x,t)$. Of course many other choices are possible, not necessarily related by gauge transformations; the only important requirement is that the equations of motion remain satisfied.}
\begin{equation}
\begin{aligned}
A_0 &\;\longmapsto\;& A_0' &= V A_0 V^{-1}\,, \\
A_M &\;\longmapsto\;& A_M' &= V A_M V^{-1} - i V \partial_M V^{-1}\,,\quad M=1,2,3,z \,,
\end{aligned}\label{gaugerot}
\end{equation}
with $V \to \bvec{a}$ as $z\to\pm \infty$. After these transformations the $M$ components of the equations of motion for the gauge field remain untouched, while equation \eqref{duez} gives the ``modified Gauss law constraint"
\begin{equation}
\begin{aligned}
&-\kappa \Bigl(h(z)D_\nu {F^{0\nu}} + D_z(k(z){F^{0 z}})\Bigr)^a{\big|_\mathrm{mass}} +  \left(\mbox{\footnotesize CS terms}\right) +\\ 
&+  \kappa \Bigl(h(z)D_\nu D^\nu \Phi + D_z (k(z) D_z\Phi )\Bigr)
=0\,,
\end{aligned}
\label{mgl}
\end{equation}
where $\Phi = -i V^{-1}\dot{V}$ and the time dependence of the moduli $\rho$, $Z$ and $\mvec{X}$ has been neglected. The first row is automatically zero on the solution for $A_0^\mathrm{mass}$ given in (\ref{azab}). Since, to compute the currents, we just need the asymptotic behavior for $z\to \infty$, we can just linearize the remaining term as
\begin{equation}
\partial_\nu \partial^\nu \Phi +h(z)^{-1} \partial_z (k(z) \partial_z\Phi)=0\,.
\end{equation}
Neglecting $\partial_0^2$ terms (as we are interested in slowly moving instantons), the asymptotic solution, at any time, can be given as a series expansion in the $\psi_n$
\begin{equation}
\Phi(r,z) \underset{z \gg 1}{\sim} \sum_{n=1}^\infty  -i\bvec{a}^{-1}\dot{\bvec{a}}  \, c_n(r)\psi_n(z)\,, \label{Phiasy}
\end{equation}
where we have implemented the boundary condition $\Phi \to -i\bvec{a}^{-1}\dot{\bvec{a}}$ as $z\to \infty$. Actually, the whole sum must be independent on $r$ when $z\to \infty$, but it  is not necessary for the present discussion to impose this requirement explicitly. The functions $c_n(r)$ contain all the information about the near core behavior of the instanton and of course they depend on the mass. At $m_q = 0$ the solution can be found explicitly and reads (reintroducing the $Z$ modulus dependence only for now)
\begin{equation}
c_n(r) =  \pi \kappa \rho^2 \frac{e^{-\sqrt{\lambda_n}r}}{r}\psi_n(Z)\,.
\label{cennemasszero}
\end{equation}
This just implies that $\Phi \propto G(\mvec{x},z)$ as defined in \eqref{GHgreen}.
\subsection{The holographic computation of the NEDM}
The electric dipole moment is evaluated using the definition \eqref{nedmdef}
\begin{equation}
\mvec{\mathcal{D}}_{B,s} = \int d^3 x\,\mvec{x}\,\langle B,s|\left( J_{V}^{0,\,a=3} + \frac{1}{N_c}\widehat{J}_{V}^0\right)|B,s\rangle\,,
\end{equation}
where the operator in parenthesis is the quantum version of the time component of the electromagnetic charge (\ref{currentem}). 

Let us first notice that the Abelian $\widehat{J}_{V}^0$ piece actually does not contribute to the NEDM since: 1) $\partial_0 {\widehat A}_z^{\rm{mass}}=0$; 2) $[k(z)\partial_z {\widehat A}^0_{\rm{mass}}]^{z\rightarrow\infty}_{z\rightarrow-\infty}$ is a function of $r$, from (\ref{hatazero}), and thus $\int d^3 x\,\mvec{x}[k(z)\partial_z {\widehat A}^0_{\rm{mass}}]^{z\rightarrow\infty}_{z\rightarrow-\infty}=0$ by parity. Let us thus concentrate on the contribution from the non-Abelian field.

After the transformation \eqref{gaugerot} the non Abelian field strength $F_{0z}$ becomes
\begin{equation}
F'_{0z}  = - V(D_z\Phi)V^{-1} - V (D_z A_0) V^{-1}\,,
\end{equation}
where, again, we have neglected $\dot{X}^I$ term. At first sight both $A_0^\mathrm{mass}$ and $\Phi$ may contribute to the NEDM. The current is easily computed from the definition \eqref{JVA}
\begin{equation}
J_V^0 = \kappa\left[k(z)V (\partial_z A^0_\mathrm{mass} + \partial_z \Phi)V^{-1}\right]^{z\to \infty}_{ z\to -\infty}\,,
\end{equation}
where the covariant derivatives have been replaced by ordinary derivatives because when $z\to \infty$ the fields $A^\mathrm{inst}$ and $A^\mathrm{mass}$ are suppressed by powers of $z^{-1}$, so the commutators disappear when the limit is taken. The gauge structure is very simple: we have
\begin{equation}
\begin{aligned}
&&\mbox{\footnotesize for $A^0_\mathrm{mass}$\;:} && V(\mvec{x}-\mvec{X})\cdot \mvec{\tau} V^{-1} & \underset{z\to \pm \infty}{\longrightarrow}(x^j-X^j) \,\bvec{a}\, \tau^j \, \bvec{a}^{-1}\,,\\
&&\mbox{\footnotesize for $D_z\Phi$\;:} && V \bvec{a}^{-1}\dot{\bvec{a}}  V^{-1} & \underset{z\to \pm \infty}{\longrightarrow} - \bvec{a}\dot{\bvec{a}}^{-1}\,.  \end{aligned}
\end{equation}
At this point it is rather obvious that $\Phi$ cannot contribute to the NEDM: the form \eqref{Phiasy} depends only on $r$, so the integral is odd in $\mvec{x}$ and hence it is vanishing. 

The matrix element is evaluated using the identity (see e.g. \cite{SS-chicur})
\begin{equation}
\langle B',s' | \Tr(\bvec{a} \tau^i \bvec{a}^{-1} \tau^a)|B,s\rangle = -\frac{2}{3}(\sigma^i)_{s's}(\tau^3)_{I_3'I_3}\,,
\end{equation}
where $\sigma$ and $\tau$ are Pauli matrices for spin and isospin respectively and the subscripts indicate the matrix elements in the standard representation. 
Using the above expression, we get the following formula for the ``semi-classical" part of the NEDM (i.e. the result before including the $\rho,Z$-dependent parts of the neutron wave function) 
\begin{equation}
\mvec{\mathcal{D}}^{s.c.}_{n,s} = \frac{8\pi}{9} \int_0^\infty d r\,r^4 \,\kappa[ k(z) \,\partial_z W(r,z)]^{z\to\infty}_{z\to-\infty}\,\langle s |\mvec{\sigma}|s\rangle = -\mvec{\mathcal{D}}^{s.c.}_{p,s}\,,
\end{equation}
where the relation with the proton dipole moment comes from the fact that the neutron has isospin $-1/2$ which is the opposite for the proton. As we can see, the dipole moment is proportional to the spin of the particle, as one would expect, and the dipole moment of the neutron has an opposite sign w.r.t. the dipole moment of the proton. 

Factorizing the tensorial structure, we define the ``semi-classical" NEDM $d^{s.c.}_n$, i.e. the leading order contribution in the $1/N_c$ expansion to the NEDM, as
\begin{equation}
d^{s.c.}_n = \frac{8\pi}{9} \int_0^\infty d r\,r^4 \,\kappa[ k(z) \,\partial_zW(r,z)]^{z\to\infty}_{z\to-\infty}\,. \label{finaldipole}
\end{equation}
In the following we present the numerical analysis for this quantity as a function of $\lambda$ for $N_c=3$.

\begin{figure}[h!]
\centering
\includegraphics[scale=.7]{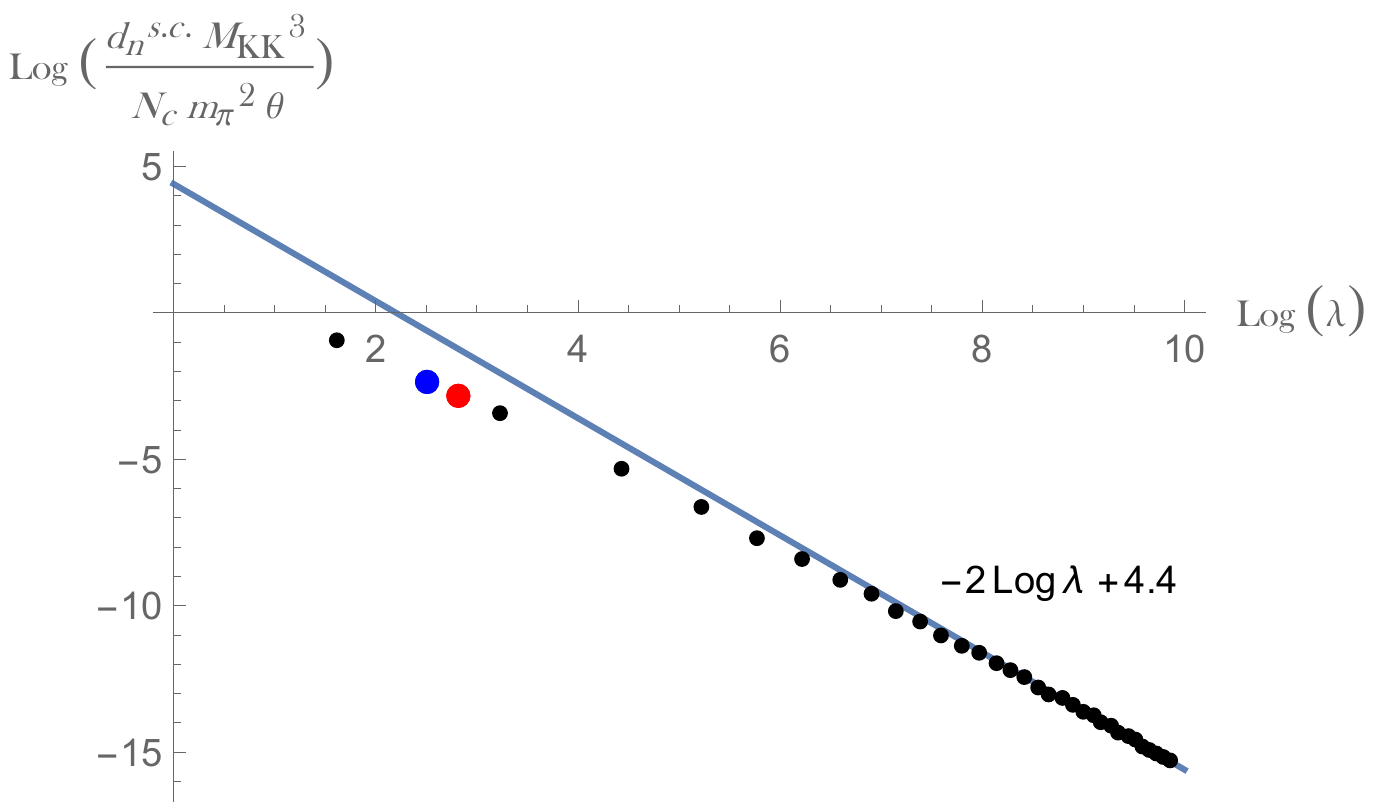}
\caption{Logarithmic plot of the NEDM  as a function of $\lambda$. The points are our numerical results, while the straight line is the $\log\left(d_n^{s.c.} M_{KK}^2/m_q \theta\right) = -2 \log\lambda + 4.4 $ one. 
The red and blue big dots correspond respectively to $\lambda =  16.63, 12.44$.
}
\label{logDipolezmfig}
\end{figure}

The equation for  $W(r,z)$ (\ref{eqW}) can be solved via standard methods of numerical integration, using for example {\fontfamily{ppl}\selectfont \textit{Mathematica}}. 
The dipole is then computed using formula (\ref{finaldipole}).
The result for the NEDM as a function of $\lambda$ is plotted in Figure \ref{logDipolezmfig}. This is a log-log plot of the dimensionless quantity $d_n^{s.c.} M_{KK}^3 / N_c m_{\pi}^2 \, \theta$. 
For large values of $\lambda$ we observe a scaling $d_n^{s.c.} \propto \lambda^{-2}$, specifically
\be
\label{largelambdadipole}
d_n^{s.c.} \simeq 82 \, \frac{N_c m_{\pi}^2 \theta}{\lambda^2M_{KK}^3 }  \,.
\ee
Notice the scaling with $N_c m_{\pi}^2$, a feature in common with the result obtained in the Skyrme model \cite{Dixon}. The above result can be expressed in terms of the quark mass $m_{q}$ using the GMOR relation (\ref{gmor1}) and (\ref{cdef})
\be
\label{largelambdadipolemq}
d_n^{s.c.} \simeq  395  \,  \frac{ m_q \, \theta}{ \lambda^{3/2} M_{KK}^2}\ .
\ee

\begin{figure}[h!]
\centering
\includegraphics[scale=.8]{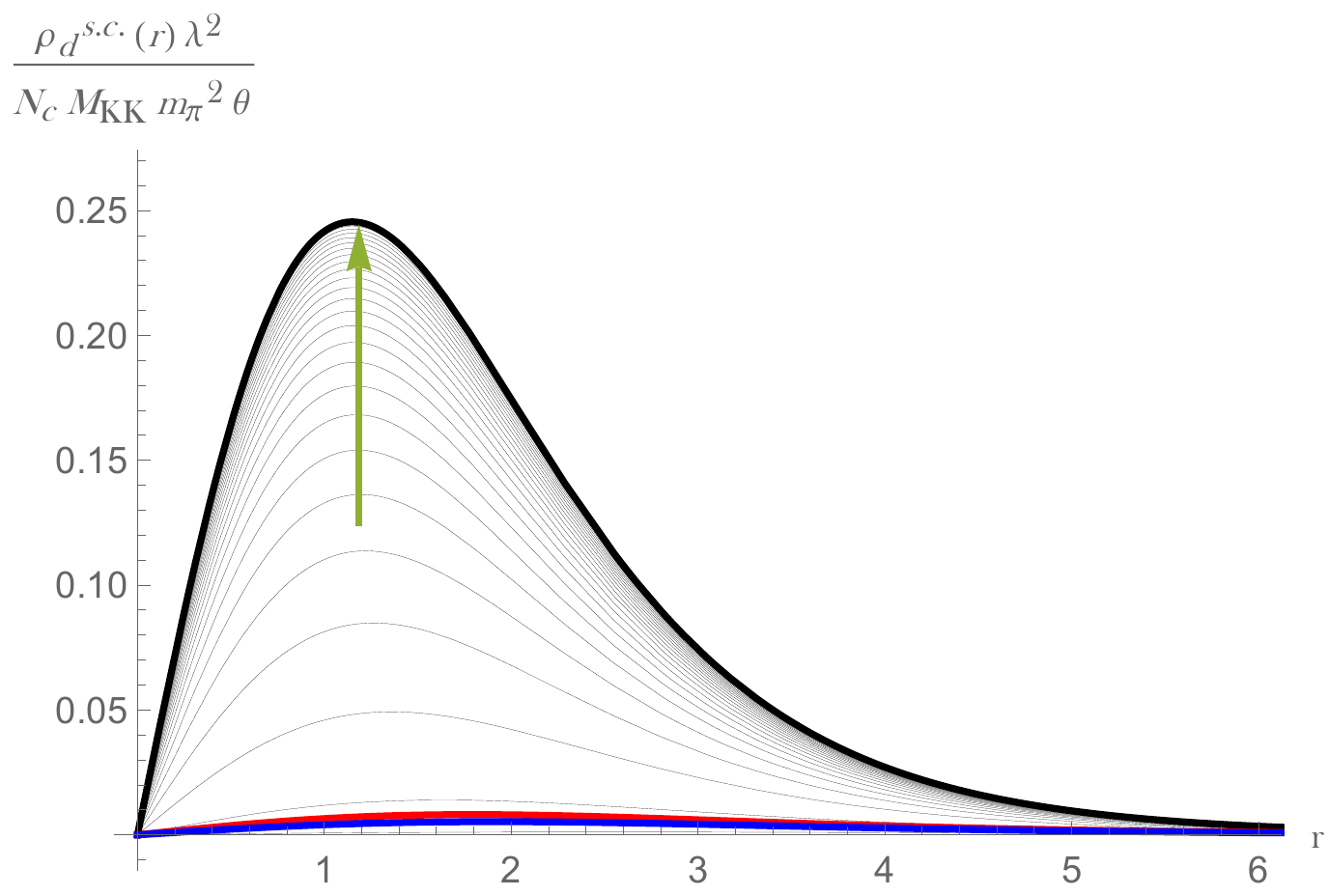}
\caption{Radial distribution of the dipole charge rescaled by $\lambda^{2}$. The arrow indicates the limit as $\lambda$ becomes large. The red and blue thick plots correspond respectively to $\lambda =  16.63, 12.44$.
}
\label{radialdistribution}
\end{figure}

The NEDM can also be written as a dipole moment of a certain charge distribution
\be
\vec{d}_n^{s.c.} = \int d^3r \,  \rho_d^{s.c.}(\vec{r}) \, \vec{r} \ ,
\ee
where $\rho_d^{s.c.}(\vec{r})$ is
\be
\rho_d^{s.c.}(\vec{r}) = (\hat{r} \cdot \hat{s})   \, \frac{16}{9 \pi}  r \,\kappa\,[ k(z) \,\partial_zW(r,z)]^{z\to\infty}_{z\to-\infty}\,,
\label{dipcharge}
\ee
where $\hat{s}$ is the spin direction. 
One advantage of the holographic computation is the possibility to compute also the full charge distribution and not only its dipole moment.
The radial charge distribution, factoring out the angular and the $\theta$ dependence and rescaling by a factor $\lambda^{2}$, is plotted in 
Figure \ref{radialdistribution} for various values of $\lambda$. 
We see that in the large $\lambda$ limit it converges to a certain distribution. The factor $\lambda^{-2}$ of the dipole (\ref{largelambdadipole}) is thus due to an overall scaling of the charge distribution by the same factor; the charge remains always distributed over a length scale of order $\sim 1/M_{KK}$. This interesting feature is shared by other static properties of the WSS baryons, like the size of the baryon number distribution \cite{SS-form}, which is governed by the vector meson inverse mass rather than by the instanton radius $\rho_{cl}\sim{\cal O}(\lambda^{-1/2})$. 

We then perform the numerical analysis with the parameters that are most commonly used in the literature to compare the WSS model with real QCD:
\begin{equation}
\label{parameters}
M_{KK}=949\ {\rm MeV}\,,\qquad \lambda=16.63\,,\qquad m_q=2.92\ {\rm MeV} .
\end{equation}
These parameters are fitted with the experimental observables $f_\pi = 92\, \mathrm{MeV}$ and $m_\rho = 776\, \mathrm{MeV}$. The parameter $m_q$ has been chosen to reproduce correctly the pion mass $m_\pi = 135\, \mathrm{MeV}$ via the GMOR relation (\ref{gmor1}) (using (\ref{cdef}) for $c$). It is a physically acceptable value being in between the up and the down masses. 
These values of the parameters yield
\begin{equation}
d_n^{s.c.} = 0.78 \cdot 10^{-16}\, \theta\;e\cdot \mathrm{cm}\,.
\label{fituno}
\end{equation}
Note that the model allows to automatically include a class of $1/\lambda$ corrections to the leading form of the result (\ref{largelambdadipole}), by solving numerically equation (\ref{eqW}).

Using the solution found in Section \ref{azeromass} the dipole moment can be also expressed as an infinite sum over vector meson modes. Taking into account the mode expansion (\ref{espa}) and the relations \eqref{gavn}, the CP violating part of the non Abelian vector current reads
\begin{equation}
J_{V\,\cancel{CP}}^0 = \kappa\left[k(z)V \partial_z A^0_\mathrm{mass} V^{-1}\right]^{z\to \infty}_{ z\to -\infty}= -\sum_{n=1}^\infty g_{v^n} R_{2n-1}(r) (x^j-X^j) \,\bvec{a}\, \tau^j \, \bvec{a}^{-1}\,.
\end{equation}
Notice that only vector mesons contribute to the NEDM (axial-vector mesons give no contribution): this accounts for the complete vector meson dominance of the model also in the CP-breaking sector.

\begin{figure}[h!]
\centering
\begin{tabular}{cc}
\includegraphics[scale=.6]{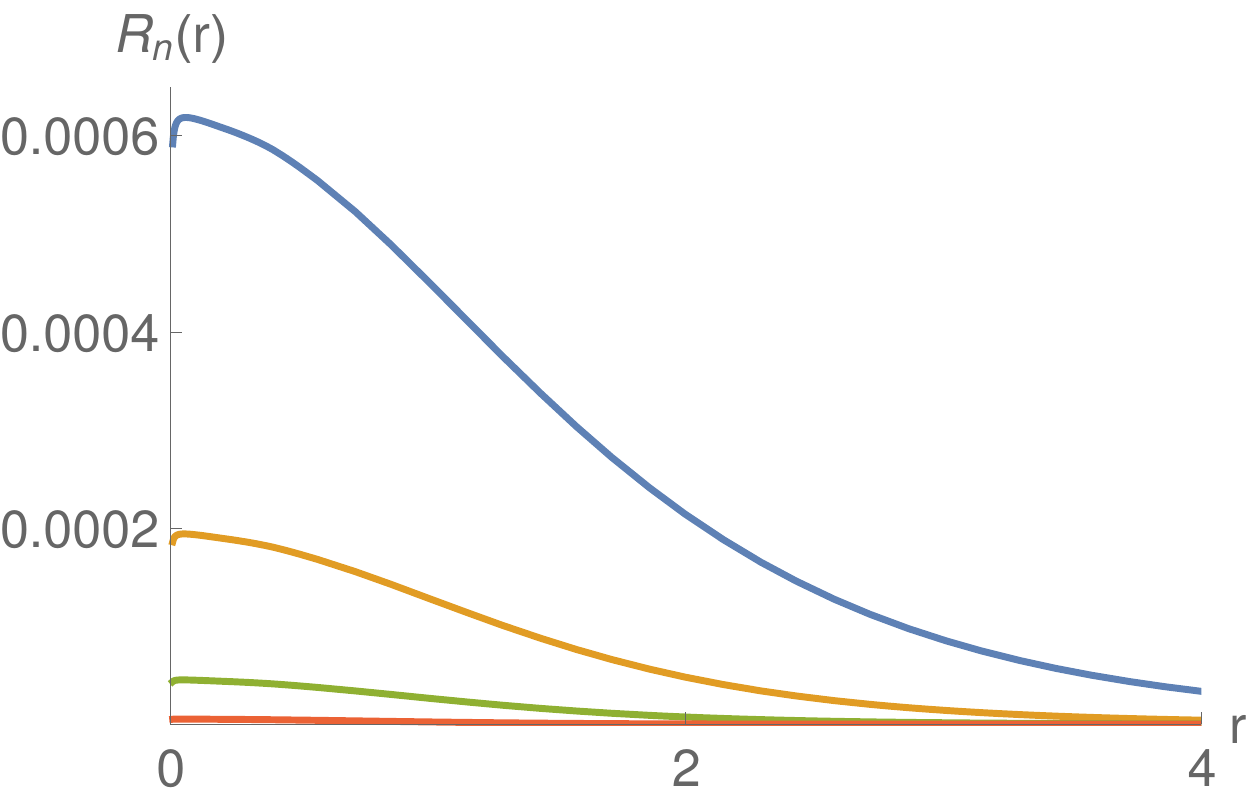} \ &\  \includegraphics[scale=.6]{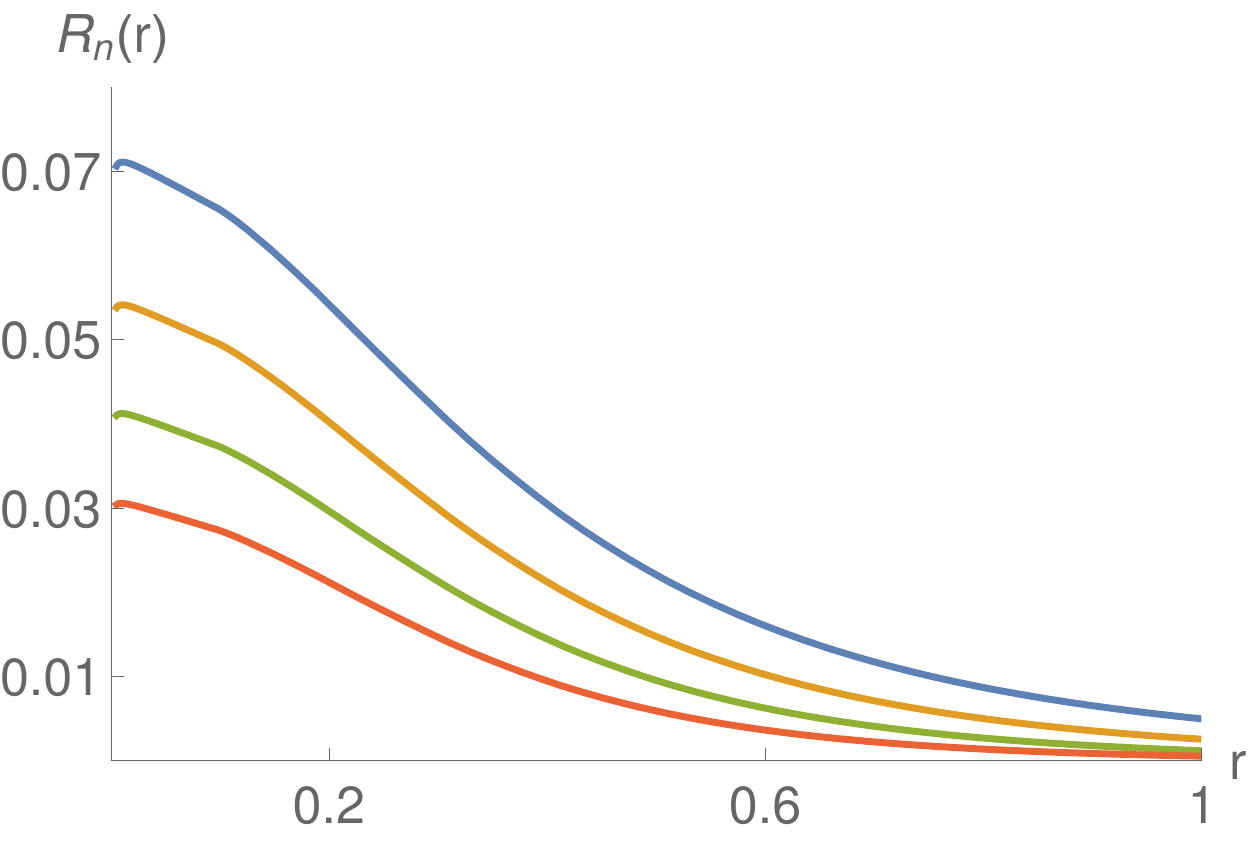} \\
\end{tabular}
\caption{The profile functions $R_{1,3,5,7}(r)$ (from top to bottom), for $\lambda=16.63$ (on the left) and $\lambda=500$ (on the right).}
\label{eigenexpansion}
\end{figure}

From the definition \eqref{nedmdef} it thus follows that
\begin{equation}
d_n^{s.c.} = -\frac{8\pi}{9}\sum_{n=1}^\infty g_{v^n} \int_0^\infty d r\,r^4 R_{2n-1}(r)\,.
\end{equation}
The functions $R_{1,3,5,7}(r)$ for the numerical solution obtained above are given in Figure \ref{eigenexpansion}.
The mode expansion method neatly indicates how all the meson tower is actually contributing to the NEDM. 
Calculating the latter including the first one, two and three modes gives
\begin{equation}
d_n^{s.c.}=
\begin{aligned}
\nonumber
\begin{cases}
& 1.09 \cdot 10^{-16}\, \theta\;e\cdot \mathrm{cm} \qquad {\rm with\ one\ mode}\,, \\
& 0.68 \cdot 10^{-16}\, \theta\;e\cdot \mathrm{cm} \qquad {\rm with\ two\ modes}\,,  \\
& 0.76 \cdot 10^{-16}\, \theta\;e\cdot \mathrm{cm} \qquad {\rm with\ three\ modes} \,,
\end{cases}
\end{aligned}
\end{equation}
to be compared to the full result (\ref{fituno}).\footnote{The numerical approximations involved in both computations do not allow us to reach a better precision with our desktop computational power.}
The first mode approximates the full result with an error of about 40\%.
This highlights the advantage of the holographic model, which allows to include the contribution of the whole tower of vector mesons.
The inclusion of the second and third modes give significant corrections.
The fourth mode is already essentially irrelevant (less than 1\% correction) for $\lambda \sim {\cal O}(10)$.
Finally, as $\lambda$ increases the higher massive vector mesons become more and more important in their contribution to the NEDM.

The value of the NEDM above is extracted from the model at leading order in $N_c$. 
The model actually allows to calculate the $1/N_c$ corrections coming from the quantization of the baryonic spectrum, providing their wave functions \cite{SS-barioni}, as reviewed in Section \ref{barquantization}.
Clearly, these do not constitute all the possible $1/N_c$ corrections.
Nevertheless, they represent important corrections to the result when extrapolating the model formulae, valid at large $N_c$, large $\lambda$, to the values $N_c=3, \lambda \sim {\cal O}(10)$.
We use the neutron wave function defined in (\ref{pnwave}). The electric current has an explicit dependence on the moduli $\rho, Z$, as can be seen from equations (\ref{eqforu}), (\ref{eqW}).\footnote{Remember that $\xi^2=(\vec x - \vec X)^2 + (z-Z)^2$.}
So, considering the full wave function rather than the classical approximation has a non-trivial effect.
Noticing that the ``semi-classical" dipole moment, as given in eq. (\ref{finaldipole}), is a function of $\rho, Z$, the NEDM is calculated as\footnote{Technically, 
we solve the differential equations numerically for a suitable grid of values of $\rho, Z$, interpolating the obtained results.}
\begin{equation}\label{nedmwf}
d_n\equiv \langle d^{s.c.}_n \rangle_{\rho,Z} \equiv \frac{\int \rho^3 R(\rho)^2\, \psi_Z^2(Z)\, d_n^{s.c.}\, d\rho\, dZ}{\int \rho^3 R(\rho)^2\,\psi_Z^2(Z)\, d\rho\,  dZ}\,.
\end{equation}
Using the standard value of parameters (\ref{parameters}), we obtain our best estimate for the NEDM value \cite{nedmshort}
\begin{equation}\label{nedm0}
d_n = 1.8 \cdot 10^{-16}\, \theta\;e\cdot \mathrm{cm}\,.
\end{equation}
The quantum $1/N_c$ correction to the semiclassical value (\ref{fituno}) is thus substantial for these phenomenological values of the parameters.

It is also known that the standard values for the parameters $\lambda, M_{KK}$ used above do not perform extremely well for baryonic observables, see e.g. \cite{SS-barioni}.
So, it is interesting to consider a different choice obtained by fitting against data such as the form factors calculated in \cite{SS-form}. 
In Appendix \ref{appendix:parameters} we give the details on how this fit is performed.
The best fit gives $\lambda=12.44, M_{KK}=790$ MeV. The value of $m_q=4.06$  MeV is taken from the GMOR relation (\ref{gmor1}) by fixing $m_\pi = 135$ MeV, and the resulting semi-classical value of the NEDM is 
\begin{equation}
d_n^{s.c.} =  2.1 \cdot 10^{-16}\, \theta\;e\cdot \mathrm{cm}\,.
\end{equation}
The difference of this value w.r.t. the one obtained with the standard values of the parameters (\ref{fituno}) highlights the importance of a proper choice of fitted observables.
For the quantum corrections we use (\ref{pnwave}) to calculate also the observables used for the fit (see \cite{SS-form}).
The parameter values obtained are $M_{KK}=785\ {\rm MeV},\ \lambda=19.38,\ m_q=3.27\ {\rm MeV}$.
Using these values, (\ref{nedmwf}) gives for the NEDM
\begin{equation}\label{nedm2}
d_n = 2.5 \cdot 10^{-16}\, \theta\;e\cdot \mathrm{cm}\,.
\end{equation}
\subsection{The electric dipole form factor}\label{sec:FF}
As we have recalled at the beginning of the present Section, the nucleon electric dipole moment is related to the dipole form factor at zero momentum $F_3(0)$. Remarkably, the WSS holographic model allows to extract the full momentum dependence of the dipole form factor. 

Working in Breit frame, where $k^{\mu}=(0,{\vec k})$, we can see, from the defining expression in eq. (\ref{defFF}) and following similar steps as in \cite{SS-form}, that the electric dipole form factor of the neutron is given by
\be
\frac{F_3(k^2)}{2M_N} = -\frac{2}{3k}\partial_k \int d^3y\, e^{-i \vec k \cdot \vec y}\kappa \langle[k(z)\partial_z W]^{z\rightarrow\infty}_{z\rightarrow-\infty}\rangle_{\rho,Z}\,,
\ee
with $\vec y \equiv \vec x -\vec X$, $k\equiv|\vec k|$.\footnote{Not to be confused with the function $k(z)$.} This formula can be also deduced from the Fourier tranform of the dipole charge distribution (\ref{dipcharge}). Since $\langle[k(z)\partial_z W]^{z\rightarrow\infty}_{z\rightarrow-\infty}\rangle_{\rho,Z}$ is a function of  $r\equiv |\vec y|$, the expression above reads
\be
\frac{F_3(k^2)}{2M_N} = -\frac{8\pi}{3}\int_0^{\infty} dr\, r^2 \left[\frac{\cos(kr)}{k^2}-\frac{\sin(kr)}{k^3r}\right] \kappa \langle[k(z)\partial_z W]^{z\rightarrow\infty}_{z\rightarrow-\infty}\rangle_{\rho,Z}\,.
\label{f3f}
\ee
Thus $F_3(k^2)$, as expected, can be expanded in even powers of $k$, around $k=0$. At $k=0$, $F_3(0)/2M_N$ precisely reproduces the NEDM as given in eq. (\ref{nedmwf}) (see (\ref{finaldipole})). Notice that in our setup with $N_f=2$ degenerate quarks, only the isovector part of the electric dipole form factor is turned on. 

The complete vector meson dominance of the dipole form factor is manifest once we implement the mode expansion for $\langle[k(z)\partial_z W]^{z\rightarrow\infty}_{z\rightarrow-\infty}\rangle_{\rho,Z}$
\be
\frac{F_3(k^2)}{2M_N} = \frac{8\pi}{3}\sum_{n=1}^{\infty}g_{v^n}\int_0^{\infty} dr\, r^2 \left[\frac{\cos(kr)}{k^2}-\frac{\sin(kr)}{k^3r}\right] \langle R_{2n-1}(r)\rangle_{\rho,Z}\,.
\label{f3m}
\ee

In order to extract the explicit functional dependence of $F_3(k^2)$ on the momentum, we need to compute the integral in eq. (\ref{f3f}). Focusing on the $k\rightarrow0$ behavior, it is easy to realize that if the function 
\be
q(r)\equiv \kappa \langle[k(z)\partial_z W(r,z)]^{z\rightarrow\infty}_{z\rightarrow-\infty}\rangle_{\rho,Z}\,,
\ee
is power-like suppressed at large $r$, the integral in (\ref{f3f}) gives generically divergent coefficients for the series expansion of $F_3(k^2)$. 
Actually, using the instanton solution found in Section \ref{sec:newsol}, we have that $q(r)\sim r^{-7}$ at large $r$. 
That solution has been found neglecting subleading corrections in the small parameters $\theta, m_q/M_{KK}$ and $\epsilon_f$ (see eq. (\ref{epsilonf})). 
In particular, working to leading order in the latter parameter, which weighs the flavor backreaction, is what justifies the fact that we have neglected the $\eta'$ mass contribution (recall that the squared Witten-Veneziano mass (\ref{mWV}) scales like $\epsilon_f$) to the equation for ${\widehat A}^{\rm{mass}}_z$ in Section \ref{sec:abfield}. 
At subleading order that contribution is generically present as it can be easily deduced starting from the effective action (\ref{effaz}). 
In order to consistently account for that, one should also include, to this order, at least also the flavor backreaction on the background (see \cite{smearedSS}). This would produce $\epsilon_f$-corrected functions $k(z)$ and $h(z)$. The 
equation of motion for ${\widehat A}_z^{\rm{mass}}$ could still possibly be solved by the ansatz ${\widehat A}_z^{\rm{mass}}= u(r)/k(z)$ with $u(r)$ now being solution of the equation\footnote{We are considering the $N_f=2$, $\varphi\sim\theta/2\ll1$ case.}
\begin{equation}
\frac{1}{r^2}\partial_r(r^2 \partial_r u(r)) - m^2 u(r)= \frac{c m_q }{\kappa}\theta (\cos \alpha +  1)\,, \label{eqforumass} 
\end{equation}
with $m=m_{WV}$ given in (\ref{mWV}). The related Green's function is now modified by the mass term 
and it reads
\begin{equation}
u_G(r,r') = \left\lbrace \begin{aligned}
& - \frac{r'}{m\, r}e^{-m\, r'} \sinh(m\, r)\,, & \qquad & r<r' \,,\\
& - \frac{r'}{m\, r}e^{-m\, r} \sinh(m\, r')\,,  & \qquad & r>r' \,,
\end{aligned}\right.
\end{equation}
to be compared with eq. (\ref{ugf}) which is obtained in the $m\rightarrow0$ limit.  The solution to (\ref{eqforumass}) is thus given by
\begin{equation}
u(r) = \frac{cm_q}{\kappa}\theta \int_0^\infty d r'\,u_G(r,r')\left(1+\cos \frac{\pi}{\sqrt{1+\rho^2/{r'}^2}}\right) \ . \label{defupiccolomass}
\end{equation}
This function closely resembles, in form, the expression for the $\eta'$ VEV obtained within the Skyrme model \cite{Dixon}. 
Crucially, $u(r)$, whose derivative enters the source term for the function $W(r,z)$ (see equation (\ref{eqW})), is now exponentially suppressed for large $r$. 
This in turn provides an exponential suppression to the function $q(r)$ at large $r$ and gives a way to regularize the computation of the form factor.

\begin{figure}[h!]
\centering
\includegraphics[scale=.6]{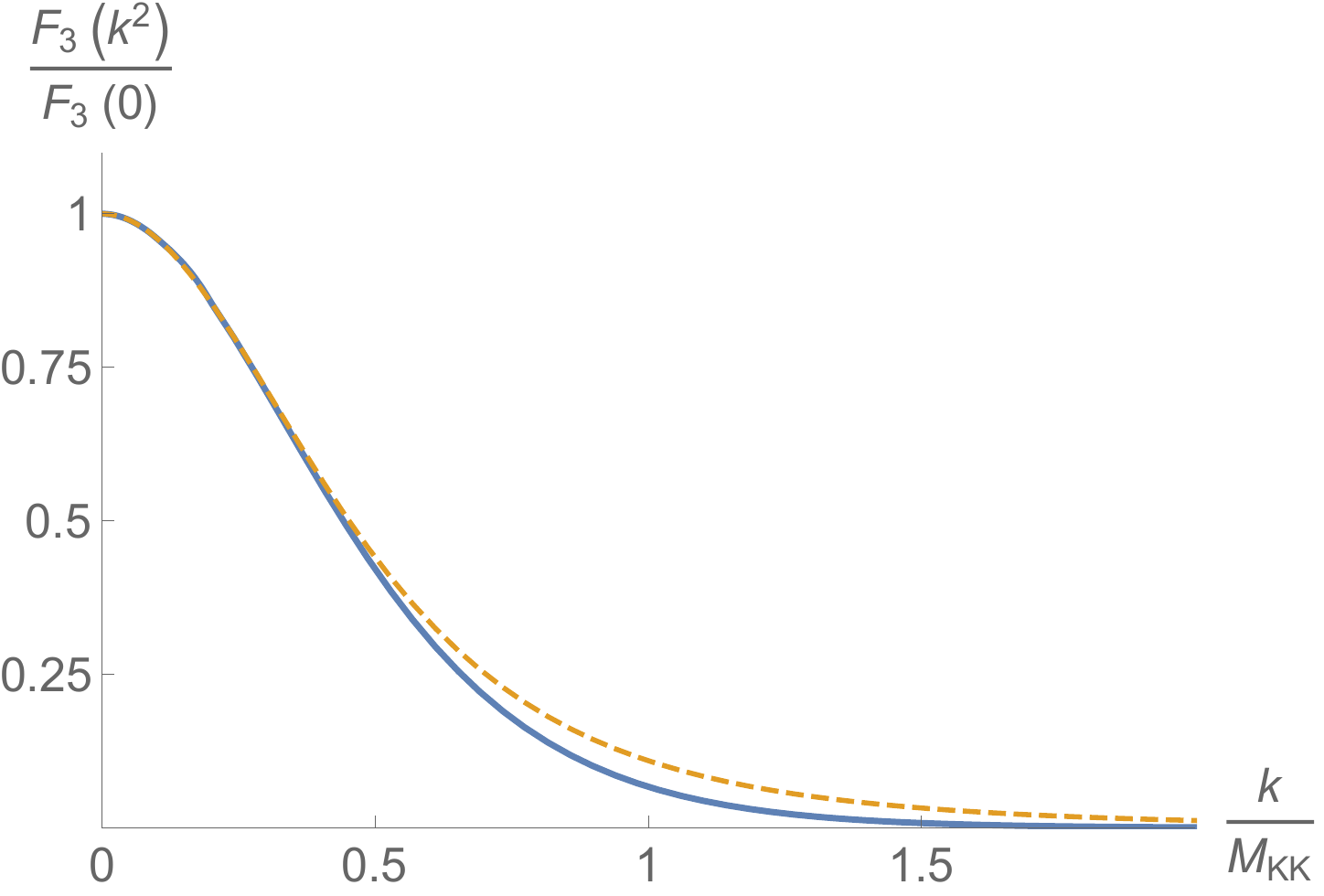}
\caption{The electric dipole form factor (solid line) and the dipole fit in (\ref{dipolefit}) (dashed line).
}
\label{figF3}
\end{figure}

We perform this computation numerically, setting $Z=Z_{cl}=0$ for simplicity (wave function corrections related to the $Z$ modulus only give small corrections to the whole result) and adopting the standard ``mesonic" choice of paramenters $N_c=3$, $\lambda=16.63$. The final outcome is the plot shown in Figure \ref{figF3}. 
Numerically, for small $k$ we find (reinserting the dependence on the scale $M_{KK}$)
\be
F_3(k^2)\approx F_3(0)\left[1 - 4 \frac{k^2}{M_{KK}^2}  + 13 \frac{k^4}{M_{KK}^4} -{\cal O}(k^6/M_{KK}^6)\right]\,.
\ee
Actually, the dipole form factor at small momenta (i.e. for $k<M_{KK}$) is fitted quite well by a dipole behavior
\be\label{dipolefit}
F_3(k^2) \approx F_3(0) \left[1-2\frac{k^2}{M_{KK}^2}\right]^{-2}\,,
\ee
just as it happens, both in QCD and in the WSS model \cite{SS-form}, for the standard electric and magnetic Sachs form factors of the nucleons. The dipole behavior is quite naturally induced in models with complete vector meson dominance, thus its occurrence in the present case is not totally surprising.  
 
For $k\gg M_{KK}$, the form factor $F_3(k^2)$ neatly deviates from the dipole behavior. Numerically, we find that it is actually exponentially suppressed with $k$. This feature, which turns out to show up also from a numerical analysis of the nucleon Sachs form factors studied in \cite{SS-form}, could be related to the very peculiar UV completion of the WSS model which, by construction, is expected to depart from perturbative QCD.


The plot in Figure \ref{figF3} indicates that the scale of momentum variation of the dipole form factor is set by $M_{KK}$. This observation can be complemented by defining, in analogy with the electric charge radius, the (isovector) electric dipole radius for the neutron
\be
\langle r_{ED}^2\rangle = - 6 \left(\frac{dF_3(k^2)}{d k^2}\right)_{k^2=0} = \frac{16}{15}\pi\, M_N\int_0^{\infty} dr\, r^6 \kappa \langle[k(z)\partial_z W]^{z\rightarrow\infty}_{z\rightarrow-\infty}\rangle_{\rho,Z}\,.
\label{r2ed}
\ee
With the parameters chosen as above we numerically get
\be
\langle r_{ED}^2\rangle \approx 48 \frac{d_n\,M_N}{M_{KK}^2}\,.
\ee
Finally, we notice that the NEDM, modified by the contribution of the Witten-Veneziano mass is now given by $d_n\equiv F_3(0)/2M_N\approx 2.6\cdot 10^{-17} \theta\,e\,cm$, which is smaller than the value reported in eq. (\ref{nedm0}). 

It is interesting to compare our findings with those obtained in chiral perturbation theory \cite{EDFFChpt}. 
There, the pion cloud dominates the physics and the scale of momentum variation of the electric dipole form factor is set by $m_{\pi}$. 
Correspondingly the dipole square radius scales like $m_{\pi}^{-2}$. These results are in line with the already noticed differences between the large $N_c$ approach and the chiral one. 
\section{The CP-breaking pion-nucleon coupling}
\setcounter{equation}{0}
\label{gpNNsec}
As we have previously discussed, there are essentially two different approaches to compute the NEDM in phenomenological models: one is based on the Skyrme model \cite{Dixon}, the other one on chiral perturbation theory \cite{CdVVW-NEDM}. 
This last method involves the computation of the CP breaking cubic coupling $\overline{g}_{\pi NN}$ between baryons and pions. 
As we will show in the following, within the limiting regimes where the holographic computations have been performed, this coupling turns out to be zero, at leading order in the $1/N_c$ expansion, in the Witten--Sakai--Sugimoto model. 
This statement actually allows for a  CP breaking coupling which is subleading in the $1/N_c$ expansion. 

We will give two different proofs of this claim; the first one, based on the form factor formalism, is given below.
\subsection{The axial form factors} 
In the $\theta=0$ case, the matrix element for the axial current between nucleon states 
\begin{equation}
\langle p', s' | \mathcal{J}_A^{\mu,C}(0)|p,s\rangle = (2\pi)^{-3} \frac{(\tau^C)_{I_3'I_3}}{2}\overline{u}(p',s') \Gamma^{(C)}_\mu(k^2) u(p,s)\,,
\label{axfor}
\end{equation} 
where $C=0,1,2,3$ and $\tau^0 = \unit_2$, is given in terms of the following expansion
\begin{equation}
\Gamma^{(C)}_\mu(k^2) = i\gamma_5\gamma_\mu g_A^{(C)}(k^2) + \frac{1}{2M_N}k_\mu \gamma_5 g_P^{(C)}(k^2)\,.
\end{equation}
The form factors $g_{A,P}^{(C)}(k^2)$ are not independent in the massless theory because current conservation imposes
\begin{equation}
g_P^{(C)} = \frac{4M_N^2}{k^2}g_A^{(C)}\,.
\label{noecons}
\end{equation}
However when the quark masses are non zero $\partial_\mu \mathcal{J}_A^\mu \neq 0$ and this relation no longer holds.

When we allow for a strong CP violation also other terms may arise. These look as the previous ones, without the $\gamma_5$ insertion. 

\begin{figure}[H]
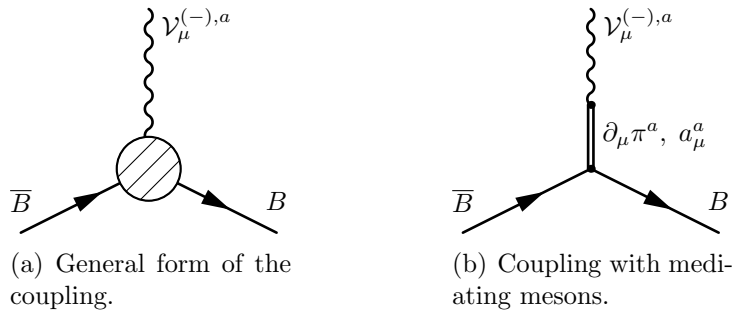

\centering
\subfigure[General form of the coupling.\label{cubic}]{\includegraphics[scale=1]{gpiNN2.mps}}\hspace{2cm}
\subfigure[Coupling with mediating mesons.
\label{cubic2}]{\includegraphics[scale=1]{gpiNN1.mps}}
\caption{Cubic vertex corresponding to the matrix element $\langle p',s'|\mathcal{J}_A^\mu|p,s\rangle$ between baryonic states.}
\end{figure}

The matrix element (\ref{axfor}) describes a cubic interaction between two nucleon states and the external source coupled to the field ($\mathcal{V}^{(-)}_\mu$ in this case), so it actually computes diagrams of the type in \figurenamev \,\ref{cubic}. 
However we can imagine that the mesons are mediating this interaction and we already know their coupling \eqref{SJ} with the external field. 
Hence we find something of the form shown in \figurenamev\,\ref{cubic2}.
Diagrams such as that in \figurenamev\,\ref{cubic2} arise from effective interactions between mesons and nucleons described by
\begin{equation}
\begin{aligned}
\mathcal{L}_\mathrm{eff} &= \sum_{n\geq 1}\left(\widehat{g}_{a^nNN}\widehat{a}^n_\mu \overline{N} i \gamma_5 \gamma^\mu \frac{\unit_2}{2}N + g_{a^nNN} a^{n\,c}_\mu \overline{N} i \gamma_5 \gamma^\mu \frac{\tau^c}{2}N\right) + \\ &\;\;\;+ 2i\left(\widehat{g}_{\pi NN}\widehat{\pi} \overline{N}  \gamma_5  \frac{\unit_2}{2}N + g_{\pi NN}\pi^{c}  \overline{N}  \gamma_5   \frac{\tau^c}{2}N\right)\,. 
\end{aligned}
\end{equation}
This is only the CP conserving part: the CP breaking one is the same but without $i\gamma_5$. For example the coupling $\overline{g}_{\pi NN}$ appears as
\begin{equation}
\mathcal{L}_\mathrm{eff\cancel{CP}} =  2\left(\widehat{\overline{g}}_{\pi NN}\widehat{\pi} \overline{N}      \frac{\unit_2}{2}N + \overline{g}_{\pi NN}\pi^{c}  \overline{N}     \frac{\tau^c}{2}N\right) +(\mbox{\footnotesize vector mesons})\,.
\end{equation}
Since the $\eta'$ is very massive we expect that the low energy physics is dominated by the isovector coupling $\overline{g}_{\pi NN}$. 

Let us proceed to write down the amplitude of \figurenamev \ref{cubic2} retaining only the CP conserving terms of $\mathcal{L}_\mathrm{eff}$ plus the CP breaking $\overline{g}_{\pi NN}$. The propagators can be read from the kinetic terms for the mesons \eqref{kineticmesons}, namely a Proca propagator and a scalar propagator (not massless in this case because the pions acquire a mass)
\begin{equation}
\langle p',s'|\mathcal{J}_A^{\mu\,C}|p,s\rangle = \sqrt{2p^0}\sqrt{2{p'}^0}\overline{u}(\mvec{p}',s')\,\left[\mathscr{A}^{\mu\,C}\right]\, u(\mvec{p},s) \,,
\end{equation}
where
\begin{equation}
\footnotesize
\begin{aligned}
&\mathscr{A}^{\mu\,C} =\left[
\frac{\delta^{C0}\delta_{I_3'I_3}}{2}\left(i\gamma_5\gamma_\nu \sum_{n\geq 1}\frac{\eta^{\mu\nu}+k^\mu k^\nu /\lambda_{2n}}{k^2+\lambda_{2n}}g_{a^n}\widehat{g}_{a^nNN} + 2k^\mu f_\pi (\gamma_5\widehat{g}_{\pi NN} - i\widehat{\overline{g}}_{\pi NN} ) \frac{1}{k^2+m_\pi^2}
\right)\right.\\&+\left.
\frac{\delta^{Ca}(\tau^a)_{I_3'I_3}}{2}\left(i\gamma_5\gamma_\nu \sum_{n\geq 1}\frac{\eta^{\mu\nu}+k^\mu k^\nu /\lambda_{2n}}{k^2+\lambda_{2n}}g_{a^n}g_{a^nNN} + 2k^\mu f_\pi (\gamma_5g_{\pi NN} - i\overline{g}_{\pi NN} )  \frac{1}{k^2+m_\pi^2}\right)\right] \\&  + (\mbox{\footnotesize $\cancel{CP}$ vector mesons}).
\end{aligned}
\end{equation}
The Lorentz tensor structure can be readily compared with the general form factor: take for instance only the isoscalar CP conserving part
\begin{equation}
\begin{aligned}
\widehat{g}_A(k^2)&= \sum_{n\geq 1}\frac{g_{a^n}\widehat{g}_{a^n NN}}{k^2+\lambda_{2n}} \,,\\
\widehat{g}_P(k^2) &= 2 M_N \frac{2 f_\pi \widehat{g}_{\pi NN}}{k^2 + m_\pi^2} - 4M_N^2 \sum_{n\geq 1}\frac{g_{a^n}\widehat{g}_{a^nNN}}{\lambda_{2n}}\frac{1}{k^2+\lambda_{2n}}\,.
\end{aligned}
\end{equation}
It is worth noticing the following feature: the relation \eqref{noecons}, that holds only when $m_\pi^2 = 0$, implies that the residue at the pole of $g_P$ in $k^2=0$ is proportional to $g_A$, more precisely
\begin{equation}
g_A(0) = \frac{f_\pi g_{\pi NN}}{M_N}\,.
\end{equation}
This is known as the Goldberg--Treitman relation. However when the pion is massive the pole of $g_P$ is displaced and the conservation of the axial current is broken also at the classical level, so this relation no longer holds.

In order to have a non zero $\overline{g}_{\pi NN}$ in the theory, we would need a term in the form factor proportional to 
\begin{equation}
(\tau^a)_{I_3'I_3} \delta_{s's} k^\mu\,,
\end{equation}
which means in the current a term like
\begin{equation}
J^{\mu\,a}_V = I^a \frac{\partial}{\partial X^\mu}f(Z,\mvec{x}-\mvec{X}) \;,\quad \langle n_z,n_\rho | f(Z,\mvec{x}-\mvec{X})|n_z,n_\rho\rangle \neq 0\,.
\end{equation}
The derivative with respect to $X$ can be traded for a derivative with respect to $x$, which in Fourier transform yields $k^\mu$. The isospin operator defined in (\ref{spiniso}) is explicitly given by
\begin{equation}
I^a= -4i\pi^2 \kappa \rho^2 \Tr(\tau^a \bvec{a}\,\dot{\bvec{a}}^{-1})\,.
\end{equation}
Clearly this term, which contains an $\dot{\bvec{a}}$, can only appear in the field $F_{0z}^a$, as a result of the modified Gauss Law constraint (\ref{mgl}). Indeed we have
\begin{equation}
F'_{0z} = - V (D_z \Phi) V^{-1} - V(D_z A_0)V^{-1}\,.
\end{equation}
The relevant term is the first one, indeed
\begin{equation}
V \partial_z \Phi V^{-1} \underset{z\gg 1}{\sim} i \bvec{a}\dot{\bvec{a}}^{-1} \sum_{n=1}^\infty c_n(r) \partial_z \psi_n(z)\,,
\end{equation}
as we argued in \eqref{Phiasy}. In the axial current, as it is easy to see from the definition, only the terms with \emph{even} $n$ contribute. 
Clearly $c_n$ for even $n$ has to vanish for $\theta = 0$, as a consequence of CP conservation and as it can also be inferred by \eqref{cennemasszero} computed at $Z = 0$. 
Now the argument is simple:  since $c_n$ solves an equation $D_M^2 \Phi = 0$, it gets contribution \emph{only} from the non Abelian fields, the CP breaking part of those is proportional to $\cos \frac{\theta}{2}$. 
The only way for $c_n$ to vanish at $\theta \to 0$ is to be identically zero.
\subsection{A more direct argument}
Let us notice, as in \cite{ANWMassSkyrme}, that a possible way to define the $g_{\pi NN}$ coupling is to take the large $r$ behavior of the pion expectation value in a nucleon state
\begin{equation}
\langle N | \pi^a | N \rangle \approx - \frac{g_{\pi NN}}{8\pi M_N}\frac{m_\pi x^i}{r^2} e^{-m_\pi r}\langle \sigma^i \tau^a \rangle\,.
\end{equation}
In the same fashion, including the CP breaking contribution gives
\begin{equation}
\langle N | \pi^a | N \rangle \approx - \frac{g_{\pi NN}}{8\pi M_N}\frac{m_\pi x^i}{r^2} e^{-m_\pi r}\langle \sigma^i \tau^a \rangle   - \frac{\overline{g}_{\pi NN}}{8\pi  }\frac{m_\pi  }{r^2} e^{-m_\pi r}\langle  \tau^a \rangle\,. 
\end{equation}
In our model this expectation value becomes
\begin{equation}
\langle N | \bvec{\pi} | N \rangle = \langle N | \int dz \,A_z'(x^M;\bvec{a})|N\rangle\,,
\end{equation}
where the moduli dependence has been explicitly indicated. 
There are essentially two reasons why this does not give a CP breaking contribution to the $\pi N N$ coupling. 
The first one is analogous to the one above: $A_z$, being a non Abelian field, contains contributions proportional to $\cos \frac{\theta}{2}$, which cannot automatically vanish in the limit $\theta \to 0$ unless $\overline{g}_{\pi NN}$ is identically zero. Secondly, we would expect a precise moduli dependence from $A_z$, namely $A_{z,\,\cancel{CP}}  \sim\bvec{a}\dot{\bvec{a}}^{-1}$.
On the contrary, we have a dependence
\begin{equation}\label{924}
A_{z,\,\cancel{CP}}\sim \bvec{a}( \mvec{x}\cdot \mvec{\tau}) {\bvec{a}}^{-1}\,.
\end{equation}
This can be explicitly checked by the solution given in Section \ref{solnonabelianApp}, but there is no need to do it since the one in (\ref{924}) is the only combination compatible with the spin--isospin symmetry with no time derivatives. This dependence gives precisely the CP conserving behavior $\langle \sigma^i \tau^a \rangle$.
 
\section{Conclusions} 

In this paper we have studied effects of the $\theta$ parameter in the Witten--Sakai--Sugimoto model \cite{witten,SS1}, the top-down holographic theory closest to QCD.
The (small) quark mass needed to make the $\theta$ parameter physical has been introduced by means of world-sheet instantons \cite{AK,Hashimotomass}.

Let us recapitulate our main results. To begin with, we have studied the vacuum structure at finite $\theta$, showing that it is identical to that of QCD, as derived from the chiral Lagrangian \cite{WittenLargeNChi}.
Then, we have analyzed the baryon spectrum, arguing that the $\theta$ parameter affects it only at subleading order (${\cal O}(\theta^2)$).
Moreover, the existing solitonic solutions corresponding to baryons have been extended to include the leading quark mass and $\theta$ parameter corrections.
We have reviewed and discussed in detail the results in \cite{nedmshort} for the neutron electric dipole moment:
we have extracted a value of the NEDM which is of the same order of magnitude as existing results in the literature based on effective models;
we have discussed the dependence of the NEDM and the associated charge distribution on the theory parameters;
exploiting the advantage of the holographic model on the effective theories for QCD, we have analyzed the dependence of the NEDM on higher vector mesons, showing that the first few modes are important to obtain the result at percent accuracy level. Moreover, we have presented a novel study of the full electromagnetic dipole form factor. Finally, we have argued that the CP-violating pion-nucleon coupling constant is subleading in the $1/N_c$ expansion.

Along the way, we have also pointed out a Horava-Witten-like solution to the anomalous Bianchi identity in the WSS model, which as far as we know was not present in the literature.

Given the qualitative and quantitative success of the WSS model in comparing with phenomenology, it is certainly worth extending the results of this paper on the $\theta$ dependence of QCD physics.
An obvious generalization concerns the calculation of the NEDM with three quarks of different masses.
But it would also be worth studying nuclear observables in the same setting.
 
\vskip 15pt \centerline{\bf Acknowledgments} \vskip 10pt \noindent We are grateful to Francesco Becattini, Claudio Bonati, Massimo D'Elia, Jordy de Vries, Luca Martucci, Haralambos Panagopoulos, Domenico Seminara, Shigeki Sugimoto and Ettore Vicari for relevant comments and discussions. FB and ALC would like to thank the organizers of the workshops ``Applications of AdS/CFT to QCD and condensed matter physics'', CRM Montreal and ``Quantum Information in String Theory and Many-body Systems'', YITP Kyoto for their hospitality during the completion of this paper. ALC is partly supported by the Florence Univ. grant ``Fisica dei plasmi relativistici: teoria e applicazioni moderne".\\
\appendix
\section{Meson sector}
\label{mesons}
\setcounter{equation}{0}
In this appendix we give a brief review of the holographic description of mesons in the WSS model \cite{SS1}. Let us consider the Yang-Mills part of the D8-brane effective action (\ref{SYM}) setting $N_f=1$ for the moment:
\begin{equation}
S=-\frac{\kappa}{2}\int d^4x d z\, \left(\frac{1}{2}h(z)  \,  \mathcal{F}_{\mu\nu}\mathcal{F}^{\mu\nu} +   k(z)   \mathcal{F}_{\mu z}\mathcal{F}^\mu_{\;\; z}\right)\,.
\label{actT}
\end{equation}
The expansions (\ref{expA}) for the fields $\mathcal{A}_\mu$ and $\mathcal{A}_z$ imply that
\begin{equation}
\begin{aligned}
\mathcal{F}_{\mu\nu}(x^\mu,z) &= \sum_n\left(\partial_\mu B^{(n)}_\nu(x^\mu) - \partial_\nu B^{(n)}_\mu (x^\mu)\right) \psi_n(z) \\&\equiv \sum_n \mathcal{F}_{\mu\nu}^{(n)}(x^\mu)\psi_n(z)\,,\\
\mathcal{F}_{\mu z}(x^\mu,z) &= \sum_n\left(\partial_\mu \varphi^{(n)}(x^\mu)\phi_n(z) - B_\mu^{(n)}(x^\mu)\psi'_n(z)\right)\,.
\end{aligned}
\end{equation}
The functions $\psi_n$ and $\phi_n$ will be discussed in a moment and $\psi'$ means $\partial_z\psi$. 

Let us first set the $\varphi^{(n)}$ to zero. The action (\ref{actT}) then becomes
\begin{equation}
S=-\frac{\kappa}{2}\int d^4x d z\, \left(\frac{1}{2}h(z)  \,  \sum_{m,n}\mathcal{F}^{(n)}_{\mu\nu}\mathcal{F}^{\mu\nu\;(m)}\psi_n\psi_m+   k(z)   \sum_{m,n}B_\mu^{(n)}B^{\mu\;(m)}\psi'_n\psi'_m\right)\,.
\end{equation}
Imposing the conditions
\begin{equation}
\kappa \int d z\,h(z)\psi_n(z)\psi_m(z) = \delta_{mn}\;,\qquad
\kappa \int d z\,k(z)\psi'_n(z)\psi'_m(z) = \lambda_n\delta_{mn}\label{norpsi}\,,
\end{equation}
and integrating by parts (the $\psi_n$ approach zero for $z\to\pm \infty$ because of the normalization) we get the eigenvalue equations (\ref{eqforpsi}). When the $\lambda_n$ are ordered such that $\lambda_1<\lambda_2<\cdots$ it can be shown that $\psi_n$ has positive (negative) parity for $n$ odd (even) under the transformation $z\to -z$. 
The transformation $(x^\mu,z)\to (-x^\mu,-z)$ is interpreted as the holographic equivalent of the parity transformation in the boundary theory.

If we use the above relations we find a Proca action for the fields $B_\mu^{(n)}$, with masses $m_n^2 = \lambda_n$  (in units $M_{KK}=1$). These fields are interpreted as the vector mesons of the field theory. 

Now it is easy to include scalar fields $\varphi^{(n)}$ as well. As before, let us require
\begin{equation}
\kappa \int d z\, k(z) \phi_n\phi_m = \delta_{mn}  \label{norphi}\,.
\end{equation}
We can take $\phi_n$ to be just $\phi_n = \psi'_n/\sqrt{\lambda_n}$. However there is a zero mode
\begin{equation}
\phi_0 = \frac{1}{\sqrt{\kappa\pi}}\frac{1}{k(z)}\,,
\end{equation}
which is orthogonal to all the $\psi_n'$. In fact the $\psi_0$ mode whose derivative would be $\phi_0$ is proportional to $\arctan(z)$: this is not normalizable by means of the integral \eqref{norpsi}. 
The field $\phi_0$, instead, has the correct normalization with respect to \eqref{norphi}. The ${\cal F}_{\mu z}$ field strength is rewritten as
\begin{equation}
\mathcal{F}_{\mu z} = \partial_\mu\varphi^{(0)}  \frac{1}{\sqrt{\kappa\pi}}\frac{1}{k(z)} + \sum_{n\geq 1}\left(m_n^{-1}\partial_\mu \varphi^{(n)}-B_\mu^{(n)}\right)\,.
\end{equation}
The gauge transformation $B_\mu^{(n)}\mapsto B_\mu^{(n)}+m_n^{-1}\partial_\mu \varphi^{(n)}$ can be used to eliminate all the $\varphi^{(n)}$ with $n\geq 1$ from the theory; the $\varphi^{(0)}$ mode survives instead. 
All in all we get the following four dimensional action
\begin{equation}
\begin{aligned}
S &=-\kappa\int d^4x \, \left[\sum_{n\geq 1}\left(\frac{1}{4}  \mathcal{F}^{(n)}_{\mu\nu}\mathcal{F}^{\mu\nu(n)}+  \frac{1}{2} m_n^2 B_\mu^{(n)}B^{\mu(n)}\right)+  \frac{1}{2}\partial_\mu \varphi^{(0)} \partial^\mu \varphi^{(0)}\right]\,.
\end{aligned} \label{kineticmesons}
\end{equation}
The massless field $\varphi^{(0)}$ is associated to the mode $\psi_0 \propto \arctan z$ which is an odd function: it is thus a pseudoscalar field and we interpret it as the pion field, which is the Goldstone boson of the spontaneous chiral symmetry breaking.
 
A similar analysis can be performed to include also the massive scalar mesons: they arise as fluctuations of the embedding of the D8-branes in the background.

It is possible to generalize the pion effective action to $N_f>1$ flavors. The gauge fields $\mathcal{A}_\alpha$ approach zero at $z\to \pm \infty$, but we still have a residual gauge symmetry for functions that approach constants as $z\to \pm\infty$. This residual gauge symmetry is interpreted as the global symmetry of the boundary theory $G^\mathrm{glob} = U(N_f)_L\times U(N_f)_R$
\begin{equation}
\begin{aligned}
&\;\; \mathcal{A}_\alpha(x^\mu,z) \mapsto g(x^\mu,z) \mathcal{A}_\alpha(x^\mu,z) g^{-1}(x^\mu,z) - i g(x^\mu,z)\partial_\alpha g^{-1}(x^\mu,z)\,, \\
&\lim_{z\to \pm \infty} g(x^\mu,z) = g_\pm\;,\quad 
\lim_{z\to \pm \infty} \partial_\alpha g(x^\mu,z) = 0\,,
\qquad (g_+,g_-)\in G^\mathrm{glob}\,.
\end{aligned}
\end{equation}
We know that the Wilson line from a point $x_A$ to a point $x_B$ transforms with the gauge function evaluated at the two points. If in particular we consider the path given in (\ref{pionm}), then the transformation law is $\U \mapsto g_+ \U g_-^{-1}$. This is precisely the transformation law for the pion matrix. We can thus define the pion field as
\begin{equation}
\U(x^\mu) \equiv \exp \left(\frac{2i}{f_\pi}\pi^a(x^\mu) T^a\right)\,,
\end{equation}
where $T^a$ are $U(N_f)$ generators normalized to $\Tr(T^aT^b)=\frac{1}{2}\delta_{ab}$ and $f_\pi$ is the pion decay constant.\footnote{This implies that the decay constant for the singlet $f_S$ equals $f_\pi$: this is true only if, as in the present case, we work up to first order in $N_f/N_c$ in the $N_c\to \infty$ limit.}

Let us now move to a gauge where $\mathcal{A}_z=0$. This is done using a gauge function $g$ defined as
\begin{equation}
g(x^\mu,z)=\mathcal{P} \exp\left(i \int_0^z d z'\, \mathcal{A}_z(x^\mu,z')\right)\,.
\end{equation}
Under this gauge transformation also $\mathcal{A}_\mu$ changes, but now the requirement $\mathcal{A}_\mu\to 0$ as $z\to \pm\infty$ is not satisfied anymore (this is not a problem since we are dropping the CS terms). We obtain in fact
\begin{equation}
\begin{aligned}
\mathcal{A}_z & \mapsto g \mathcal{A}_z g^{-1} - i g \partial_z g^{-1} = 0\,,\\
\mathcal{A}_\mu &\mapsto g \mathcal{A}_\mu g^{-1} - i g \partial_\mu g^{-1} \;\underset{z\to\pm\infty}{\longrightarrow}
-i\xi_\pm \partial_\mu \xi_\pm^{-1}\,,
\end{aligned} 
\end{equation}
where we have defined $\xi_\pm$ as the limit for $z\to\pm \infty$ of $g$. 
As a result, the expansion in terms of the $\psi_n$ is not valid anymore (because all those functions approach zero): we have to include the non--normalizable zero mode $\psi_0 =\frac{2}{\pi}\arctan{(z)}$. This has the limit $\psi_0 \to \pm 1$ as $z\to \pm \infty$. The following expansion matches the limit properly (we have defined $\psi_\pm(z) = -\frac{i}{2}(1\pm \psi_0(z))$)
\begin{equation}
\begin{aligned}
\mathcal{A}_\mu(x^\mu,z) &= \xi_+\partial_\mu \xi_+^{-1} \,\psi_+(z) + \xi_-\partial_\mu \xi_-^{-1} \,\psi_-(z)\,. 
\end{aligned}
\end{equation}
There is a further residual gauge symmetry given by all the functions $h(x^\mu)$ that are independent on $z$: it is possible to impose $\xi_- =1$, but in this case $\xi_+$ becomes exactly the inverse of the pion matrix: $\U^{-1}$
\begin{equation}
\mathcal{A}_\mu(x^\mu,z) = \U^{-1} \partial_\mu \U \,\psi_+\,.
\end{equation}
We can finally substitute these fields in the DBI action. The field strengths read
\begin{equation}
\begin{aligned}
\mathcal{F}_{\mu\nu} &= -i [\U^{-1} \partial_\mu \U,\U^{-1} \partial_\nu \U]\psi_+\psi_-\,,\\ 
\mathcal{F}_{\mu z} &= \U^{-1} \partial_\mu \U \psi_+'\,.
\end{aligned}
\end{equation}
Using the normalization conditions given at the beginning of this section we find
\begin{equation}
S = -\kappa \int d^4x\,\Tr\left(a\,(\U^{-1} \partial_\mu \U)^2 + b\, ([\U^{-1} \partial_\mu \U,\U^{-1} \partial_\nu \U])^2\right)\,,
\end{equation}
where $a$ and $b$ are constants given by
\begin{equation}
a=  \int d z\,k(z)(\psi_+')^2 = \frac{1}{\pi}\;,\quad b =    \int d z \frac{1}{2} h(z)(\psi_+\psi_-)^2 = -\frac{1}{2\pi^4}\cdot 15.25\ldots
\end{equation}
The constant $15.25\ldots$ is the integral $\int d z \frac{1}{1+z^2}\big(\frac{\pi^2}{4}-\arctan^2 (z)\big)^2$. 

We see that we have obtained the Skyrme model (see \cite{ZB} for a review) with parameters (\ref{fpaie}).
\section{The $C_7$ and $\tilde F_2$ action}
\label{applu}
\setcounter{equation}{0}
Let us consider the following off-shell action\footnote{We are grateful to Luca Martucci for a relevant discussion about this section.} 
\be\label{lulu}
S_l = -\frac{1}{4\pi (2\pi l_s)^6} \int \tilde F_2 \wedge ^{\star}\tilde F_2 + \frac{1}{2\pi} \int C_7 \wedge ({\rm Tr} {\cal F}\wedge \omega_y - d \tilde F_2) \, , 
\ee
where $\int dy\,\omega_y =1$.
The three fields $\tilde F_2, C_7, {\cal F}$ are all independent.

The equation of motion for $\tilde F_2$ gives the usual duality relation\footnote{Note that we could start with a plus sign for the term $d \tilde F_2$ in the action; this just amounts to a different convention of the sign of the Hodge dual of $\tilde F_2$.} 
\begin{equation}\label{hodge}
^{\star}\tilde F_2 =-(2\pi l_s)^6 dC_7\,, 
\end{equation}
which gives the on-shell action (remembering that $ ^{\star}(^{\star}\tilde F_2) = - \tilde F_2$)
\be
S_{(1)} = -\frac{1}{4\pi} (2\pi l_s)^6 \int dC_7 \wedge ^{\star}dC_7 + \frac{1}{2\pi} \int C_7 \wedge  {\rm Tr}{\cal F}\wedge \omega_y \, , 
\ee
in terms of the well-defined field $C_7$, supplemented by the Hodge duality relation (\ref{hodge}).

Alternatively, the equation of motion for $C_7$ gives 
\begin{equation}\label{modb}
d \tilde F_2={\rm Tr}{\cal F}\wedge \omega_y\,, 
\end{equation}
which gives the on-shell action used in \cite{SS1}
\be
S_{(2)} = -\frac{1}{4\pi (2\pi l_s)^6} \int \tilde F_2 \wedge ^\star \tilde F_2  \, , 
\ee
supplemented by the modified Bianchi for $\tilde F_2$ (\ref{modb}).

Working directly with the action (\ref{lulu}), we see that the modified Bianchi are not imposed but actually arise among the equations of motion
\bea \label{first}
&& ^{\star}\tilde F_2 =-(2\pi l_s)^6 dC_7 \,,\\
&& d \tilde F_2 = {\rm Tr}{\cal F}\wedge \omega_y \,, \label{second} \\
&& \frac{\delta S_l}{\delta {\cal A}} =  \frac{1}{2\pi}\int dC_7 \wedge \omega_y \,. \label{third}
\eea
Obviously eq. (\ref{first}) implies
\be \label{fourth}
d ^{\star} \tilde F_2=0\,.
\ee
\section{Alternative choice of parameters}\label{appendix:parameters}
\setcounter{equation}{0}
The standard choice for the parameters in the WSS model, $M_{KK}=949$ MeV, $\lambda=16.63$, is fixed to reproduce the mass of the $\rho$ meson and the pion decay constant \cite{SS1}.
This choice performs very well against phenomenology for mesonic observables.
Nevertheless, it is known to produce large discrepancies when applied to some baryonic observables, e.g. the mass spectrum \cite{SS-barioni}.
For this reason, in order to provide an estimate of the NEDM, it is perhaps more suitable to obtain the parameters of the model by fitting directly baryonic observables.
In particular, in \cite{SS-form} a number of observables have been calculated, which are quite close to the NEDM in nature.
Here we extract the values of the parameters $M_{KK}, \lambda$ from the best fit of these observables.

To be more precise, we consider the following observables:\footnote{Cfr. the first table in Section 5 of \cite{SS-form}.} the mean squared radius of the isoscalar state and 
of the excited states, the mean charge squared radius of the isovector state (proton)
and of the excited states,\footnote{We do not consider the (ground state) neutron because the model result is automatically null.}  the axial radius, the isovector and isoscalar $g$-factors,\footnote{In this context it is more appropriate to consider these observables rather than their combinations,
namely the neutron and proton magnetic moments, since they are of different order in $N_c$.} the axial and $\pi NN$ couplings.\footnote{We do not consider the $\rho NN$ coupling
since the experimental value is not fixed with sufficient precision.}
We perform the best fit of our two parameters by minimizing 
\begin{equation}
\chi = \frac{1}{n-2} \sum_{i=1}^{n} \left(\frac{O_i^{(m)}-O_i^{(e)}}{O_i^{(m)}}\right)^2 \,,
\end{equation}
where $n$ is the number of fitted observables and $O_i^{(m)}, O_i^{(e)}$ are the values of the observables provided by the model and by experiments.\footnote{We weight with the model values
rather than the errors because the experimental ones are tiny, while the theoretical ones are difficult to estimate reliably.}
By using the leading $N_c$, leading $\lambda$ results in the model (i.e. not considering the full wave function of the baryons), the result is\footnote{Note
that a very similar value for $\lambda$ was advocated in \cite{rebhan} to have a good fit of the ratio of the $\rho$ meson mass and the (square root of) the string tension.}
\begin{equation}\label{parameters1}
M_{KK}=790\ {\rm MeV},\qquad \lambda=12.44.
\end{equation}
Including the baryon wave functions gives instead
\begin{equation}\label{parameters2}
M_{KK}=785\ {\rm MeV},\qquad \lambda=19.38.
\end{equation}

\end{document}